%% file: jeffers_vsini.tex
\documentclass[oldversion]{aa} 
\usepackage{aalongtable} 
\usepackage{amsmath}
\usepackage{amssymb}
\usepackage{natbib}
\bibpunct{(}{)}{;}{a}{}{,}
\usepackage{graphicx}
\usepackage{txfonts}
\usepackage[english]{babel}
\usepackage{hyperref}
\usepackage{longtable}
\def\imagetop#1{\vtop{\null\hbox{#1}}}

\hypersetup{colorlinks=true, linkcolor=blue, citecolor=blue, urlcolor=blue}

\begin{document}

\title{CARMENES input catalogue of M dwarfs}
\subtitle{III. Rotation and activity from high-resolution spectroscopic observations\thanks{Based on observations made at the Calar Alto Observatory, Spain, the European Southern Observatory, La Silla, Chile and McDonald Observatory, U.S.A.}}

\titlerunning{CARMENES: High-resolution spectroscopy}

\author{
        S.~V.~Jeffers\inst{1}
        \and
        P.~Sch\"{o}fer\inst{1}
        \and
        A.~Lamert\inst{1}
        \and
        A.~Reiners\inst{1}      
        \and
        D.~Montes\inst{2}
        \and
        J.~A.~Caballero\inst{3,4}
        \and
        M.~Cort\'es-Contreras\inst{2}
        \and
        C.~J.~Marvin\inst{1}
        \and
        V.~M.~Passegger\inst{1}
        \and
        M.~Zechmeister\inst{1}
        \and
        A.~Quirrenbach\inst{3}
        \and
        F.~J.~Alonso-Floriano\inst{2,5}
        \and
        P.~J.~Amado\inst{6}
        \and
        F.~F.~Bauer\inst{1}
        \and 
        E.~Casal\inst{6}
        \and
        E. Diez Alonso\inst{2}
        \and
        E.~Herrero\inst{7}
        \and
        J.~C.~Morales\inst{7}
        \and
        R.~Mundt\inst{8}
        \and
        I.~Ribas\inst{7}
        \and
        L.~F.~Sarmiento\inst{1}
                }


\institute{Institut f\"{u}r Astrophysik, Georg-August-Universit\"{a}t, Friedrich-Hund-Platz 1, 37077 G\"{o}ttingen, Germany
\and
Departamento de Astrof\'{\i}sica y Ciencias de la Atm\'osfera, Facultad de Ciencias F\'{\i}sicas, Universidad Complutense de Madrid, 28040 Madrid, Spain
\and
Landessternwarte, Zentrum f\"ur Astronomie der Universit\"at Heidelberg, K\"onigstuhl 12, 69117 Heidelberg, Germany
\and
Departamento de Astrof\'isica, Centro de Astrobiolog\'ia (CSIC--INTA), PO~Box~78, 28691 Villanueva de la Ca\~nada, Madrid, Spain
\and
Leiden Observatory, Universiteit Leiden, PO Box 9513, NL-2300 RA Leiden, The Netherlands 
Instituto de Astrof\'isica de Andaluc\'ia (CSIC), Glorieta de la Astronom\'ia s/n, 18008 Granada, Spain 
\and
Institut de Ci\`encies de l'Espai (IEEC-CSIC), Can Magrans s/n, Campus UAB, 08193 Bellaterra, Spain
\and
Max-Planck-Institut f\"ur Astronomie, K\"onigstuhl 17, 69117 Heidelberg, Germany 
}

   \date{\today}

 
  \abstract  
   {CARMENES is a spectrograph for radial velocity surveys of M dwarfs with the aim of detecting Earth-mass planets orbiting in the habitable zones of their host stars.  To ensure an optimal use of the CARMENES Guaranteed Time Observations, in this paper we investigate the correlation of activity and rotation for approximately 2200 M dwarfs, ranging in spectral type from M0.0\,V to M9.0\,V.  We present new high-resolution spectroscopic observations with FEROS, CAFE, and HRS of approximately 500 M dwarfs.  For each new observation, we determined its radial velocity and measured its H$\alpha$ activity index and its rotation velocity.  Additionally, we have multiple observations of many stars to investigate if there are any radial velocity variations due to multiplicity.  The results of our survey confirm that early-M dwarfs are H$\alpha$ inactive with low rotational velocities and that late-M dwarfs are H$\alpha$ active with very high rotational velocities.  The results of this high-resolution analysis comprise the most extensive catalogue of rotation and activity in M dwarfs currently available.  }


\maketitle


\section{Introduction} 

Current exoplanet research is driven by the detection of small planets with an emphasis on rocky planets orbiting in the habitable zones (HZ) of their host stars.  The HZ\ of a star is defined as the range in star-planet separation, where the flux incident on the planet results in temperatures in which water could be liquid.  Such planets are difficult to detect around F, G, and K dwarfs with the radial velocity (RV) technique because the induced Doppler shift due to the gravitational pull of the planet on the star is beyond current (and expected) instrumental precision.  However, the induced RV is comparatively greater for a planet of the same mass orbiting a less massive star.  Additionally for less massive stars, such as M dwarfs, the HZ of the star is located much closer to the star compared to the HZ of a more massive G dwarf.  Planets located in the HZ of M dwarfs consequently have a much stronger gravitational pull on their host stars, making it comparatively easier to detect these planets using the RV technique.  Furthermore, M dwarfs are the most common type of star comprising at least three-quarters of all stars in the Galaxy.  

While M dwarfs appear to be ideal targets for searching for small planets orbiting in the HZ of their host stars, a caveat is that M dwarfs are among the most magnetically active stars.  This is evidenced by the presence of coronal X-ray emission and chromospheric H$\alpha$ emission, which is present in approximately 5\,\% of M0\,V stars and increases dramatically to 80\,\% of M7\,V stars \citep{West2008}.  It is also well established that the increase in H$\alpha$ emission as a function of spectral type is directly correlated with an increase in rotational velocity.  Additionally, from a small sample of M dwarfs, \cite{Barnes2014}  showed that there is a correlation of H$\alpha$ emission with RV variations.

The presence of magnetic activity typically induces distortions or asymmetries in the shape of the spectral line profiles of the host star \citep{Desort2007,Reiners2010, Barnes2011, Jeffers2014, Barnes2015, Barnes2017}. This distortion varies on the same timescale as the individual activity features, which range from seconds to years and can result in an apparent RV variation that can mimic the presence of an exoplanet.  However, observations using a broad wavelength range can distinguish between a real and an apparent RV measurement.  This is because any planetary RV measurement will be wavelength independent, whereas an apparent activity induced RV measurement should be wavelength dependent.

\begin{figure*}
\def\imagetop#1{\vtop{\null\hbox{#1}}}
\begin{center}
\begin{tabular}[h]{c c}               
  \imagetop{\includegraphics[angle=270,width=0.48\textwidth]{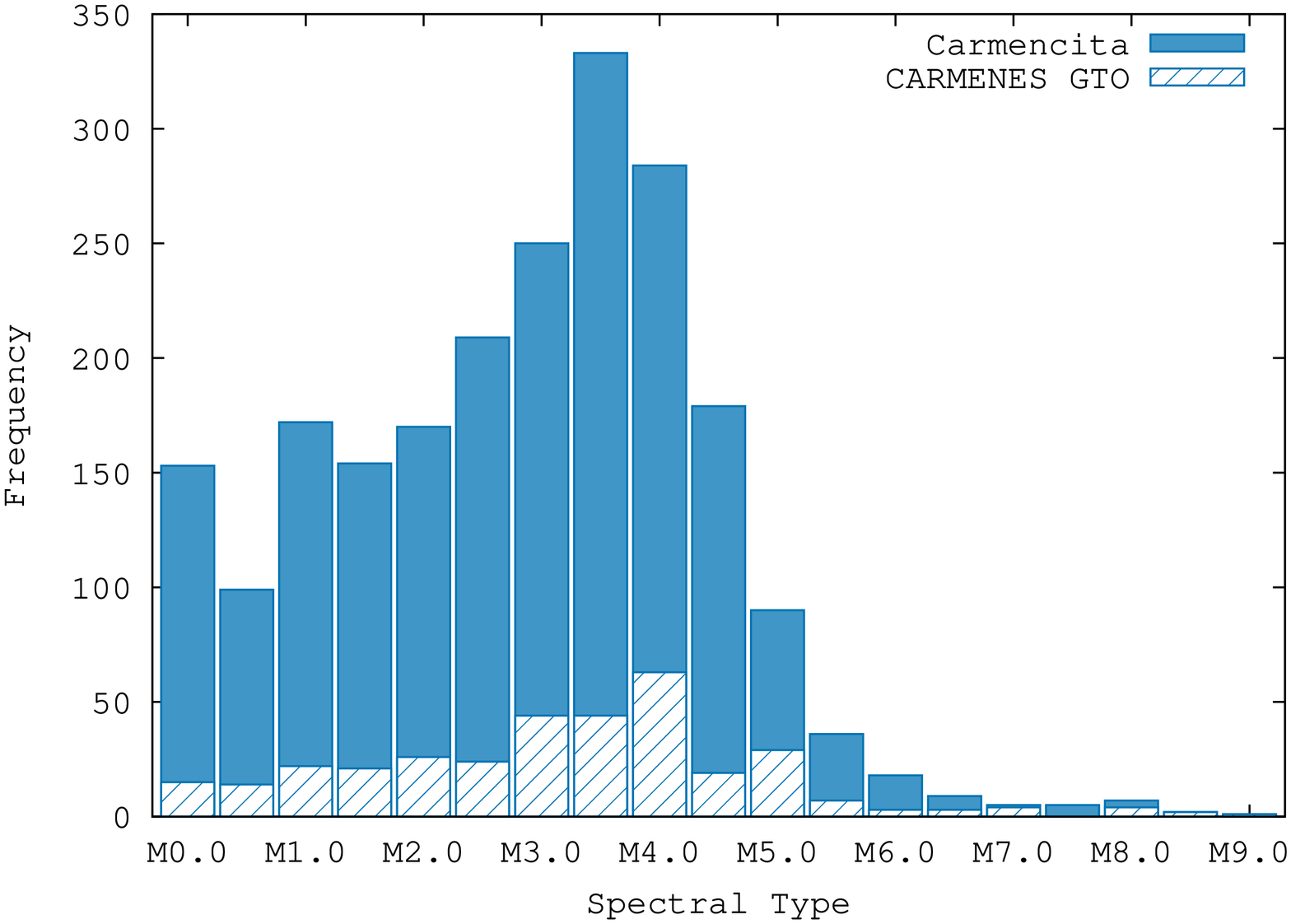}} &
  \imagetop{\includegraphics[angle=270,width=0.48\textwidth]{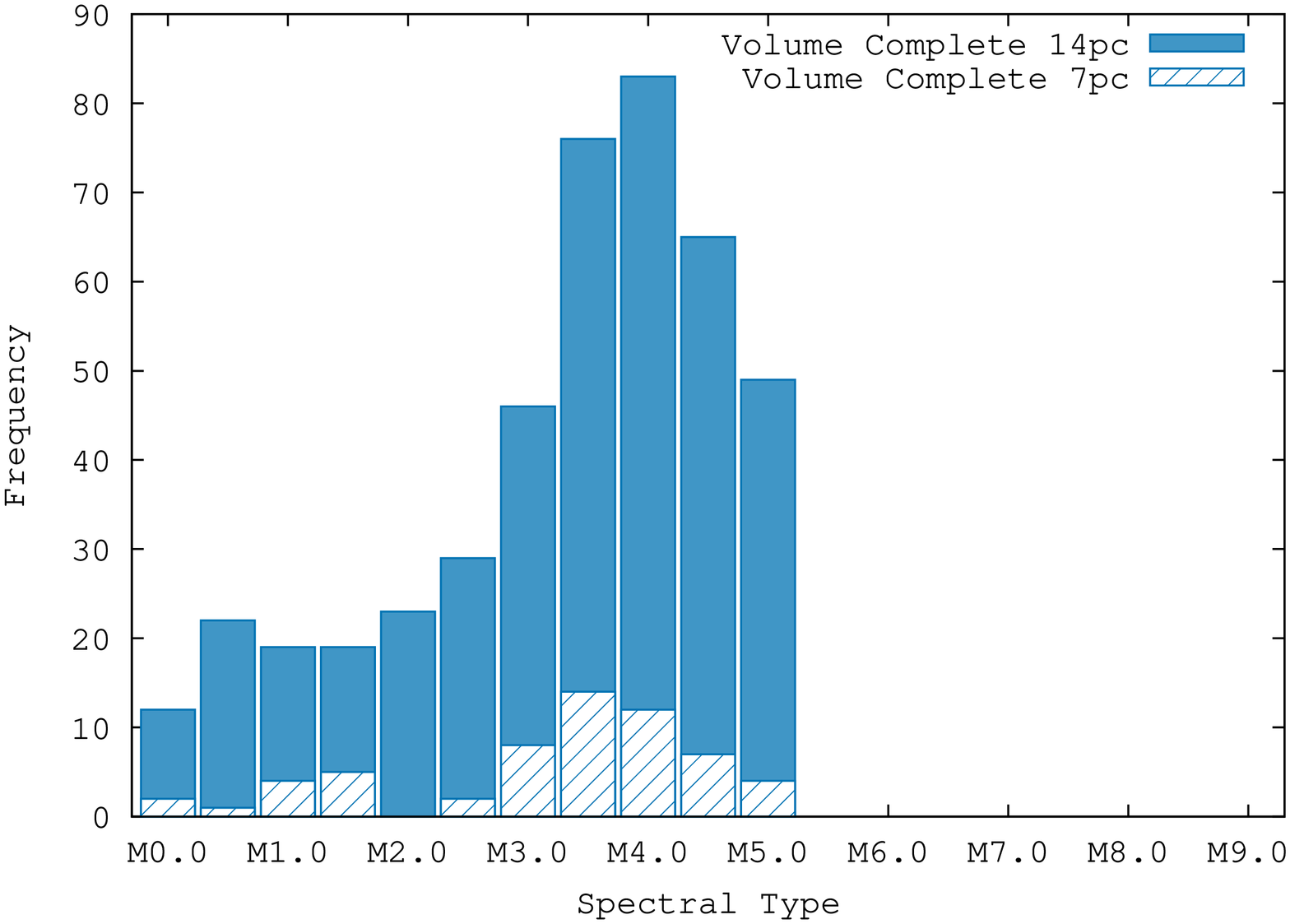}}  \\    
\end{tabular} 
\caption{Spectral type distribution of the total Carmencita sample and target list for the CARMENES GTO survey (left panel) and the 14\,pc and 7\,pc samples (right panel). Note: each plot has a different range of the y-axis.}
\protect\label{f-carmenes+CARMENCITA}
\end{center}
\end{figure*}

Current progress in instrumental development is focussed on obtaining high precision measurements at near-infrared wavelengths.  The first exoplanet hunting instrument that operates at both optical and near-infrared wavelengths is CARMENES \citep[Calar Alto high Resolution search for M dwarfs with Exoearths with Near-infrared and optical Echelle Spectrographs;][]{Quirrenbach2016}. This instrument comprises two highly stable, fibre-fed spectrographs covering the wavelength ranges 0.52 to 0.96\,$\mu$m and from 0.96 to 1.71\,$\mu$m with a spectral resolution of $>$80\,000. CARMENES was commissioned at the end of 2015 at the 3.5\,m Calar Alto telescope where it has achieved a stability of $\sim$1\,m\,s$^{-1}$.  
Other exoplanet hunting spectrographs currently being built or commissioned include IRD~\citep{Kotani2014}, HPF \citep{Mahadevan2014}, and SPIRou \citep{Artigau2014}.   The advantage of the very large wavelength coverage of instruments such as CARMENES is that the RV of the star can be measured using different wavelength regions of the spectrum to clearly identify the wavelength independent planetary RV signature.  Moving towards infrared wavelengths is also advantageous for detecting planets orbiting M dwarfs.  Currently a total of approximately 50 planets have been detected around early-M dwarfs using the RV technique \citep[e.g.][\footnote{See also {\tt carmenes.caha.es/ext/science} and {\tt www.exoplanet.eu}.}]{Bonfils2013}.  
However, only very few mid- to late-M dwarfs have been included in exoplanet hunting surveys, as they emit most of their flux at near-infrared wavelengths, i.e. at 1.0--1.2 $\mu$m \citep{Reiners2010}.  Since exoplanet surveys, such as HARPS \citep[High Accuracy Radial velocity Planet Searcher;][]{Mayor2003}, operate at optical wavelengths, mid- to late-M dwarfs are too faint to achieve a sufficiently high signal-to-noise ratio (S/N) using these instruments.   

The main scientific objective of CARMENES is the search for very low-mass planets 
(i.e. earth-like to super-earths) orbiting M dwarfs.  The CARMENES survey, which began in January 2016 and will last for at least three years, aims to observe approximately 300 M stars, spread over the complete M spectral range.  To maximise the scientific return of the CARMENES Guaranteed Time Observations (GTO)  survey, it is necessary to select the most promising targets with low levels of magnetic activity and low rotational velocities that are not spectroscopic binaries. This is because firstly, the broader the spectral lines of a star are, the more difficult it is to accurately measure its RV, and secondly because rotation plays a key role in the generation of magnetic activity, which can lead to a false planetary detection.  While the vast majority of the targets in the CARMENES sample are of low activity, there is a small subsample of moderately active M dwarfs.  For the last few years the CARMENES consortium has performed an extensive data compilation from the literature in addition to many new low- and high-resolution spectroscopic and high-resolution imaging observations to identify a well-studied sample for the exoplanet search.

In this paper we present the stellar rotation and activity of the targets in the Carmencita (CARMENES Cool dwarf Information and daTa Archive) catalogue.  This comprises data compiled from the literature as well as new measurements  derived from the high-resolution spectroscopic observing campaign.  Firstly we describe the Carmencita catalogue and its subsamples in Section ~\ref{sec:target}, followed by the description of the high-resolution observations and data reduction in Section ~\ref{sec:observations}, which includes the radial velocity and frequency of binary systems in the new observations.  The analysis of the new observations is presented in Section~\ref{sec:analysis}, including the chromospheric activity measured by the H$\alpha$ line, the Ca~{\sc ii} R$'_{\rm HK}$ , and the method for determining the rotation velocity of a star.  The rotation-activity relationship is investigated in Section~\ref{sec:rotact} for all stars in the Carmencita catalogue.  The results are summarised in Section~\ref{sec:magactivityM}.  This is the third paper in a series that aims to describe the selection and characterisation of the CARMENES sample; the first paper described the low-resolution spectroscopic observations ~\citep{Alonso2015} and the second paper investigated close multiplicity of the targets in the Carmencita catalogue ~\citep{Cortes2017}.  


\section{Samples}
\label{sec:target}

The global catalogue that forms the basis of this work is the Carmencita catalogue of M dwarfs.  We also investigate the subsamples comprising the CARMENES GTO sample and the volume complete sample out to (i) 7\,pc and (ii) 14\,pc. 

\begin{itemize}

\item{\em Carmencita catalogue:} The CARMENES input catalogue, Carmencita, contains the stellar parameters of approximately 2200 M dwarfs.  This includes all published M dwarfs that are, firstly, observable from Calar Alto, i.e. with declinations $>-23$\,degrees and, secondly,  the brightest stars for each spectral subtype (measured from spectroscopic observations) with the requirement that all stars are brighter than $J$ =11.5 mag.  The stellar parameters of these stars, which have been taken from the literature or determined by the CARMENES science team from new data, include accurate astrometry and distance, spectral type, photometry in 19 bands from the ultraviolet to the mid-infrared; rotational, radial, and Galactocentric velocities; H$\alpha$ emission; X-ray count rates; hardness ratios and fluxes;  close and wide multiplicity data; membership in open clusters and young moving groups; target in other radial velocity  surveys;  exoplanet  candidacy; radii; and masses. A more detailed description of the Carmencita catalogue is provided by \cite{Caballero2016} and \cite{Alonso2015}.  The final target list of the CARMENES survey as a function of spectral type is shown in Fig.~\ref{f-carmenes+CARMENCITA}.

In the Carmencita database, the values for the stellar rotational velocities, or $v \sin{i}$, were taken from the literature where available 

\footnote{\cite{Barnes2014,Browning2010,Christian2002,Delfosse1998,Deshpande2012,Duquennoy1988,Gizis2002,Griffin1985,Hartmann1987,Hartmann1989,HMR14,Houdebine2010,
Houdebine2012,Jenkins2009,Lopez2010,Malo2014,Marcy1992,Mochnacki2002,Mohanty2003,Morales2009,Reid2002,Reiners2012,Schlieder2010,Schlieder2012b,
Stauffer1986,Stauffer1997,Tokovinin1992,Torres2002,Torres2006,White2007,Zboril1998}}.

From the total of nearly 2200 M stars in the Carmencita catalogue, there are a significant number of stars that have $v \sin{i}$ measurements (721 stars), pEW(H$\alpha$) values (2129 stars), rotation periods from photometry (353 stars), and X-ray detections (715 stars).  \\

\item{\em CARMENES GTO sample:} The CARMENES GTO sample comprises over 300 M dwarfs with predominant spectral types at mid-M as shown in the left-hand panel of Figure~\ref{f-carmenes+CARMENCITA}.  The detailed analysis of activity presented in this work is an important reference for studying the environments of planets that are being found around these stars. \\

\item{\em Volume-limited samples:} The completeness of the Carmencita catalogue, for spectral types M0--5, is 100\,\% within 7\,pc, comrprising 60 stars, and 86\,\% within 14\,pc, comprising over 440 stars.  In this paper we investigate the correlation of activity and rotation for these two samples separately and refer to them as the Volume7 and the Volume14 samples in the text.   The distribution of these two subsamples as a function of spectral type is shown in the right-hand panel of Figure~\ref{f-carmenes+CARMENCITA}. \\

\end{itemize}

\begin{table}[]
\centering
\caption{Wavelength ranges used to calculate the RV measurement.}
\protect\label{tab-rvwav}
\begin{tabular}{ccc}
\hline
\hline
   \noalign{\smallskip}
Set A   & Set B                 & Set C         \\
{[\AA]}         & [\AA]                 & [\AA]                 \\
   \noalign{\smallskip}
\hline
   \noalign{\smallskip}
6565:6620 & 7300:7350 & 6200:7000\\
8200:8320 & 7675:7725 & 7000:7800\\
8440:8470 & 8350:8400 & 7800:8600\\
8500:8550 & 8400:8450\\
8620:8670 & 8500:8550\\
   \noalign{\smallskip}
\hline
\hline
\end{tabular}
\end{table}


\begin{table}[]
\centering
\caption{Candidate spectroscopic multiple star systems from this work.}
\protect\label{Tab_binarystars}
\begin{tabular}{l l l l}
\hline
\hline
   \noalign{\smallskip}
Karmn   & Name  & Spec.         & Remark$^{a}$ \\
                &               & bin.  &  \\
   \noalign{\smallskip}
\hline
   \noalign{\smallskip}
J01466-086 & LP 708-416 & SB1 & Astrom. $^1$ \\
J03526+170 & Wolf 227 & SB1 & SB2$^2$ \\
J05078+179  & G 085-041  & SB1 & Astrom.$^3$\\
J07119+773  & TYC 4530-1414-1  & SB1 & New \\
J07545+085  & LSPM J0754+0832  & SB1 & New \\
J09143+526  & HD 79210  & SB1 & New \\
J10143+210  & DK Leo & SB1 & Astrom.$^4$\\
J11036+136  & LP 491-051  & SB1 & New \\
J13195+351E  & BD+35 2436B  & SB1 & New\\
J19354+377  & RX J1935.4+3746  & SB1 & New\\
J21442+066  & G 093-033  & SB1 & New \\
   \noalign{\smallskip}
\hline
   \noalign{\smallskip}
J00428+355 & FF And  & SB2 & SB2$^5$\\
J00502+086  & RX J0050.2+0837  & SB2 & New \\
J02033-212  & G 272-145  & SB2 & New \\
J02289+120  & [R78b] 140  & SB2 & New \\
J03346-048  & LP 653-008  & SB2 & New \\
J04252+080S & HG 7-206  & SB2 & New \\
J04488+100  & 1RXS J044847.6+100302  & SB2 & New \\
J05032+213  & HD 285190 A & SB2 & New \\
J05322+098  & V998 Ori  & SB2 & SB2$^6$\\
J05466+441  & Wolf 237  & SB2 & Astrom.$^3$\\
J07418+050  & G 050-001  & SB2 & New \\
J09011+019  & Ross 625  & SB2 & Astrom.$^3$ \\
J09120+279  & G 047-028  & SB2 & New \\
J09506-138  & LP 728-070 & SB2 & New \\
J09531-036  & GJ 372  & SB2 & SB2$^7$\\
J12191+318  & LP 320-626  & SB2 & New \\
J12290+417  & G 123-035 & SB2 & Astrom.$^4$\\
J12299-054W  & LP 675-076  & SB2 & New \\
J13143+133  & NLTT 33370 AB & SB2 & Astrom.$^8$\\
J14130-120 & GQ Vir & SB2 & New \\
J14171+088 & 2M J14170731+0851363 & SB2 & New \\
J14368+583 & LP 098-132 & SB2 & New \\
J16255+260 & LTT 14889 & SB2 & New \\
J16487+106 & LSPM J1648+1038 & SB2 & Astrom.$^3$\\
J17136-084 & V2367 Oph & SB2 & New \\
J20301+798 & GSC 04593-01344 & SB2 & New \\
J20445+089N & LP 576-039 & SB2 & New \\
J20568-048 & FR Aqr & SB2 & New \\
J23096-019 & G 028-044 & SB2 & Astrom.$^3$\\
J23174+382 & G 190-017 & SB2 & New \\
J23302-203 & GJ 1284 & SB2 & SB2$^{9}$\\
J23438+325 & G 130-006 & SB2 & Astrom.$^1$ \\
J23573-129W & LP 704-014 & SB2 & SB2$^{9}$\\
\noalign{\smallskip}
\hline
\noalign{\smallskip}
J03346–048 & LP 653–008 & SB3 & SB3 \\
   \noalign{\smallskip}
\hline
\hline
\end{tabular}
\begin{list}{}{}
\item[$^{a}$] References: 1, \cite{Jodar2013}; 2, \cite{Bonfils2013}; 3, ~\cite{Cortes2017}; 4, \cite{Bowler2015}; 5, \cite{Bopp1977}; 6, \cite{Shkolnik2010}; 7, \cite{Harlow1996}; 8, \cite{Law2006}; 9, \cite{Gizis2002}.
\end{list}
\end{table}

\section{Observations and data reduction}
\label{sec:observations}

\subsection{Target list}

In addition to the values for the rotational velocities contained in the Carmencita catalogue, we also secured high-resolution observations of stars based on the following criteria:

\begin{itemize}
\item single stars without any previous $v \sin{i}$ measurement 
\item stars with accurately measured $v \sin{i}$ values (for comparison with our results)
\item stars with published $v \sin{i}$ values that are less than the minimum achievable value given the resolution of the spectrograph (as discussed in Section~\ref{section.6.4}) 
\item stars with measured $v \sin{i}$ values not corresponding to stellar X-ray to J-band luminosity ratio, $L_{\rm X}$/$L_{\rm J}$ \citep{Gonzalez2014}
\item stars with very different measured $v \sin{i}$ values in the literature
\item stars with poor or absent uncertainties on a previous $v \sin{i}$ measurement 
\item stars with unclear or conflicting multiplicity status
\end{itemize}

The final target list comprised 480 M dwarfs after applying these criteria to the stars listed in the Carmencita catalogue.  The spectral types used in this analysis are taken from Carmencita.    


\subsection{Spectrographs}

A total number of 1374 new spectra of 480 M dwarfs were observed using the Calar Alto Fiber-fed Echelle (CAFE), Fiber-fed Extended Range Optical Spectrograph (FEROS), and High-Resolution Spectrograph (HRS) spectrographs.  The observations were secured over a time span ranging from September 2011 to September 2014.  In general, stars with declinations $>$ +20\,degrees were observed with the CAFE spectrograph and stars with declinations between --23 and +20\,degrees were observed with FEROS.  The HRS observed the faintest M dwarfs in our sample. A total of 222 stars were observed more than once to check for variability in the measured RV values, and a small sample were observed with different spectrographs to check for consistency of the results.  A detailed journal of the observations is presented in Table~\ref{tab-journal_obs}, where the parameters S/N and RV are also listed.  

\subsubsection{CAFE}

CAFE \citep{Aceituno2013} is located at the 2.2\,m telescope at the Calar Alto Observatory in Spain. The wavelength coverage of CAFE is from 3960 \AA{}  to 9500 \AA{}  comprising 84 orders with a resolution of 62,000.   The CAFE spectrograph is fibre-fed and stabilised to a precision of 10--20\,m\,s$^{-1}$.  The observations of 927 spectra of 297 stars were secured over a time period from 21-01-2013 to 26-09-2014 over 99 observing nights.  

Data reduction was performed in the usual manner for bias subtraction, flat fielding and wavelength calibration using a modified version of the {\sc reduce} package \citep{Piskunov2002} that includes the Flat-field Optimal eXtraction (FOX) algorithm developed by \cite{Zechmeister2014}.  The reduced data shows overlapping orders at the blue end of the CCD, while the orders at the red end of the CCD ($>$ 7000 \AA) have gaps between the orders.  
 
\subsubsection{FEROS}

The FEROS spectrograph is at the 2.2\,m telescope of the European Southern Observatory located at La Silla observatory in Chile.  The resolving power of FEROS is 48,000, and it covers the wavelength range 3600 \AA{}  to 9200 \AA{}  over 39 orders.    The RV precision of FEROS is 21\,m\,s$^{-1}$ \citep{Kaufer1999}.  A total of 651 spectra of 297 stars were observed with FEROS over 52 observing nights spanning from 31-12-2012 to 11-07-2014.  The FEROS data were reduced using the ESO pipeline, which is based on the data reduction programme MIDAS.  

\subsubsection{HRS}

The observations secured with HRS at the 9.2\,m Hobby-Eberly Telescope at McDonald Observatory in the USA cover a wavelength range from 4200 \AA{}  to 11000 \AA{}   with a resolving power of 40,000.  Becasue several orders are located in the gap between the two CCDs, the wavelength range 6900 \AA{}  to 7065 \AA{}  is not covered.  The RV precision of HRS is $<$ 10\,m\,s$^{-1}$ ~\citep{Tull1998}.  The HRS observations were reduced using the same procedure as for the reduction of the observations secured with the CAFE spectrograph.  


\section{Analysis of new observations}
\label{sec:analysis}

In this Section we present the analysis of the new observations, which includes measuring the stellar RV and identifying any binary stars, quantifying the chromospheric activity using normalised H$\alpha$ luminosity and R$'_{\rm HK}$ indicies, and measuring the stellar rotational velocity.

\subsection{Radial velocity and identification of binary stars}

The radial velocities were measured for each star observed in our analysis by cross-correlating several wavelength ranges of the observed spectra with a synthetic {\sc{phoenix}} model spectrum taken from ~\cite{Husser2013} and with T$_{\rm eff}$ = 3600K, $\log{g}$ = 5.00 and solar metallicity.  The results did not change significantly with slightly different sets of parameters.  The three wavelength ranges are listed in Table~\ref{tab-rvwav}.  The main peak in the cross-correlation function (CCF) is identified as the RV of the star, the location of which was measured by fitting Gaussian profiles.  

The cross-correlation peaks were measured firstly for the five wavelength ranges listed in Set A.  If these values significantly disagreed or were not valid for at least three ranges, then the wavelength ranges from Set B were used.  A valid measurement is defined as when the Gaussian fit to the peak converges after less than 20 iterations and the result is within 3$\sigma$ of the mean of all measurements.  If there were not at least three valid measurements from Set B, then the wavelength ranges from Set C were used.  In the case in which there were still not three valid measurements, it was most likely that the star is either a spectroscopic binary star or a star with an earlier spectral type, for example a K dwarf.  The RV value for each star in the sample is listed in Table ~\ref{tab-journal_obs} and the error is the standard deviation of the RVs measured in each wavelength range.

A total of 45 stars showed significant variations in their measured RV values.  These stars are classified into ($i$) single-lined spectroscopic binary systems (SB1),  ($ii$) double-lined spectroscopic binary systems (SB2), or ($iii$) triple-lined spectroscopic binary systems (SB3).  For SB2 binary systems, the CCF shows more than one visible peak.  The candidate spectroscopic binary stars are listed in Table~\ref{Tab_binarystars}.  

Of the 11 SB1 type binary systems, four systems had previously known resolved companions (one SB2 and three astrometric) and seven are new detections, while for the SB2 systems there were 12 with previously known resolved companions (five SB2 and seven astrometric) and 21 completely new detections. The triple system LP-653-008 is known as an SB3.  The previously known SBs are indicated in Table~\ref{Tab_binarystars} in the fourth column.  The astrometric binary systems in the sample were previously resolved in high-resolution imaging surveys and are labelled ``Astrom.'' in the fourth column of Table~\ref{Tab_binarystars}. Two of the stars, namely \object{Wolf~237} and \object{Ross~625}, whose astrometric companions are located at 3.1--3.7\,arcsec, could actually be hierarchical triple systems of a spectroscopic binary with a faint companion at a relatively close projected physical separation of 50--70\,au \citep{Cortes2017}.

\subsection{Chromospheric activity}
\label{sec:chromoact}

The normalised H$\alpha$ luminosity and ratio of the flux in the cores of the Ca~{\sc ii} H\& K lines to the surrounding continuum are commonly used proxies for chromospheric activity.  The H$\alpha$ line is observed as an absorption feature for non-active stars.  With increasing activity the line cores first become stronger and then starts to fill in, becoming a strong emission feature for the most active stars \citep{Stauffer1986}.  Similarly, the cores of the Ca~{\sc ii} H\&K lines are clearly observed as an absorption line for non-active stars and with increasing activity the cores start to fill in and form an emission feature in the line centre. The ratio of the flux in the line cores relative to the surrounding continuum is commonly know as the R$'_{\rm HK}$ .

\subsubsection{Normalised H$\alpha$ luminosity}

The strength of the H$\alpha$ line is determined by calculating the pseudo-equivalent width pEW(H$\alpha$).  In general, resulting values that are negative are considered to result from an active star, while positive pEW(H$\alpha$) values are considered to indicate inactive stars.  The minimum detectable pEW(H$\alpha$) depends on the spectrograph resolution.

For the new observations described in Section ~\ref{sec:observations}, the spectra are renormalised by applying a linear fit to two wavelength regions, 6455--6559 \AA{}  and 6567--6580 \AA, on either side of the H$\alpha$ line.  The value for pEW(H$\alpha$) is measured over the wavelength range 6560 \AA--6565\,\AA.  The selected wavelength range is sufficiently narrow to exclude nearby spectral lines and sufficiently broad to measure the H$\alpha$ line of a fast rotating star.  The value of pEW(H$\alpha$) is calculated from

\begin{equation}
{\rm pEW(H\alpha)} = \int^{\lambda_2}_{\lambda_1} \Bigg(1-\frac{F(\lambda)}{F_{\rm pc}}\Bigg)~d\lambda,
\end{equation}

\noindent where $F_{\rm pc}$ is the average of the median flux in the pseudo-continuum in the ranges [6545:6559] and [6567:6580], $\lambda_1$ = 6560 \AA, $\lambda_2$ = 6565 \AA, and the error is determined with the method of \cite{Vollman2006}, which is applied as follows:

\begin{equation}
\sigma_{\rm pEW} = \frac{\Delta \lambda - \rm{pEW(H\alpha)}}{\rm S/N} \sqrt{1 + \frac{F_{\rm pc}}{F_{1,2}}},
\end{equation}

\noindent where $\Delta \lambda = \lambda_2 - \lambda_1$ and $F_{1,2}$ is the mean flux in the wavelength range between $\lambda_1$ and $\lambda_2$. For H$\alpha$ active stars, i.e. where pEW(H$\alpha$) is negative, the normalised H$\alpha$ luminosity can be expressed as 

\begin{equation}
\log \Bigg(\frac{L_{{\rm H}\alpha}}{L_{\rm bol}}\Bigg) = \log \chi + \log (-{\rm pEW}({\rm H}\alpha)),
\label{Eq-logha}
\end{equation}

\noindent where log $\chi$ is given as a function of the effective temperature \citep[e.g.][]{Reiners2008}.  Lines with H$\alpha$ in absorption have positive pEW(H$\alpha$) values and lines with H$\alpha$ in emission have negative pEW(H$\alpha$) values.  The minimum level of emission in H$\alpha$ that we are sensitive to is pEW(H$\alpha$)$\leq$-0.5 \AA , which is the definition for an active star in this work and is indicated by the term H$\alpha$ active.  This detection threshold was determined by visually inspecting the spectra and is consistent with values obtained by \cite{Newton2017} and \cite{West2015}.   The values derived from the new observations are tabulated in Table~\ref{tab-rotACT}.

\subsubsection{R$'_{\rm HK}$ indicator }

The R$'_{\rm HK}$ proxy of magnetic activity is not included in the Carmencita catalogue and is here determined using the Ca~{\sc ii} H \& K lines as follows.  Firstly, the $S$-index is measured with spectra from FEROS, $S_\mathrm{FEROS}$, using the description by \citet{Duncan1991} to mimic the HKP-2 photometer installed on top of Mount Wilson Observatory with spectrographic measurements. The flux of the Ca~{\sc ii} H \& K line cores is found by centring a 1.09 \AA{}  triangular bandpass on  the H (3968.47 \AA) and K (3933.66 \AA) line cores and summing these line cores. This sum is multiplied first by 2.4, which is a proportionality constant relating the HKP-2 instrument to the original HKP-1, and then by 8, which is the relative width between line core and continuum bandpasses. This is then divided by the sum of the fluxes in the 20 \AA{}  wide $V$ and $R$ continuum bandpasses,
centred at 3901.01 \AA{}  and 4001.07 \AA, respectively.  These $S_\mathrm{FEROS}$ values are then scaled to Mount Wilson Observatory values using the following linear fit relation:
\begin{equation}
    S_\mathrm{MWO} = 1.688 S_\mathrm{FEROS} + 0.06,
\end{equation}
where $S_\mathrm{MWO}$ is the $S$-index on the Mount Wilson scale. The above relation is determined by a linear fit of $S$-index values of common stars in FEROS archival data and \citet{Baliunas1995}.  To obtain log $R'_\mathrm{HK}$, we use the 
$S_\mathrm{MWO}$-$R'_\mathrm{HK}$ conversion of (Marvin et al. in prep.) which is dependent on effective temperature rather than B-V.  The effective temperature is determined from the spectral type  of the star using the calibration of \citet{KenyonHartmann1995}. The advantage of this relation is that it is on the same scale as  FEROS spectrograph as the S/N of the CAFE data was too poor for an accurate measurement.  The values derived from the new observations are tabulated in Table~\ref{tab-rotCAR}.

\subsection{Rotational velocity} 
\label{sec:rotvel}

In this section, the method to obtain the rotational velocity, or $v \sin{i}$, for each of the new observations is described.  

\subsubsection{Method}

The rotational velocity, or $v \sin{i}$, for each spectrum of the single stars in our sample was determined using the cross-correlation method \citep{Tonry1979,Basri2000} as implemented by \cite{Reiners2012}.  The concept of the cross-correlation method is that the spectrum of a slowly rotating star is artificially broadened using a series of $v \sin{i}$ values.  The range of $v \sin{i}$ values is determined by first measuring the full width half maximum (FWHM) of the cross-correlation profile and automatically selecting a range of approximately 10 reasonable $v \sin{i}$ values.  These broadened cross-correlation profiles then provide a calibrated measure of the real value of the star's $v \sin{i}$.  In the next step, the spectrum of the target is cross-correlated with a non-broadened template spectrum.  The resulting correlation peak is then converted into a $v \sin{i}$ measurement via the calibration of the broadened template spectrum.  This final step is performed on smaller (wavelength) chunks of spectra (see Table~\ref{tab-vsini}) successively rather than the whole spectra all at once. 

\subsubsection{Template stars}

Given the significant changes in the shape of M dwarf spectra over the spectral range of our targets stars, we carefully selected the optimal template star per spectral type bin (e.g. one template for M0\,V \& M0.5\,V stars, etc.).  Our initial tests showed that selecting a template star with a spectral type in the middle of our sample (e.g. with a spectral type of M\,3.5\,V or M\,4.0\,V) provided very poor results for early-M stars and late-M stars.  

A list of potential template stars was compiled for each spectral type by selecting four stars in each spectral type bin that showed the highest S/N values and H$\alpha$ in absorption and considered to be a  H$\alpha$  inactive star.  To select the best template spectrum, each of the four template stars was used to measure the $v \sin{i}$ on a subsample of $\sim$100 target stars.  The final template star per spectral type bin was selected by comparing the results of each of the four stars.  The template that resulted in a largest number of useful chunks for each of the 100 targets stars was chosen. 

\subsubsection{Wavelength range}

The spectra of each template and target star were compared in a series of chunks that were chosen to cover a wavelength range of approximately 500 \AA.  Regions containing strong telluric lines and emission lines were removed from the analysis.  The spectral regions used in the analysis were optimised for each of the three spectrographs.  To determine if a spectral chunk was used or not in the final $v \sin{i}$ value, the accuracy of the RV of the chunk was used.  For example,  for chunks of spectra that are within a predefined value, i.e. $|{\rm RV}_{\rm chunk} - \rm{RV}_{\rm median}| \le$ 0.2\,km\,s$^{-1}$, then only the $v \sin{i}$ values determined for these chunks contribute to the final $v \sin{i}$ value for the star.  For very fast rotating stars it was necessary to use a higher threshold value. 

\subsubsection{Detection limit}
\label{section.6.4}

Obtaining an accurate measurement of the $v \sin{i}$ value of a star not only requires a precise measurement of the width of the spectral line profile, but also an understanding of the detection limit.  This limit is influenced by other broadening mechanisms that are dominated by instrumental effects.  The instrumental broadening of the spectral lines is primarily determined by the spectral resolving power of the instrument.  In this analysis we considered the minimum $v \sin{i}$ value that can be reliably measured to be 3 km s$^{-1}$ for spectrographs with a resolving power similar to that of CAFE and FEROS \citep{Reiners2012}.  Measurements below these values were not included in the analysis of the results but are listed as the minimum reliable measurement.  

For stars with multiple spectra, the measurement with the highest S/N was selected.  There are a total of 68 stars in the sample that were observed with the CAFE and FEROS instruments.  The $v \sin{i}$ measurements are consistent and independent of the spectrograph used.  From the sample of 68 stars, we also have multiple observations of the same stars taken with the same spectrograph.  From these we note that the $v \sin{i}$ measurement shows a higher dependence on the S/N of the spectra than on the instrument used to secure the observations.   The $v \sin{i}$  values derived from the new observations are tabulated in Table~\ref{tab-rotACT}.
Recently \cite{Reiners2017} determined the rotational velocities for the CARMENES GTO sample using the same method as this work.

\section{Global results}

In this Section we include the new observations in the Carmencita catalogue and present the results in the context of the four samples of (1) the complete Carmencita sample, (2) the CARMENES GTO sample, (3) the 14\,pc sample, and (4) the 7\,pc sample.   The full CARMENCITA catalogue of $v \sin{i}$ and pEW(H$\alpha$) values is presented in Table~\ref{tab-rotCAR}.

\subsection{Chromospheric activity}

\subsubsection{H$\alpha$ activity indicator}

\begin{figure*}
\def\imagetop#1{\vtop{\null\hbox{#1}}}
\begin{center}
\begin{tabular}[h]{c c}               
  \imagetop{\includegraphics[angle=270,width=0.48\textwidth]{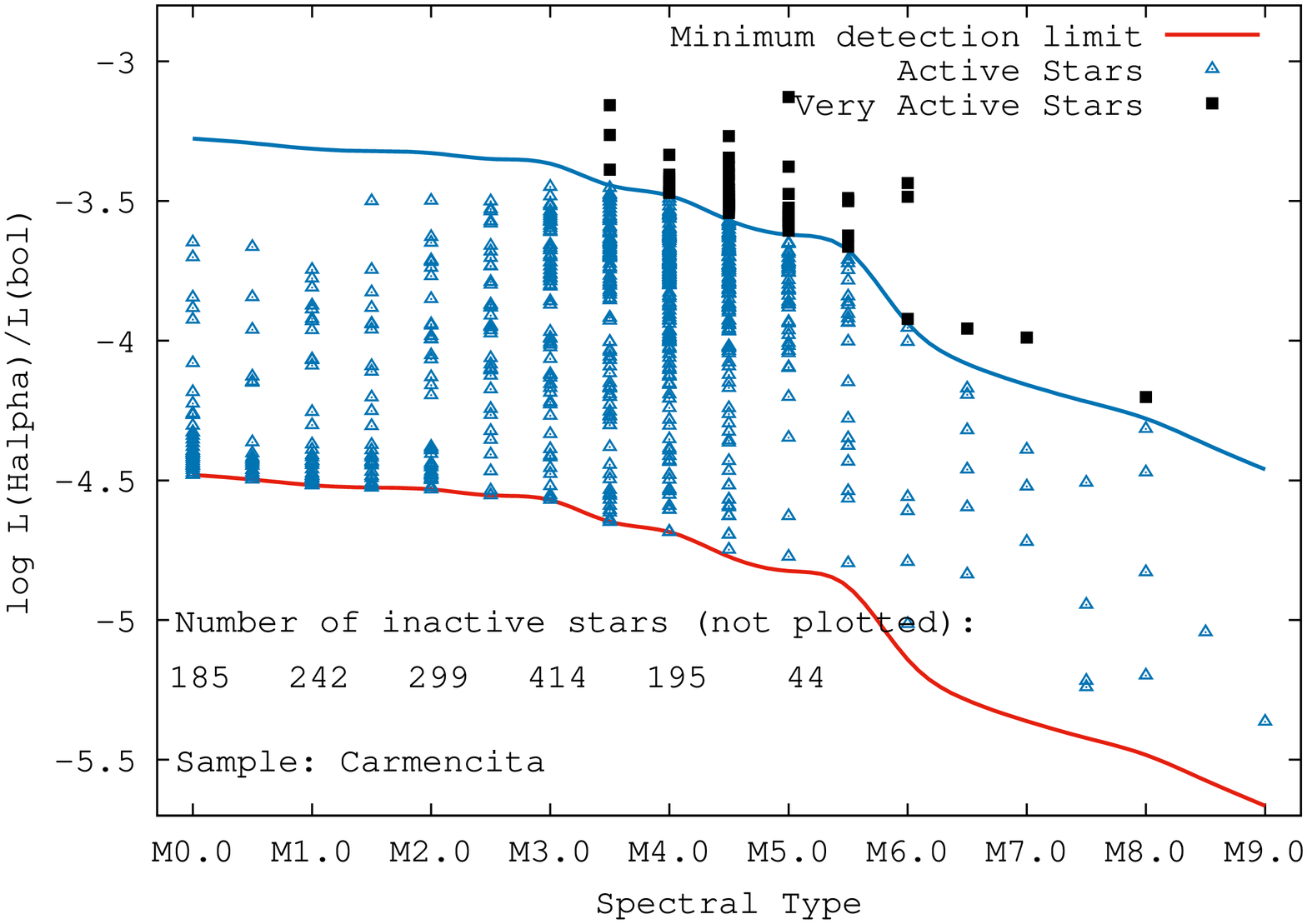}} &
  \imagetop{\includegraphics[angle=270,width=0.48\textwidth]{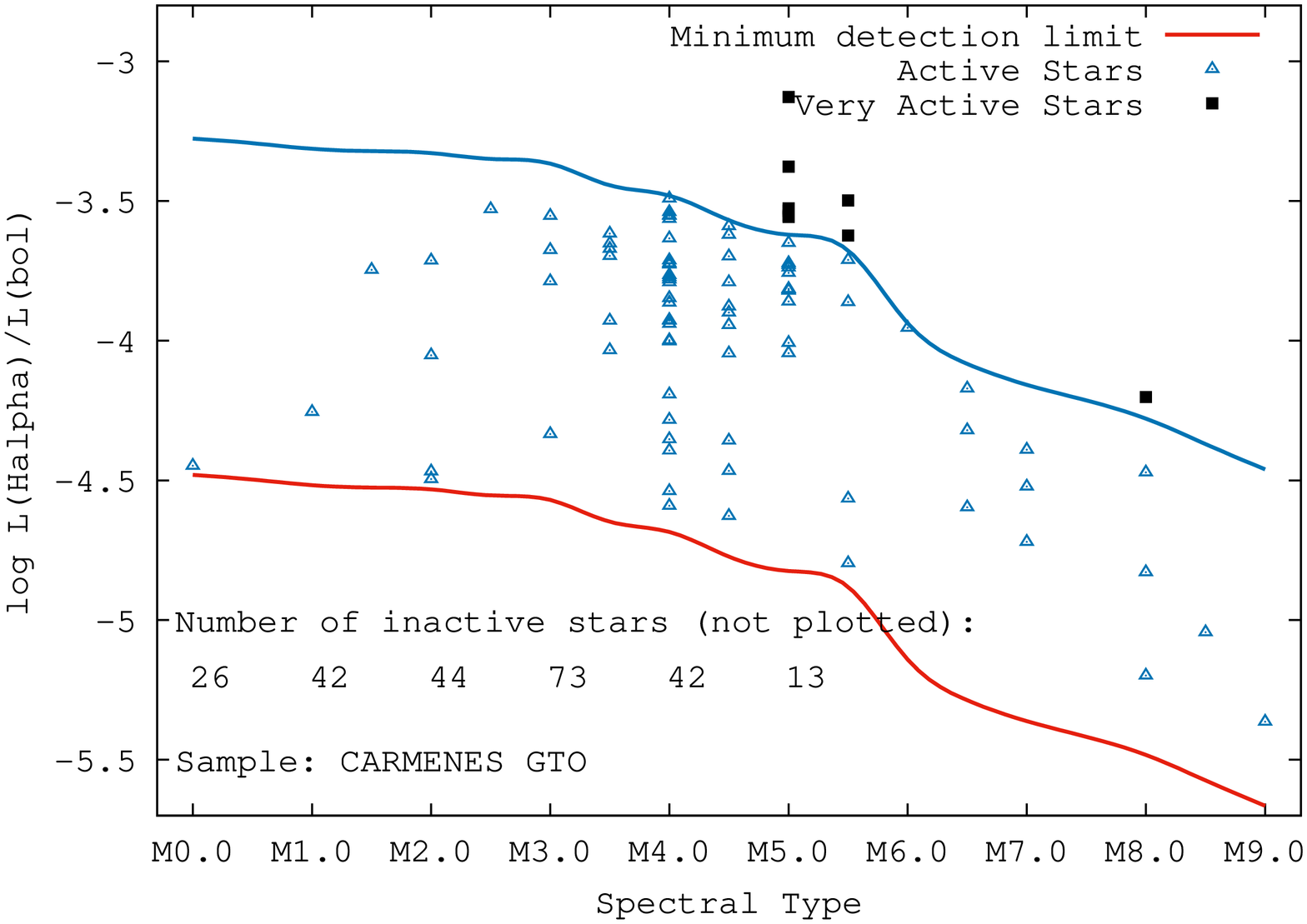}}  \\    
  \imagetop{\includegraphics[angle=270,width=0.48\textwidth]{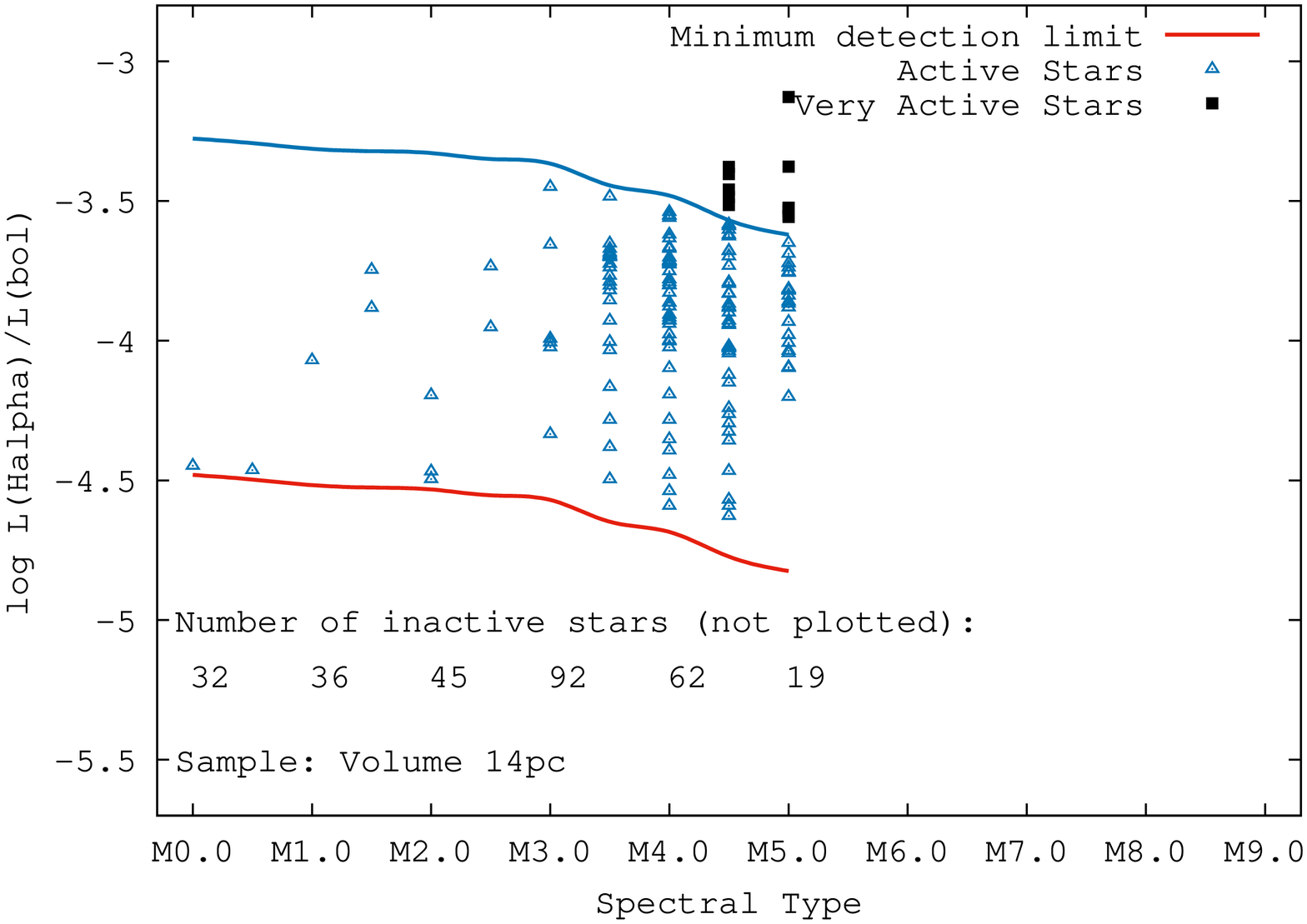}} &
  \imagetop{\includegraphics[angle=270,width=0.48\textwidth]{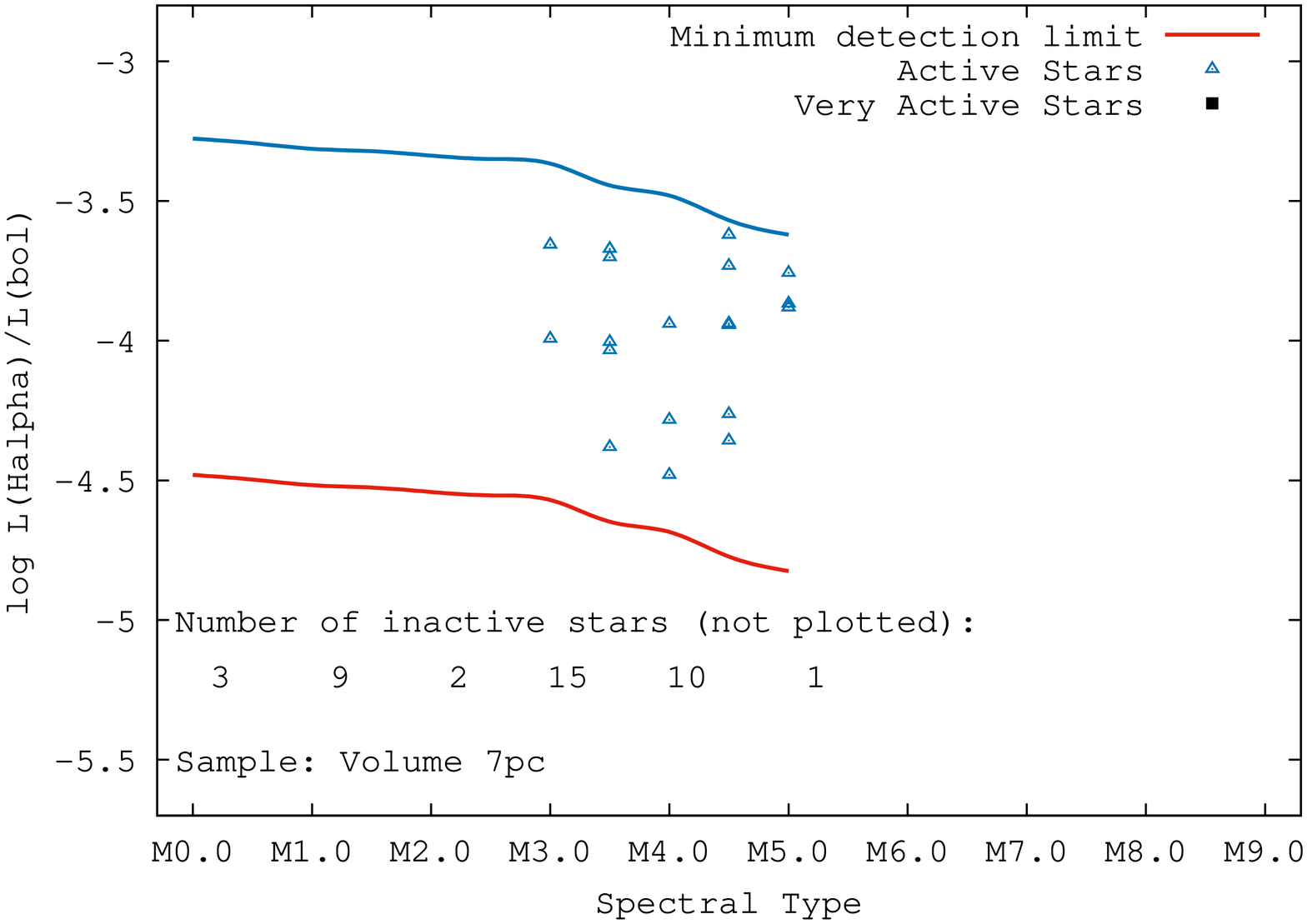}} \\ 
\end{tabular} 
\caption{Normalised H$\alpha$ luminosities as a function of spectral type for the M dwarfs contained in the Carmencita catalogue (upper left), CARMENES GTO target sample (upper right), 14\,pc sample (lower left) and 7\,pc sample (lower right).  The red line indicates the values of $H\alpha$ emission that our observations are sensitive to as a function of spectral type, which is defined as pEW(H$\alpha$)=-0.5 \AA.  Stars with values greater than this are considered to be  H$\alpha$ inactive stars.  The values below the minimum detectable emission, or  H$\alpha$ inactive stars, are not shown on the plot, but instead the total number of  H$\alpha$ inactive stars is shown.  Also indicated is a population of extremely  H$\alpha$ active stars that are defined as having pEW(H$\alpha$)$<$--8.0\,{\AA,} which are only visible at later spectral types (solid blue line).
}
\protect\label{f-pEW_SpT} 
\end{center}
\end{figure*}

In Fig.~\ref{f-pEW_SpT} the resulting normalised H$\alpha$ luminosities or log($L_{\rm H\alpha}$/$L_{\rm bol}$) values are shown as a function of spectral type for the four samples.  The red line in each sub-figure indicates the minimum level of H$\alpha$ emission that we can detect (pEW(H$\alpha$)=-0.5 \AA) as a function of spectral type.  As noted in previous studies \citep[e.g.][among others]{West2015, Reiners2012} and is evident by the slope of the red line, there is a decrease in log$\chi$ from which we calculate log($L_{\rm H\alpha}$/$L_{\rm bol}$) with increasing spectral type.  Highlighted are a small population of extremely  H$\alpha$  active stars, which are located above the solid blue line and are defined as stars with values pEW(H$\alpha$)$<$--8.0\,{\AA}.  The resulting pEW(H$\alpha$) values are tabulated in Table~\ref{tab-rotACT}.  Values from this analysis are indicated by values of pEW(H$\alpha$), where the name of the instrument (CAFE, FEROS, and HRS) is shown in the right-hand most column.  

\begin{figure*}
\def\imagetop#1{\vtop{\null\hbox{#1}}}
\begin{center}
\begin{tabular}[h]{c c}               
  \imagetop{\includegraphics[width=0.48\textwidth]{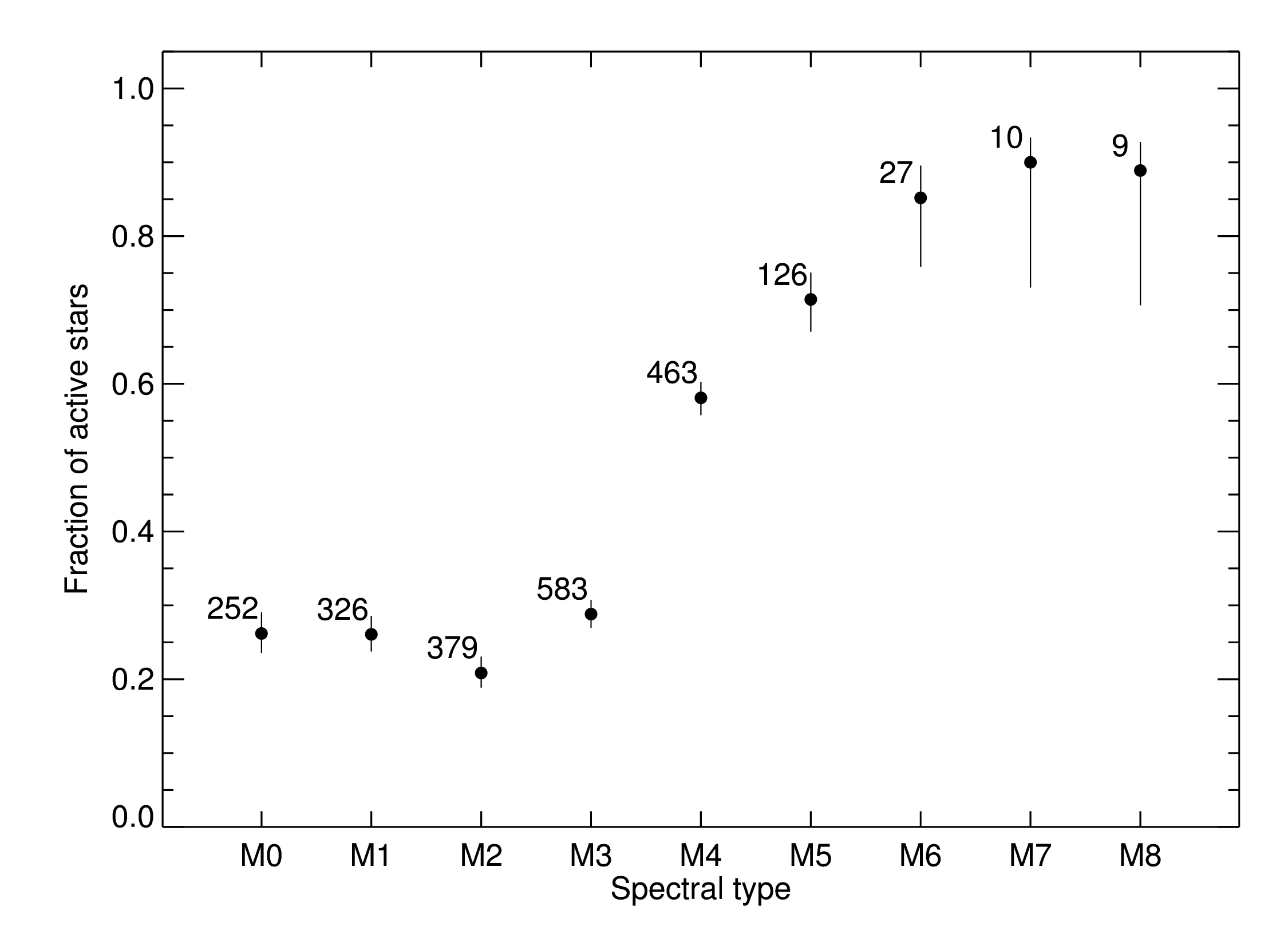}} &
  \imagetop{\includegraphics[width=0.48\textwidth]{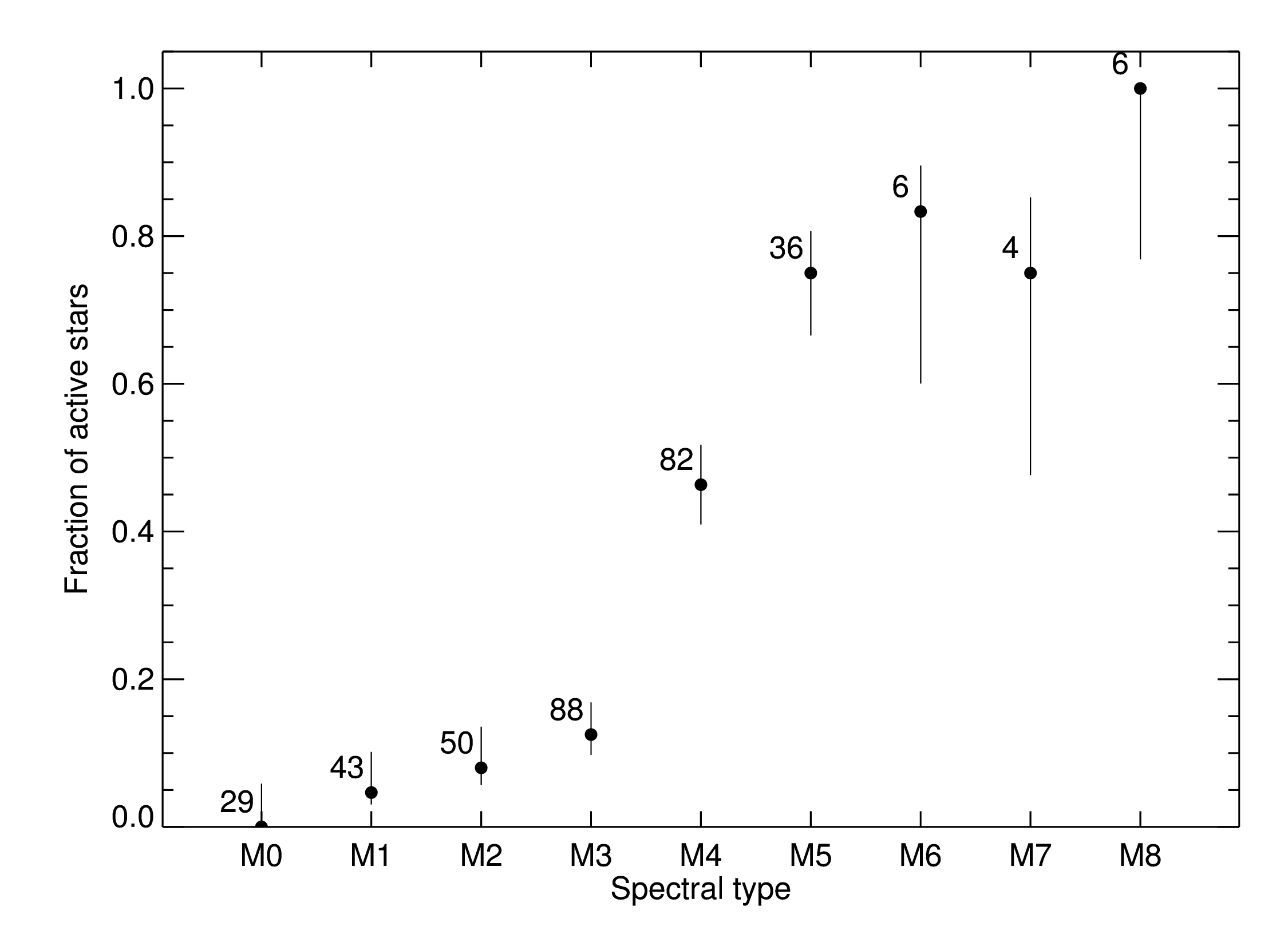}}  \\    
  \imagetop{\includegraphics[width=0.48\textwidth]{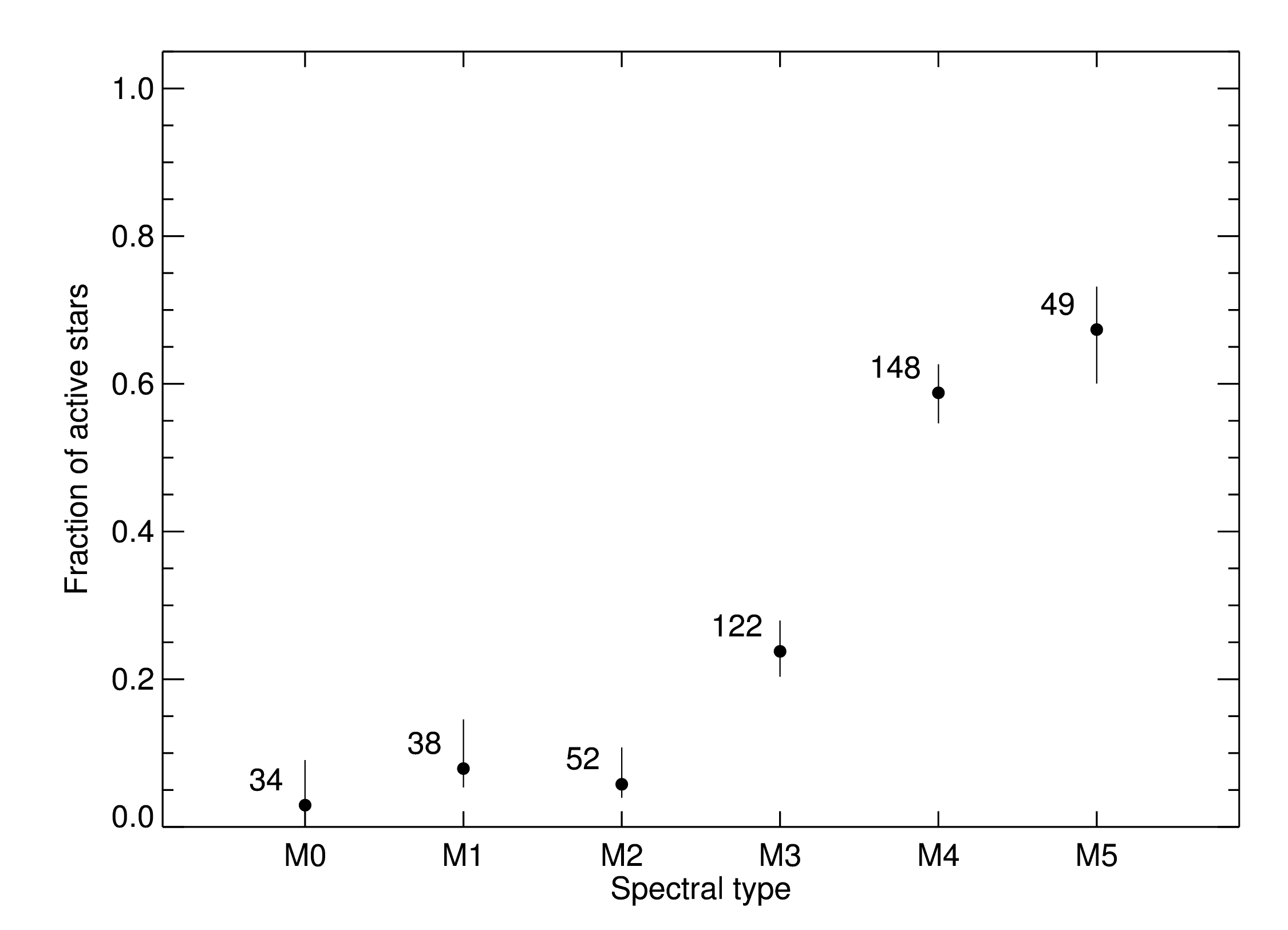}} &
  \imagetop{\includegraphics[width=0.48\textwidth]{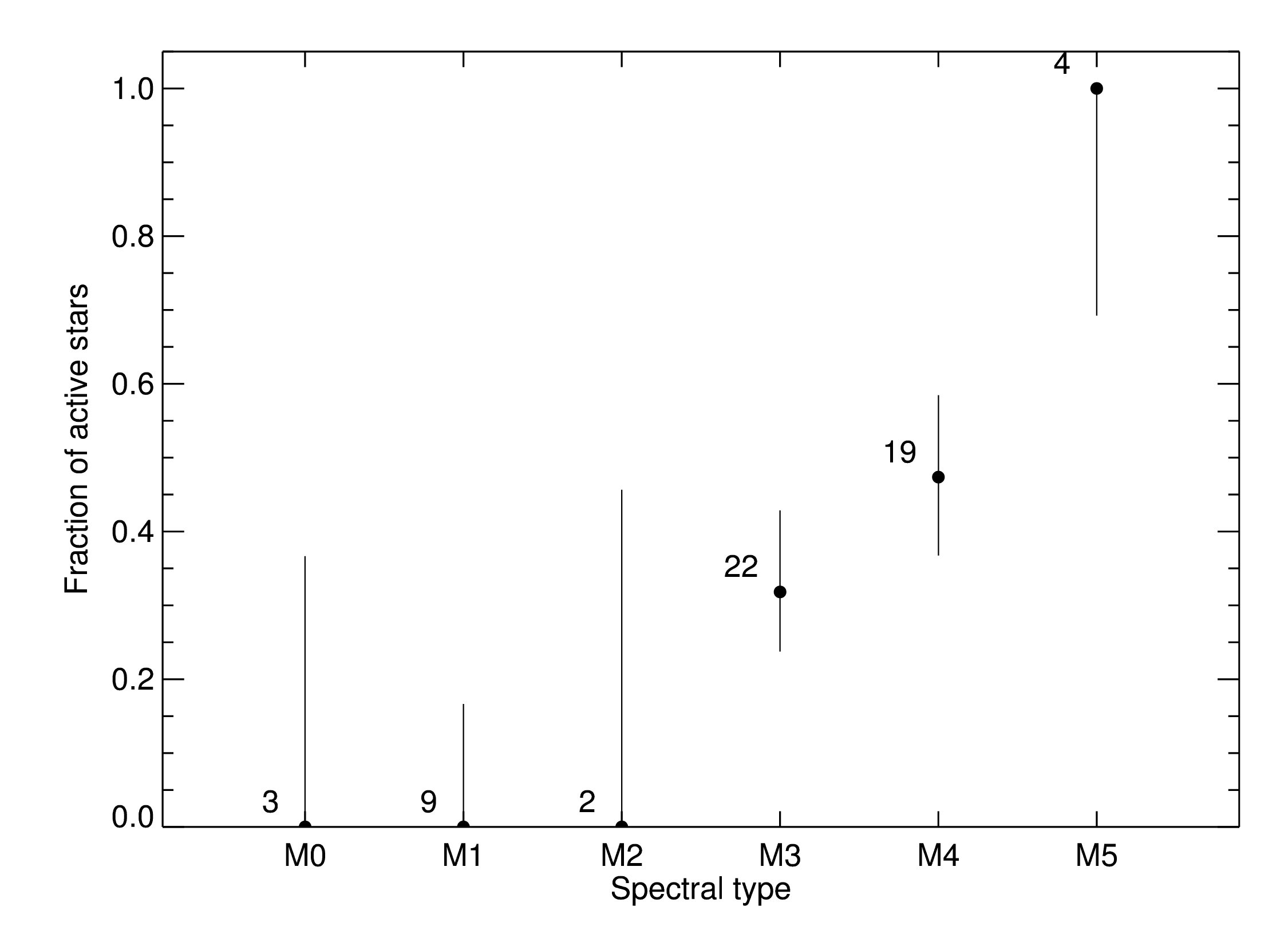}} \\ 
\end{tabular} 
\caption{Fraction of  H$\alpha$ active stars as a function of spectral type for the Carmencita sample (upper left), CARMENES GTO sample (upper right), 14\,pc sample (lower left) and 7\,pc sample (lower right).  The numbers shown on the plot are the total number of stars in the respective spectral type bin.  Error bars show 1 $\sigma$ uncertainties. The lower two panels have a smaller range in spectral type.}
\protect\label{f-frac_active}
\end{center}
\end{figure*}

There is a large range of log($L_{\rm H\alpha}$/$L_{\rm bol}$) values for each spectral type bin.  This is primarily due to including a large number of stars with a large range of ages and rotation rates.  Despite this, there are still some trends that are only apparent with the very large number of stars in the Carmencita sample.  The total sample from the Carmencita catalogue shown in Fig.~\ref{f-pEW_SpT}, comprises 2128 individual stars with H$\alpha$ measurements, where 35\,\% of the sample (750 stars) have H$\alpha$ in emission with pEW(H$\alpha$)$\leq$-0.5 \AA.  Of these 750  H$\alpha$ active stars, there is a small subsample of extremely  H$\alpha$ active stars (59 stars or 3\% of total number of stars) with pEW(H$\alpha$)$\leq$--8.0\,{\AA}.  The other subsamples shown in Fig.~\ref{f-pEW_SpT} follow the same global trends.   The spread in log($L_{\rm H\alpha}$/$L_{\rm bol}$) values is largest at intermediate spectral types (i.e. from M3.5 to M4.5), where there is an almost uniform distribution of log($L_{\rm H\alpha}$/$L_{\rm bol}$) values from the red line up to and including the extremely  H$\alpha$ active stars.  For earlier spectral types (i.e. $<$ M3.0), the log($L_{\rm H\alpha}$/$L_{\rm bol}$) values are mainly concentrated to just above the red line, and only a few points are scattered above log($L_{\rm H\alpha}$/$L_{\rm bol}$) values of $>$-3.75\,\AA.  For later spectral types between M\,4.5 and M\,5.5, the points are concentrated at just below the solid blue line, and only a few scattered points are close to the red line.  The same trend is evident in the CARMENES GTO sample (Fig.~\ref{f-pEW_SpT}, upper right), where the extremely  H$\alpha$ active stars have spectral types starting at M5.0 compared to M3.5 for the Carmencita sample.  

For the CARMENES GTO sample, the fraction of  H$\alpha$ inactive stars is much higher than in the Carmencita sample, although the largest range in values also occurs at mid-spectral types.  For the 14\,pc sample, similar global trends are apparent, although as with the CARMENES GTO sample, there are very few  H$\alpha$ active stars at early spectral types.  Similar to the Carmencita sample, the distribution shows a large range in log($L_{\rm H\alpha}$/$L_{\rm bol}$) values for mid-spectral types, however the extremely  H$\alpha$ active stars are only apparent at spectral types starting at M4.5\,V.  As this sample, along with the 7\,pc sample, only extends to spectral type M5.0\,V, there is no information at later spectral types.  The 7\,pc sample is primarily comprised of stars with mid-M spectral types.  There are no  H$\alpha$ active stars with spectral types M3.0\,V or earlier, and no extremely  H$\alpha$ active stars.  

\begin{figure*}
\def\imagetop#1{\vtop{\null\hbox{#1}}}
\begin{center}
\begin{tabular}[h]{c c}      
\imagetop{\includegraphics[angle=270,width=0.48\textwidth]{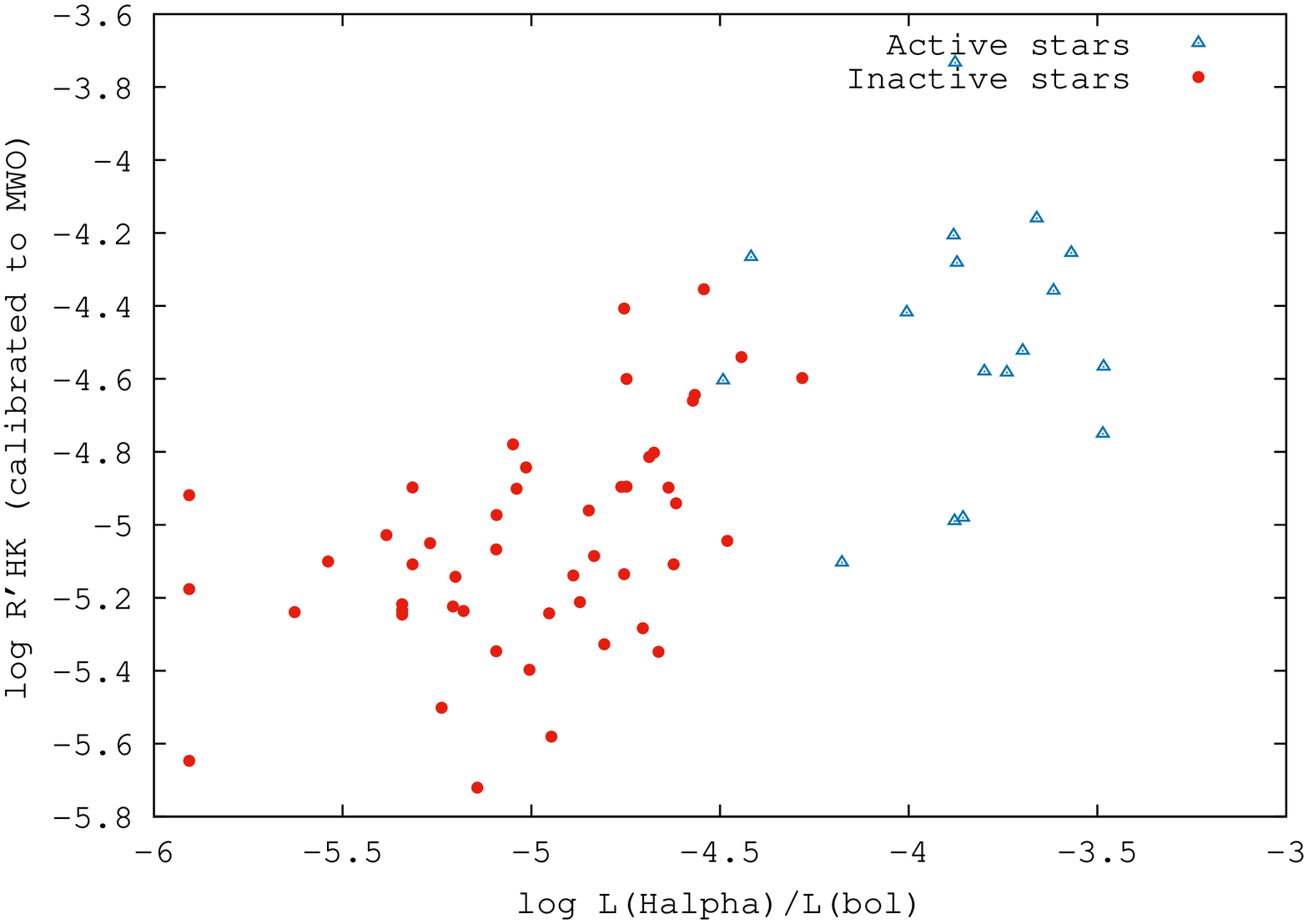}} &
\imagetop{\includegraphics[angle=270,width=0.48\textwidth]{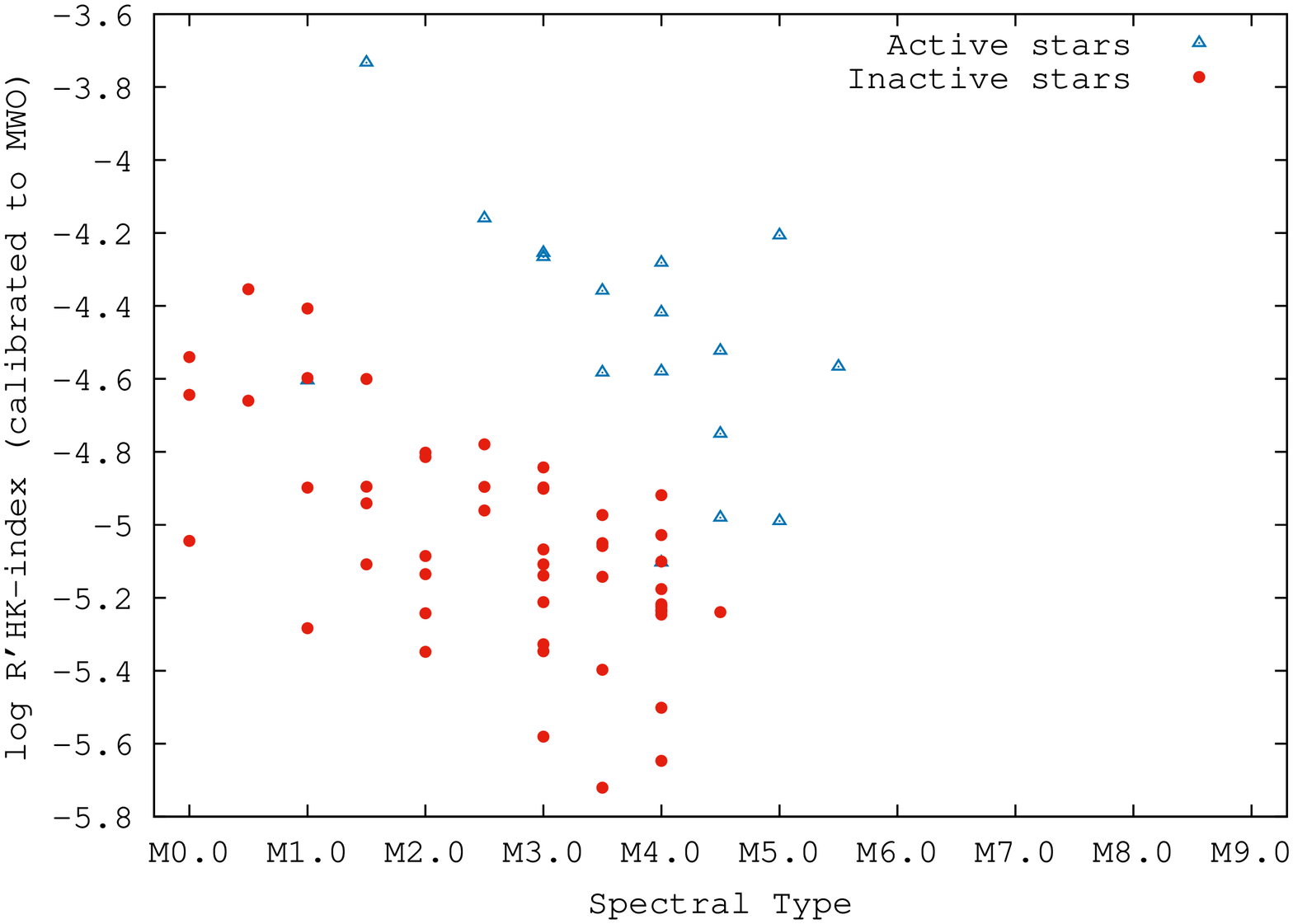}}\\
\end{tabular}     
\end{center}
\caption{R$'_{\rm HK}$ values as a function of normalised H$\alpha$ luminosity (left panel) and spectral type (right panel).  All of values are calibrated to the S-index values obtained by the Mount Wilson Observatory (MWO).}
\protect\label{f-Sindex} 
\end{figure*}

\subsubsection{Fraction of  H$\alpha$ active stars}

Later spectral types typically show much higher fractions of  H$\alpha$ active stars, as shown in Fig.~\ref{f-frac_active} for all four samples of this analysis.  For stars with early spectral types ($\leq$ M3.0\,V) in our sample, approximately 25\,\% stars in the Carmencita sample, 6\% in the CARMENES GTO sample, 8\% in the 14\,pc sample, and 9\% in the 7\,pc sample are  H$\alpha$ active.  For mid-M spectral types (between M3.5\,V and M4.5\,V), the fraction of  H$\alpha$ active stars increases to approximately 50\,\% of the Carmencita sample, 35\% of the CARMENES GTO sample, 48\% of the 14\,pc sample and 42\% of the 7\,pc sample.  The fraction of  H$\alpha$ active stars continues to increase with spectral type, where 76\,\% of the Carmencita sample, 79\% of the CARMENES GTO sample, 67\% of the 14\,pc sample, and 100\% of the 7\,pc sample with spectral types $\geq$ M5.0\,V are  H$\alpha$ active.  This result is in agreement with the conclusion of \cite{Gizis2002} and \cite{Lee2010} that H$\alpha$ strength is larger at later spectral types.

There are higher fractions of H$\alpha$ active stars in the Carmencita sample for earlier spectral types compared to the other three subsamples and the results of \cite{Reiners2012}.  These stars are typically more rapidly rotating and are consequently more magnetically active and are discussed in more detail in Section 6.3.2. 

\subsubsection{R$'_{\rm HK}$ activity indicator}

From the new observations, the R$'_{\rm HK}$ values are available for only a small subsample of 65 stars observed with the FEROS spectrograph.  This is because the S/N ration of the CAFE observations was too poor to obtain an accurate measurement.  The measured values are shown in the left panel of Fig.~\ref{f-Sindex} for the dependence of the R$'_{\rm HK}$ on the normalised H$\alpha$ luminosities.  There is a general trend of increasing R$'_{\rm HK}$ values with increasing log($L_{\rm H\alpha}$/$L_{\rm bol}$) values.  In the right-hand panel of Fig.~\ref{f-Sindex}, there is also a trend of decreasing R$'_{\rm HK}$ values moving from early to mid-M spectral types, i.e. from M0.0\,V to M4.0\,V (Fig.~\ref{f-Sindex}, right-hand plot), which is broadly in agreement with the recent results of ~\cite{Moutou2017}.  For  H$\alpha$ active stars, the R$'_{\rm HK}$ values are high and these stars typically have spectral types from M3.0\,V to M5.0\,V.  The same trend is also seen in the correlation of R$'_{\rm HK}$ with  measured log($L_{\rm H\alpha}$/$L_{\rm bol}$) values (Fig.~\ref{f-Sindex}), where more  H$\alpha$ active stars have higher R$'_{\rm HK}$ values and  H$\alpha$ inactive stars have low R$'_{\rm HK}$ values. The lack of stars at later spectral types in Fig. ~\ref{f-Sindex} is because to these stars are too faint to obtain a sufficient S/N to reliably measure R$'_{\rm HK}$. 

\begin{table}
\centering
\caption{Parameters of the new $v \sin{i}$ measurements.}
\protect\label{tab-vsini}
\begin{tabular}{l l c c c c c}
\hline
\hline
   \noalign{\smallskip}
Instrument & $\lambda_{min}$ & $\lambda_{max}$ & Chunks size & \# chunks \\
 & [\AA] & [\AA] & [\AA] &      \\
   \noalign{\smallskip}
\hline
   \noalign{\smallskip}
CAFE & 4800 & 6700 & 500 & 38 \\
FEROS & 4300 & 7200 & 500 & 58 \\
   \noalign{\smallskip}
\hline
\hline
\end{tabular}
\end{table}


\subsection{Rotational velocity} 
\label{sec:rotvelresults}

\begin{figure*}
\def\imagetop#1{\vtop{\null\hbox{#1}}}
\begin{center}
\begin{tabular}[h]{c c}               
  \imagetop{\includegraphics[angle=270,width=0.48\textwidth]{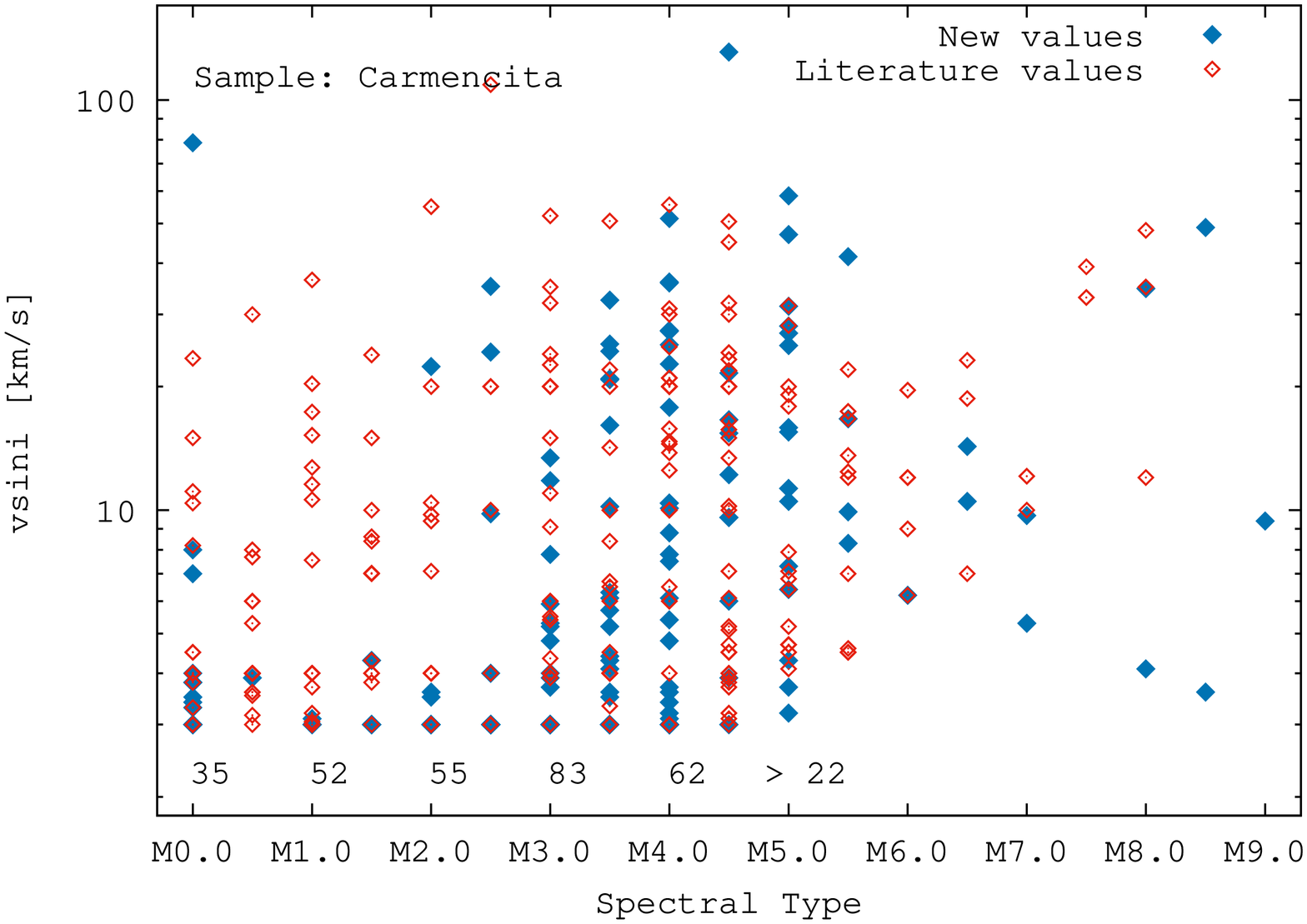}} &
  \imagetop{\includegraphics[angle=270,width=0.48\textwidth]{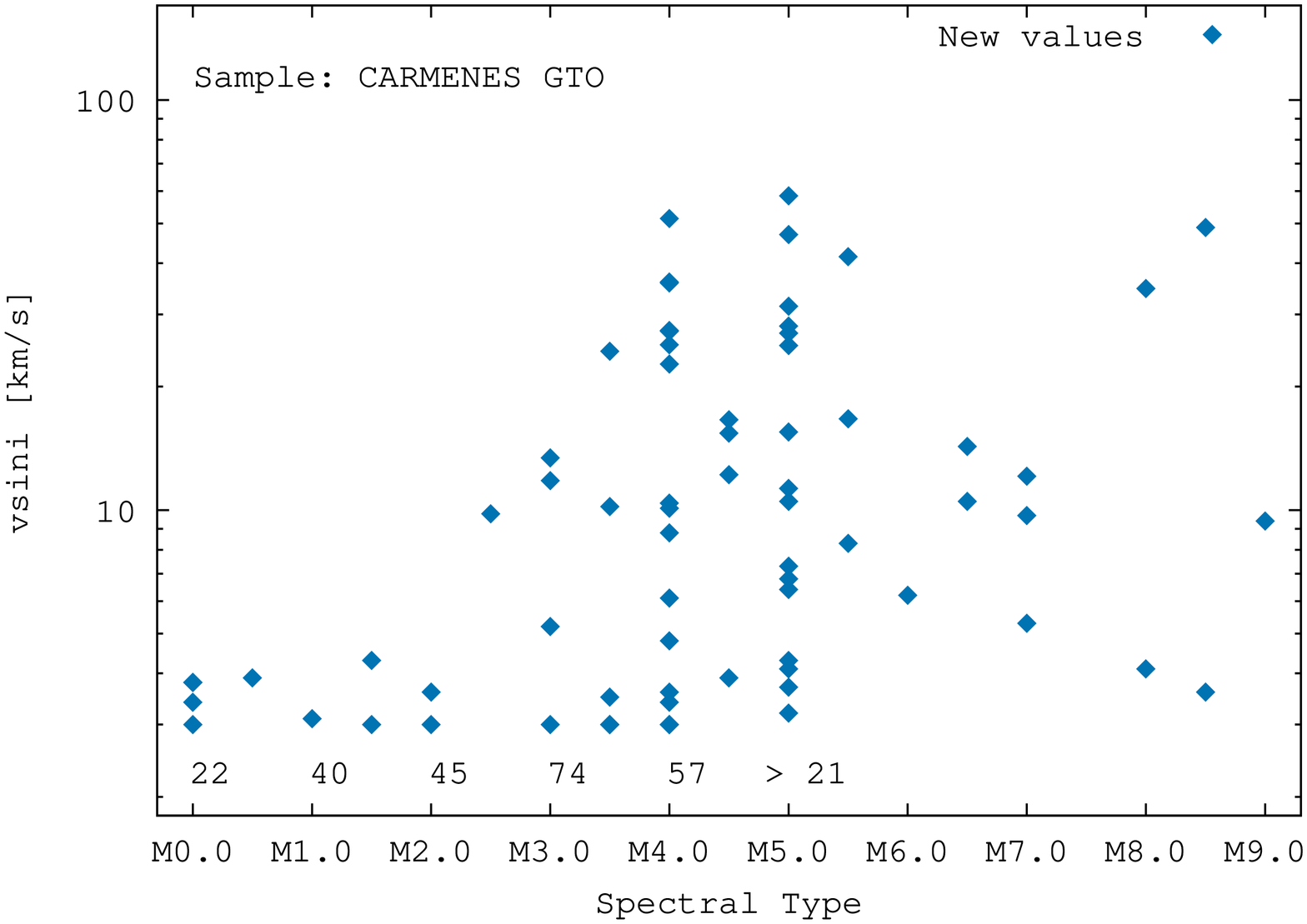}}  \\    
  \imagetop{\includegraphics[angle=270,width=0.48\textwidth]{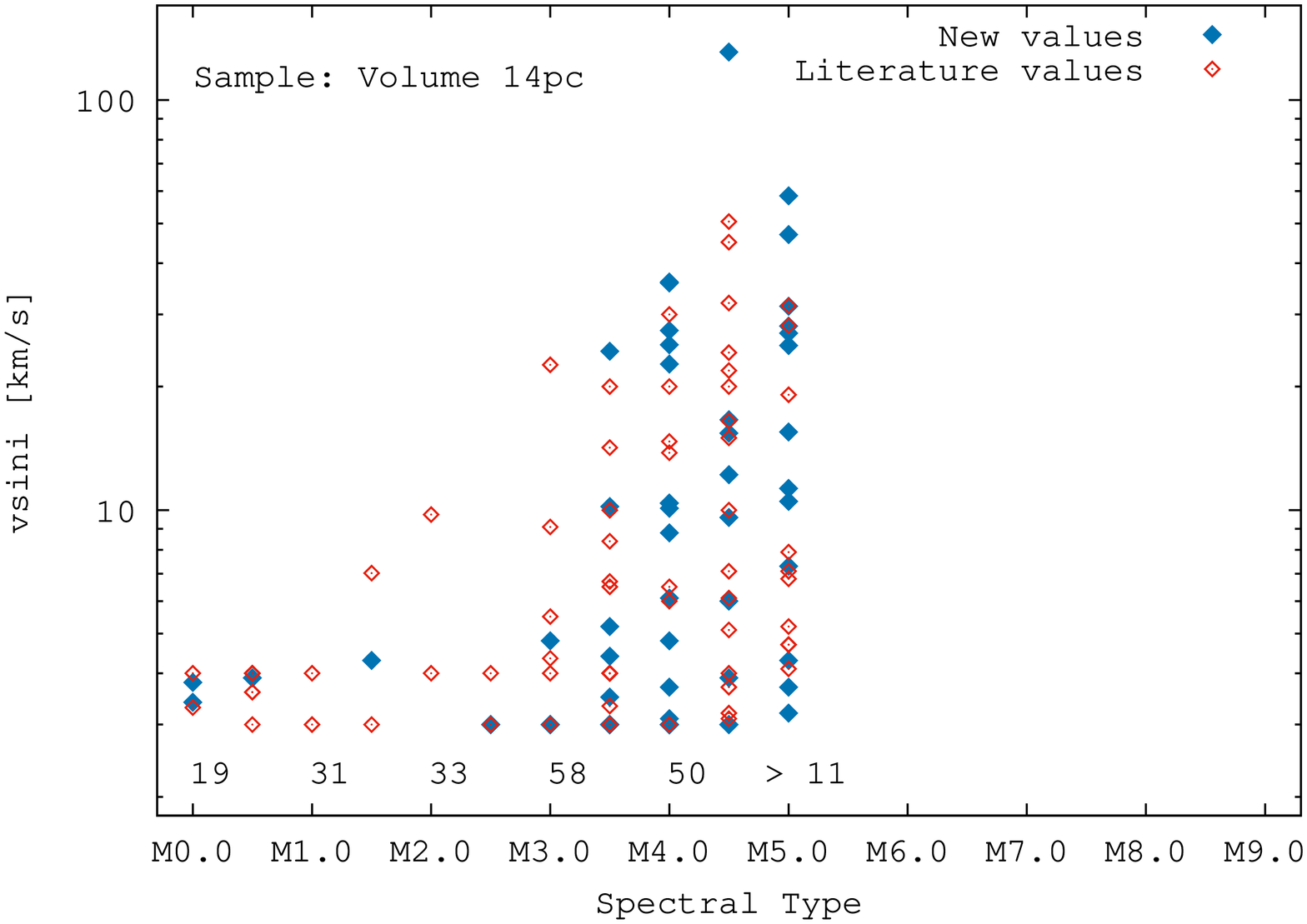}} &
  \imagetop{\includegraphics[angle=270,width=0.48\textwidth]{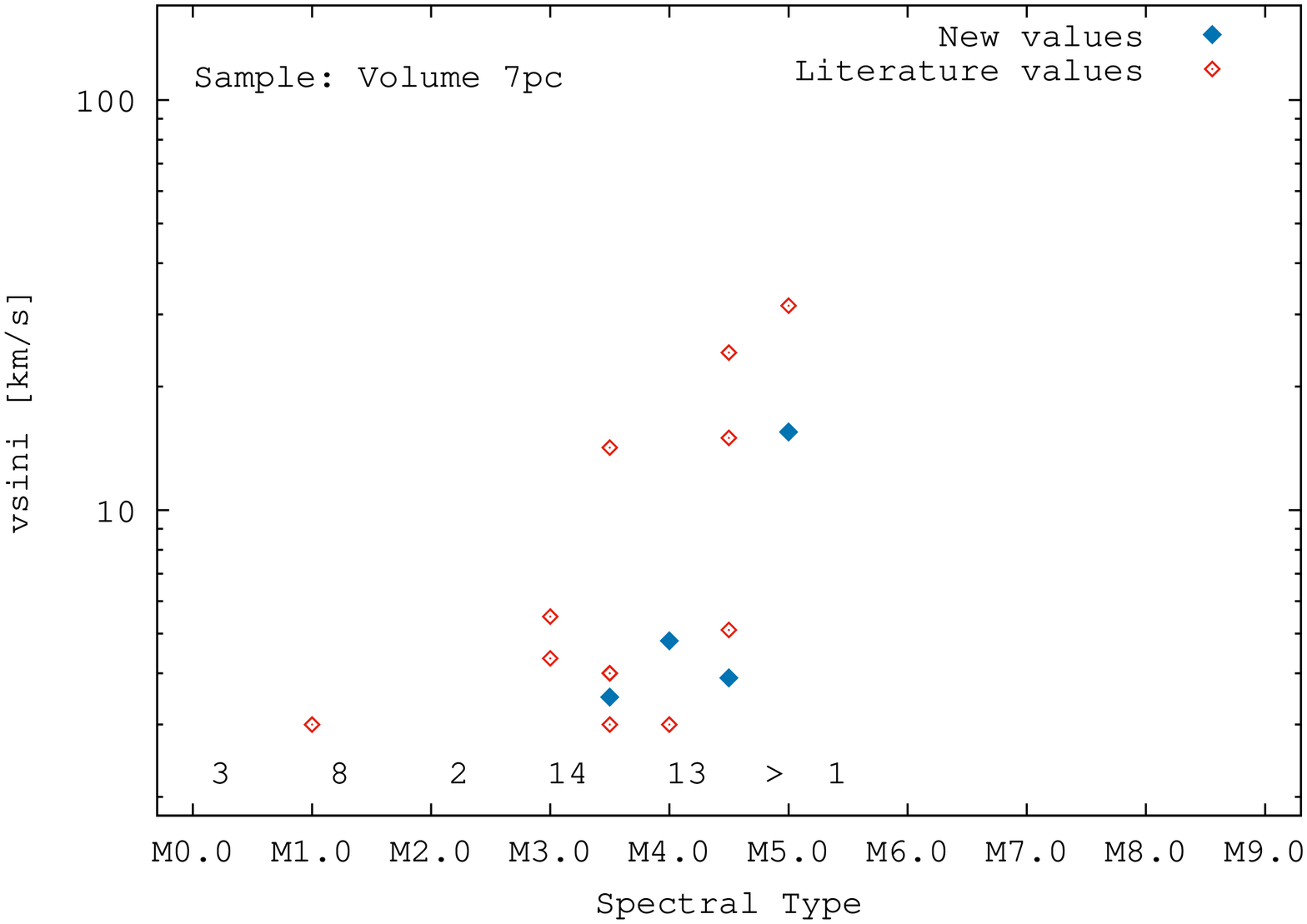}} \\ 
\end{tabular} 
\caption{{Projected rotational velocity ($v \sin{i}$) as a function of spectral type for 
the M dwarfs contained in the Carmencita catalogue (upper left), the CARMENES GTO target sample (upper right), the Volume 14 sample (lower left) and the Volume 7 sample (lower right).   The minimum value plotted is 3 km s$^{-1}$.  The total number of $v \sin{i}$ values not plotted are shown at the bottom of each plot.  All $v \sin{i}$ values in the CARMENES GTO sample are new measurements.}}
\protect\label{f-vsini_SpT} 
\end{center}
\end{figure*}

The resulting $v \sin{i}$ values are tabulated in Table~\ref{tab-rotACT}, where values from this analysis are indicated by the name of the instrument, i.e. CAFE or FEROS, in the right most column for the new observations. The $v \sin{i}$ values for the full CARMENCITA catalogue are shown in Table~\ref{tab-rotCAR}. In Fig.~\ref{f-vsini_SpT} the measured $v \sin{i}$ values are shown as a function of spectral type for (1) the complete Carmencita catalogue, (2) the CARMENES GTO survey sample, (3) the 14\,pc sample, and (4) the 7\,pc sample. 

Each of the four subsamples show a range in $v \sin{i}$ values that extend up to several tens of km s$^{-1}$ for stars with spectral types greater than M3.5\,V.  However, at earlier spectral types ($<$ M3.5\,V) only the Carmencita sample includes a population of moderately to fast rotating stars with $v \sin{i} >$ 5 km s$^{-1}$, while the other three subsamples generally show very low $v \sin{i}$ values, $<$ 5 km s$^{-1}$, at these spectral types. As many of these stars are new measurements, we investigate this sample in greater detail in later subsections.   The largest spread of measured $v \sin{i}$ values is at mid-M spectral types (i.e. M4.0\,V to M5.0\,V), which is present in all four of the samples.  From the Carmencita and the CARMENES GTO samples, there is a lack of slowly rotating stars $v \sin{i} <$ 5 km s$^{-1}$ at spectral types $>$M5.5\,V,  although this could result from the very few stars at these later spectral types.

While these trends dominate, there are additional points that are discussed in later sections.  For example, there are a significant number of  H$\alpha$ active early-M dwarfs (defined as having spectral types $<$ M3\,V) with high $v \sin{i}$ values.  The Carmencita sample also includes much later spectral types (defined as having spectral types $>$ M5\,V), where there are only a few  H$\alpha$ inactive stars. 

\begin{figure*}
\def\imagetop#1{\vtop{\null\hbox{#1}}}
\begin{center}
\begin{tabular}[h]{c c}               
  \imagetop{\includegraphics[angle=270,width=0.48\textwidth]{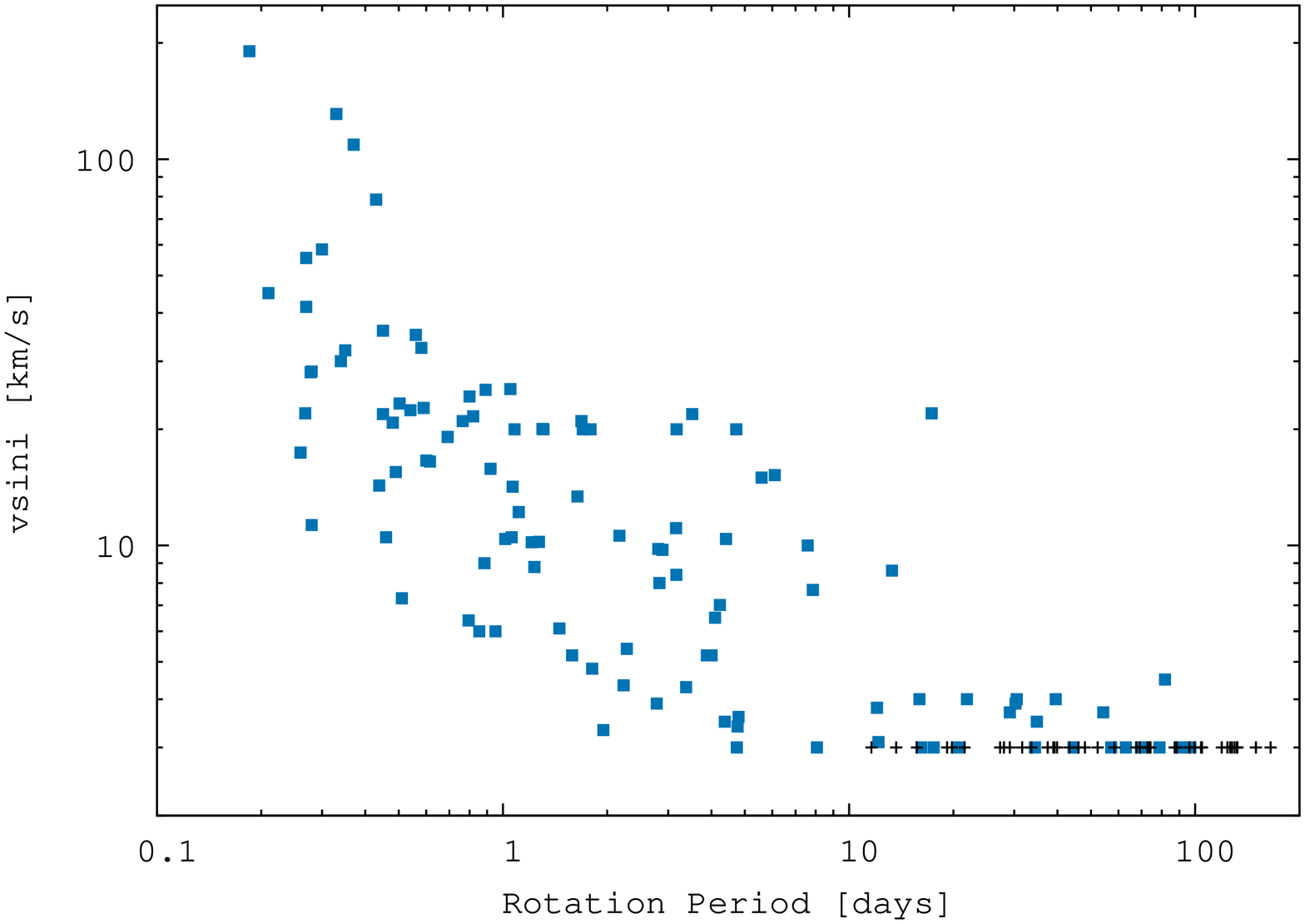}} &    
  \imagetop{\includegraphics[angle=270,width=0.48\textwidth]{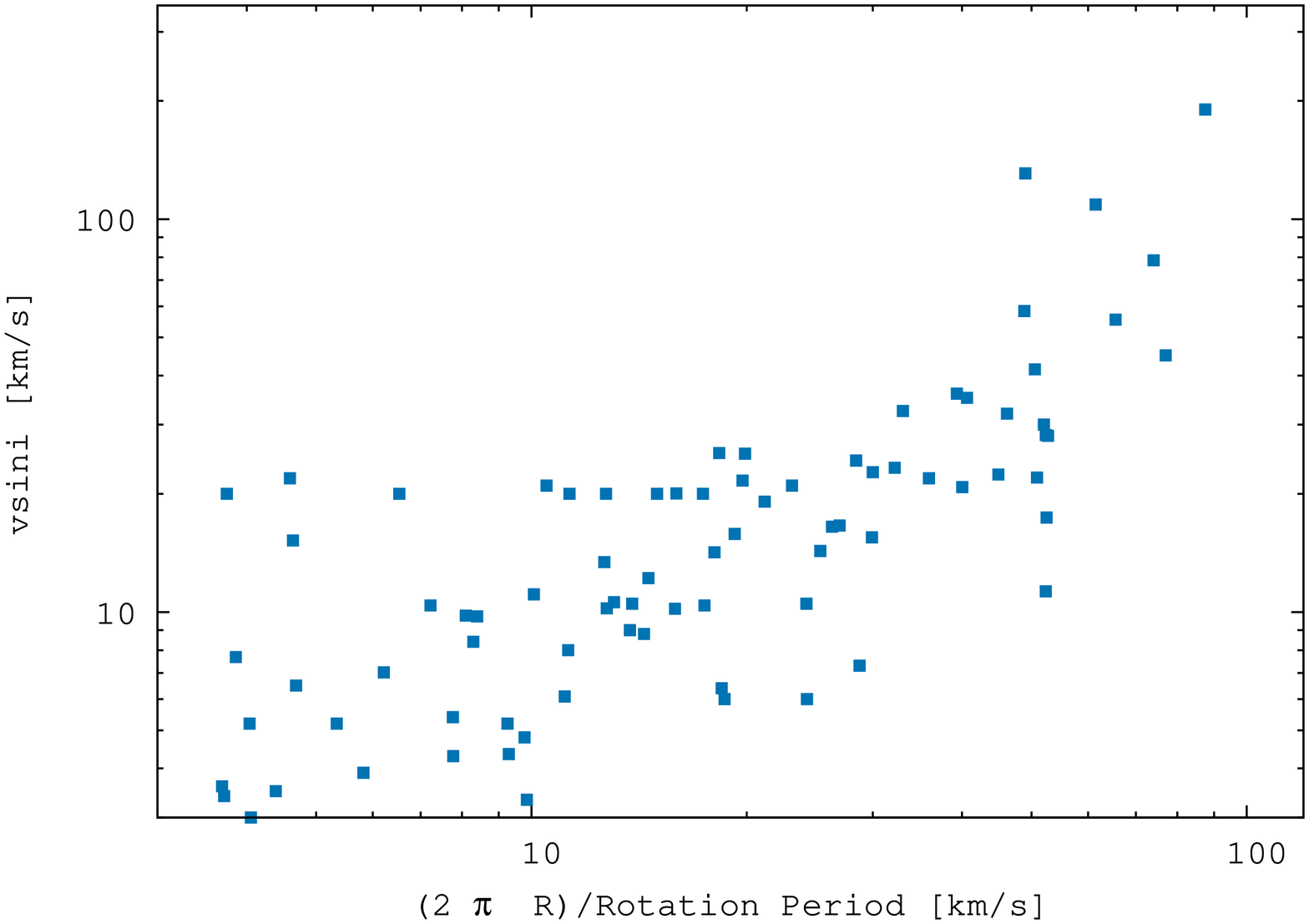}} \\
 \end{tabular} 
\caption{Measured rotational velocity as a function of (left) measured rotational period and inferred rotational velocity from stellar rotation period and radius (right). All of the values are from the Carmencita sample. The points at $v \sin{i}=$3 km s$^{-1}$ with a black cross symbol indicate upper limits. }
\protect\label{f-vsini_Period} 
\end{center}
\end{figure*}

Also listed in Carmencita are the photometric rotation periods of 353 M stars.  The correlation of measured rotation period with $v \sin{i}$ is shown in Fig.~\ref{f-vsini_Period} (left-hand figure), where there is an inverse correlation of rotational velocity with rotational periods for periods ranging from 0.2 to 100 days.  In our analysis, the observed minimim $v \sin{i}$ values of 3 km s$^{-1}$ correspond to rotation periods $<$ 9-10 days. This shows that it is possible to measure the $v \sin{i}$ of stars that are unsaturated in X-rays, which is discussed in more detail in Section 6.5.2.  Additionally, we computed an approximate rotational velocity from the period and the inferred radius of the star, which is shown in  Fig.~\ref{f-vsini_Period} (right-hand figure).   While the contribution of the inclination angle is not included, there is still a clear correlation between the measured $v \sin{i}$ and the value predicted from the rotation period of the star.  It should be noted that there are a few discrepant points with high measured $v \sin{i}$ values with respect to their rotation periods that do not influence the global trend.  For clarity, the $v \sin{i}$ measurements that are below our detection limit are not shown. The same trends are also present in the other three subsamples.

\section{Rotation-activity relation}
\label{sec:rotact}

The assumption that activity and rapid rotation are related was statistically investigated by \cite{Reiners2012} using Kolmogorov-Smirnov statistics.  The result of this test showed that rapid rotation and activity are highly correlated. In this section, we use the large amount of data in the Carmencita catalogue to investigate the correlation of activity with rotation, firstly for well-established activity indicators, and secondly for smaller subsamples of stars that exhibit noteworthy trends.   

\subsection{H$\alpha$ activity}

\subsubsection{H$\alpha$ activity versus  $v \sin{i}$}

The distributions of normalised H$\alpha$ luminosity and $v \sin{i}$ show a dependence on spectral type with earlier spectral types showing lower rotation and activity levels, while later spectral types have a much larger fraction of fast rotating and  H$\alpha$ active stars.  The minimum level of H$\alpha$ emission that we are sensitive to is pEW(H$\alpha$)=-0.5 \AA{}, which depends on the resolution of the spectrograph. Our analysis uses observations with a smaller spectrograph resolution element to detect H$\alpha$ emission than previous works, such as that of \cite{Newton2017}, where the detection limit is -0.75 \AA{}, and the work of ~\cite{West2015}, where the detection limit is -1.0 \AA.  The rotational velocities have a minimum detection level of 3 km s$^{-1}$, which also depends on the resolution of the spectrograph.  

\begin{figure*}
\def\imagetop#1{\vtop{\null\hbox{#1}}}
\begin{center}
\begin{tabular}[h]{c c}               
  \imagetop{\includegraphics[angle=270,width=0.48\textwidth]{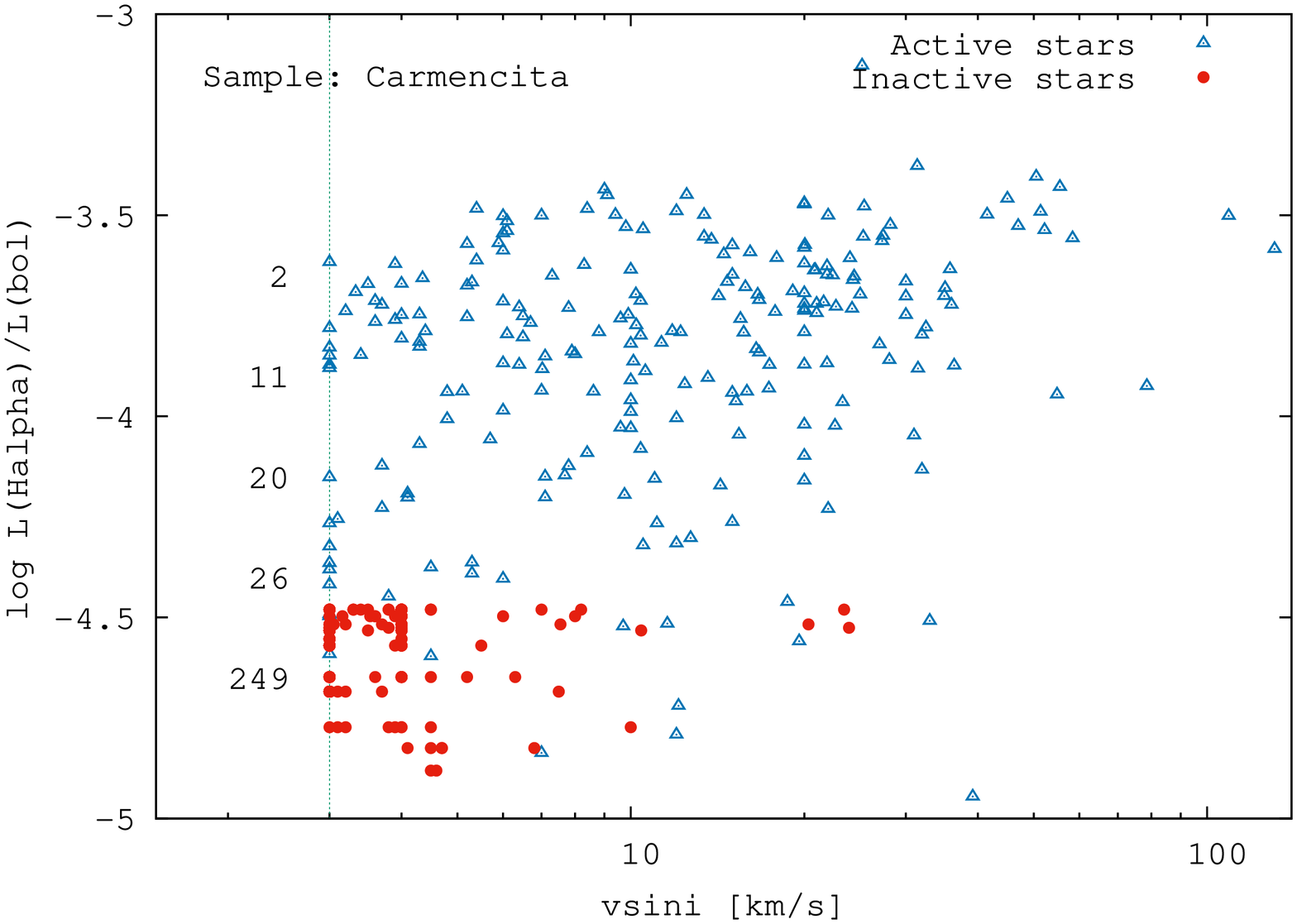}} &
  \imagetop{\includegraphics[angle=270,width=0.48\textwidth]{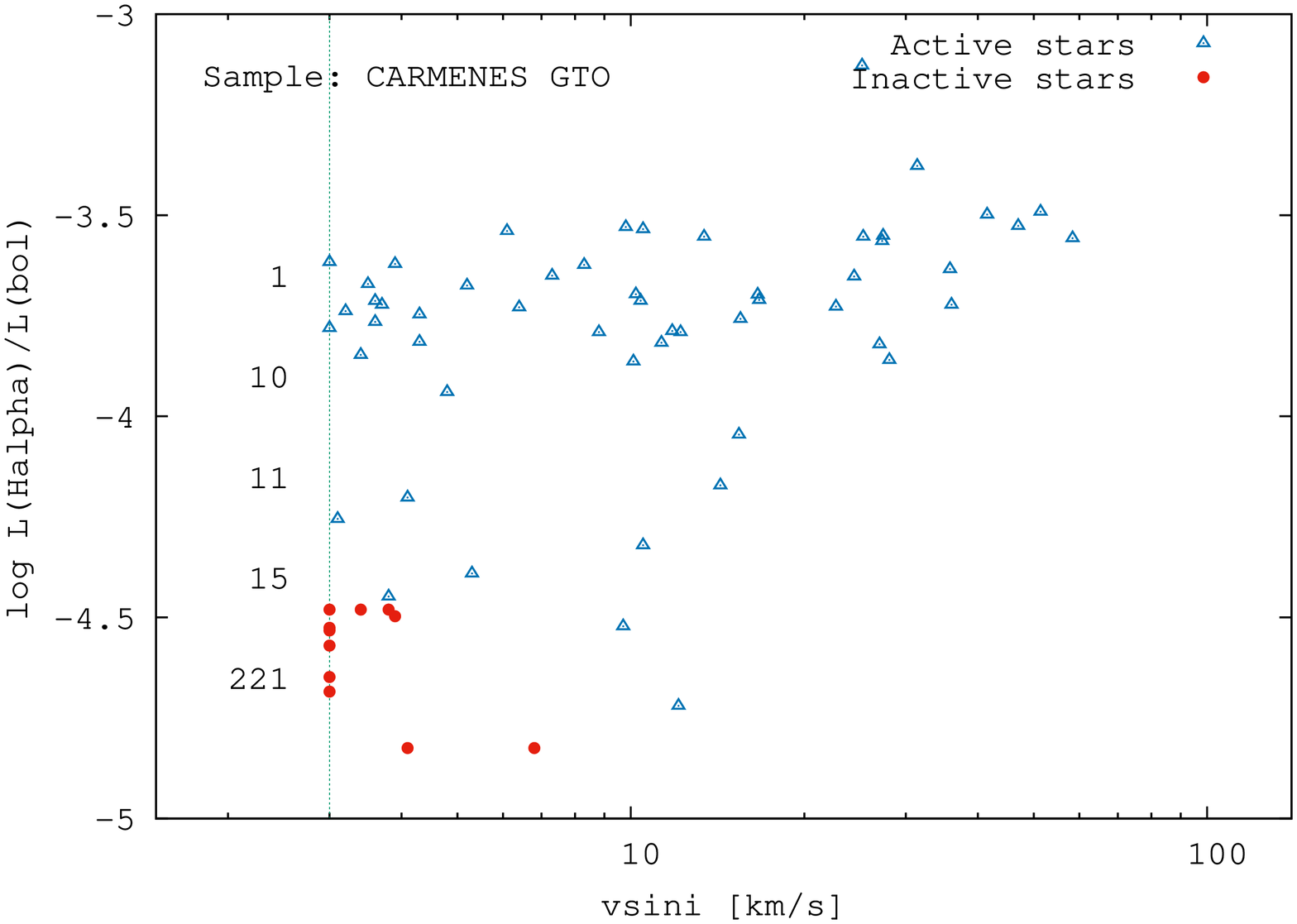}}  \\    
  \imagetop{\includegraphics[angle=270,width=0.48\textwidth]{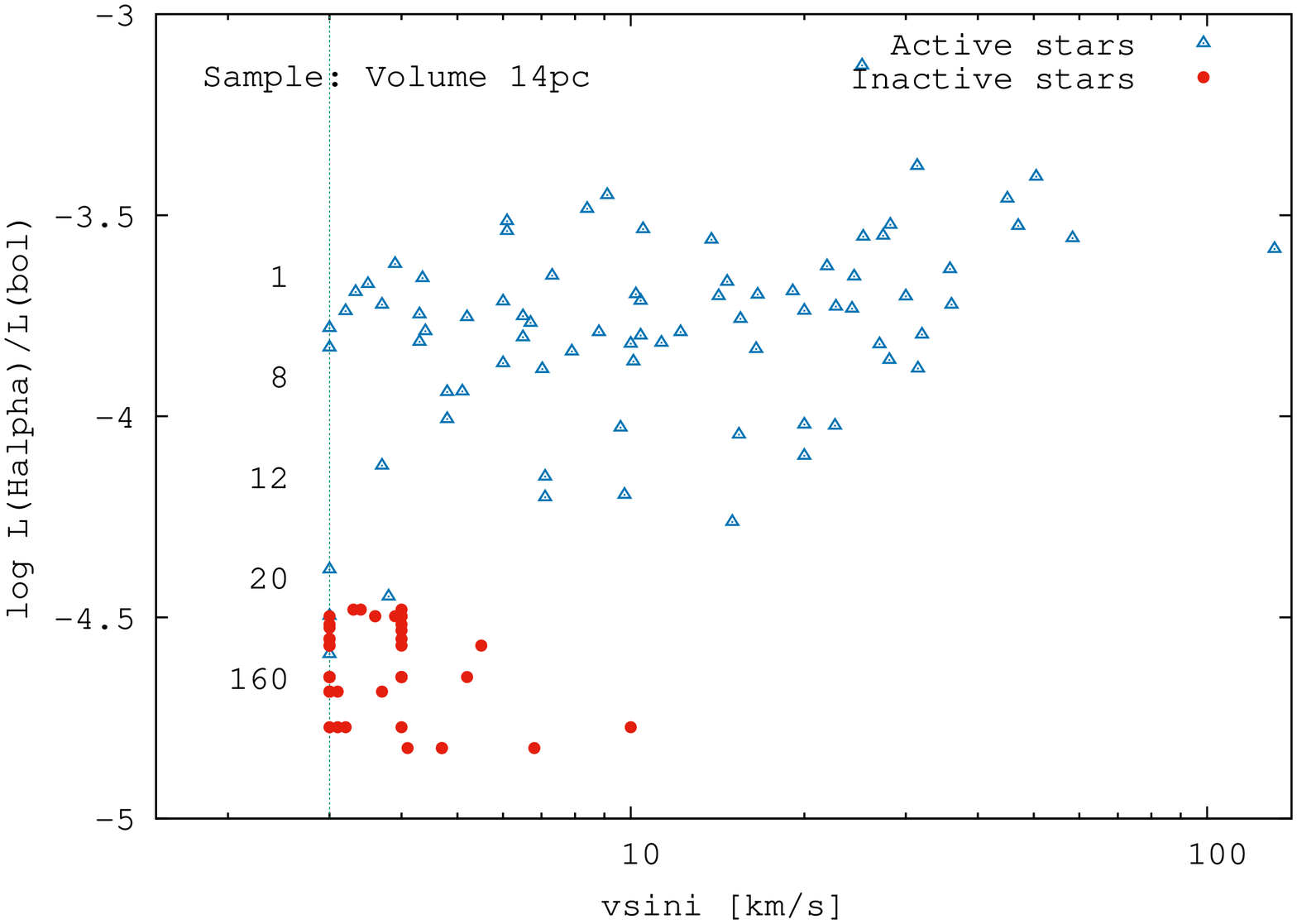}} &
  \imagetop{\includegraphics[angle=270,width=0.48\textwidth]{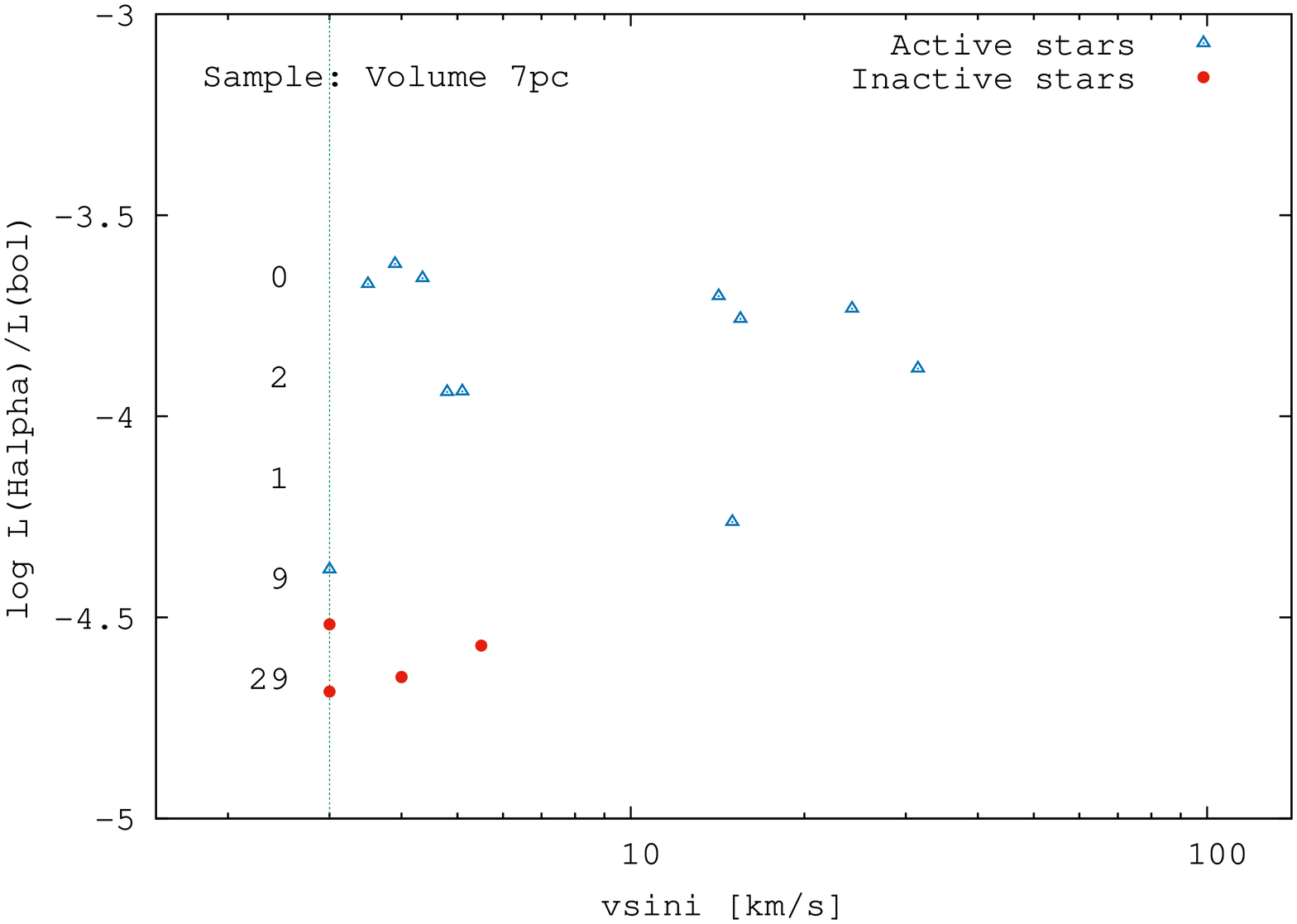}} \\ 
\end{tabular} 
\caption{Distribution of normalised H$\alpha$ luminosity as a function of $v \sin{i}$ showing the distribution for the M dwarfs contained in the Carmencita catalogue (upper left), CARMENES GTO target sample (upper right), the Volume 14 sample (lower left) and the Volume 7 sample (lower right).  Values of  $v \sin{i}$ that are $<$ 3 km s$^{-1}$ are not shown and the number of points excluded are shown to the left of the dashed line. For  H$\alpha$ inactive stars, the value of $L_{{\rm H}\alpha}/L_{\rm bol}$ where the numbers are placed in the figures are at the $L_{{\rm H}\alpha}/L_{\rm bol}$ values that correspond to the detection limit of pEW(H$\alpha$)=-0.5 \AA.}
\protect\label{f-Ha_vsini} 
\end{center}
\end{figure*}

The correlation of the normalised H$\alpha$ luminosity as a function of $v \sin{i}$ is shown in Figure~\ref{f-Ha_vsini} for the four samples of this analysis.  In each of these panels, the global trends are that  stars with  $v \sin{i}$ values that are less than or just above our minimum detection limit of 3 km s$^{-1}$ are  H$\alpha$ inactive, stars with $v \sin{i}$ values $<$ 10 km s$^{-1}$ can range in activity levels from moderate to high activity (e.g. log ($L_{\rm H\alpha}$/$L_{\rm bol}$) values between -4.0 and -3.5), while the most rapidly rotating stars with $v \sin{i}$ values $>$ 10 km s$^{-1}$ have high activity levels with log ($L_{\rm H\alpha}$/$L_{\rm bol}$) values $>$ -4.  The presence of H$\alpha$ active stars below log ($L_{\rm H\alpha}$/$L_{\rm bol}$) values of -4.5 is to due to {{the physics of the stars, which is accounted for in}} the spectral type dependence of the $\chi$ value (Section 4.2.1).  Because of the large number of stars in these samples there are a few small populations  of stars that do not follow these trends, for example stars that are H$\alpha$ inactive that show detectable rotation.  These stars are discussed in Section~\ref{sec:Ha_act_vsini}.

\subsubsection{H$\alpha$ activity versus rotation period}

\begin{figure*}
\def\imagetop#1{\vtop{\null\hbox{#1}}}
\begin{center}
\begin{tabular}[h]{c c}               
  \imagetop{\includegraphics[angle=270,width=0.48\textwidth]{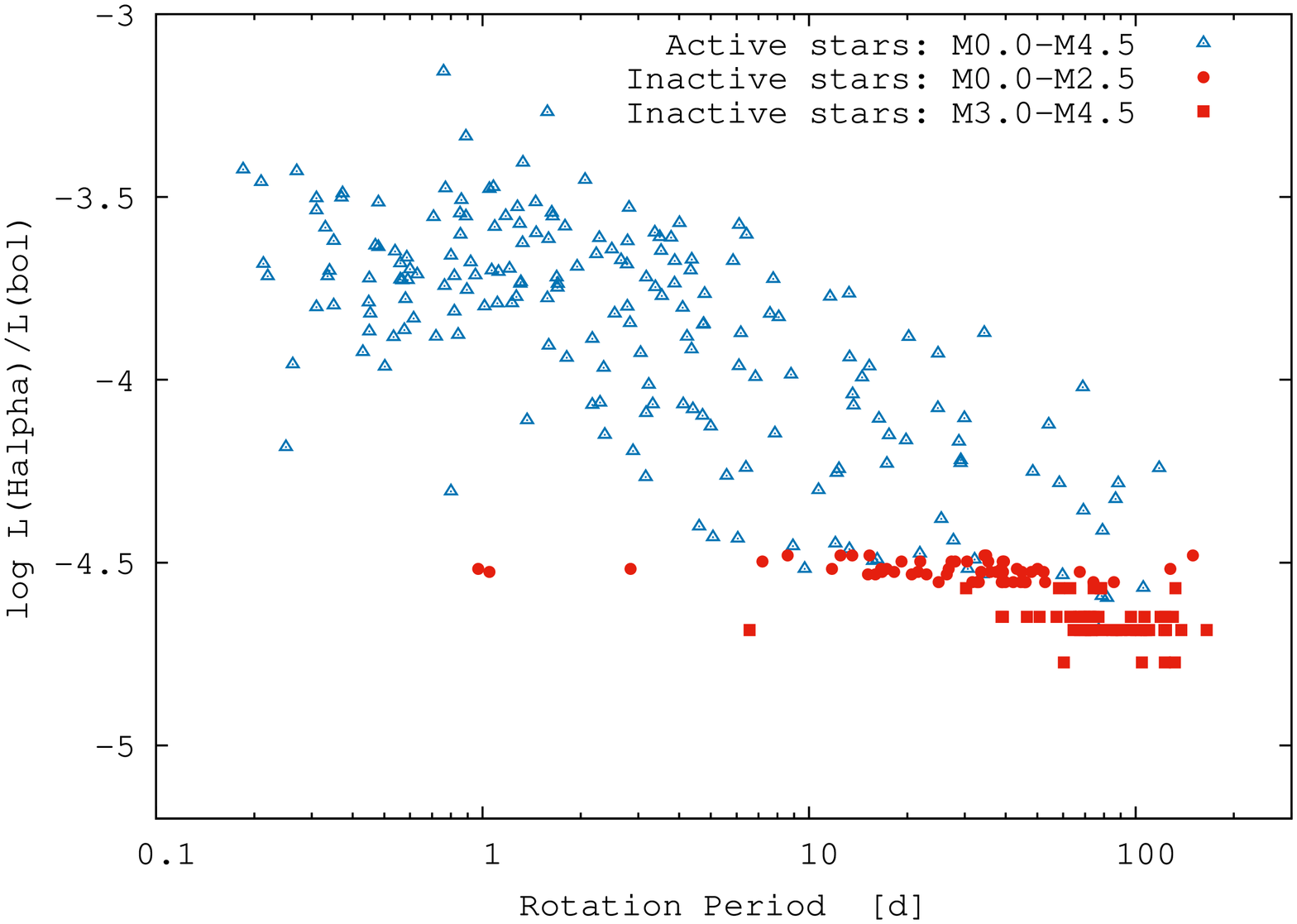}} &    
  \imagetop{\includegraphics[angle=270,width=0.48\textwidth]{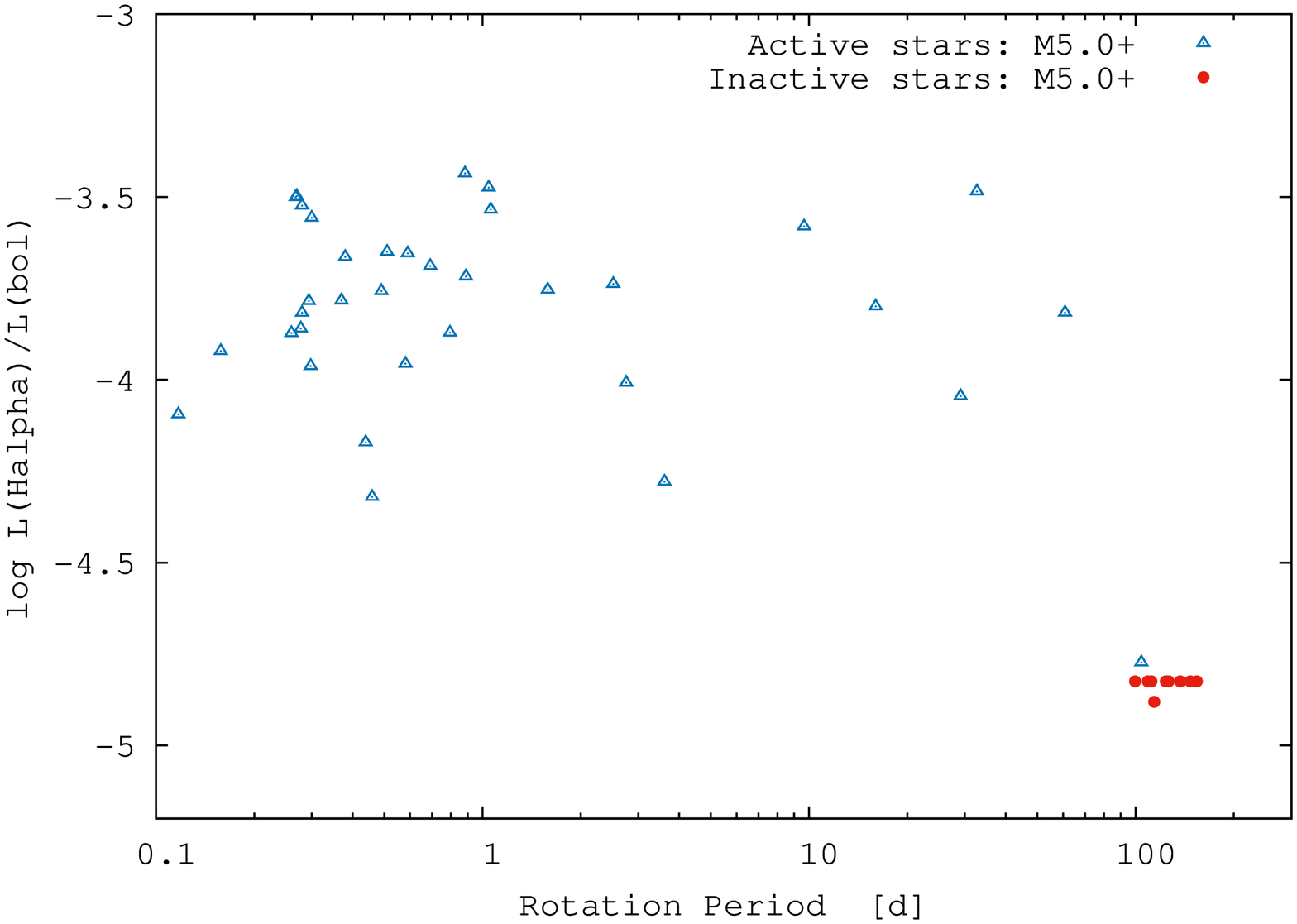}} \\
\end{tabular} 
\caption{Correlation between normalised H$\alpha$ luminosity and rotation period for early-M dwarfs (left panel) and late-M dwarfs (right panel), where  H$\alpha$ inactive stars with measured rotation periods are shown as circles. For  H$\alpha$ inactive stars, the $L_{{\rm H}\alpha}/L_{\rm bol}$ values correspond to the detection limit of pEW(H$\alpha$)=-0.5 \AA.}
\protect\label{f-RTN_Halpha} 
\end{center}
\end{figure*}

Previously, for a smaller sample of 164 M dwarfs, \cite{West2015} investigated the dependence of normalised H$\alpha$ luminosity on rotation period, where they reported a clear decrease in the strength of normalised H$\alpha$ luminosity, or magnetic activity, with increasing rotation period.  This is also evident in the large Carmencita sample as shown in Fig.~\ref{f-RTN_Halpha} for both early-M dwarfs (M0.0 to M4.5) and late-M dwarfs (M5.0 and later).  

The Carmencita sample shows that  H$\alpha$ active early-M stars can have a range of rotation periods from very short ($<$ 1 day) for very  H$\alpha$ active stars to very long ($\sim$100 days) for less  H$\alpha$ active stars.  For H$\alpha$ inactive early-M stars, the rotation periods range from 10 to 100 days as shown in Fig.~\ref{f-RTN_Halpha}.

The original sample of~\cite{West2015} did not show rotation periods $>$ 26 days for  H$\alpha$ inactive stars.  Additionally, our results show that the trend of decreasing normalised H$\alpha$ luminosity with increasing rotational period continues to long rotation periods ($\sim$100 days) for  H$\alpha$ active stars.  For stars with spectral types M5.0 and later (shown in the right-hand panel of Fig.~\ref{f-RTN_Halpha}), the Carmencita sample shows the same global trends as previously noted in Fig. 7 (right-hand panel) of ~\cite{West2015}, where decreasing normalised H$\alpha$ luminosities correlates with increasing rotation period.  The Carmencita sample at these late spectral types is approximately 50\% smaller and so this trend is more sparsely sampled.   For example, the range of rotation periods for  H$\alpha$ active stars ranges from very short ($<$ 1 d) to very long ($\sim$100 d).  Notably, the  H$\alpha$ inactive stars only show rotation periods $>$ 100 days in agreement with the results of ~\cite{West2015}.  

\subsection{Activity saturation}

\subsubsection{Rotation period}

\begin{figure*}
\def\imagetop#1{\vtop{\null\hbox{#1}}}
\begin{center}
\begin{tabular}[h]{c c}               
  \imagetop{\includegraphics[angle=270,width=0.48\textwidth]{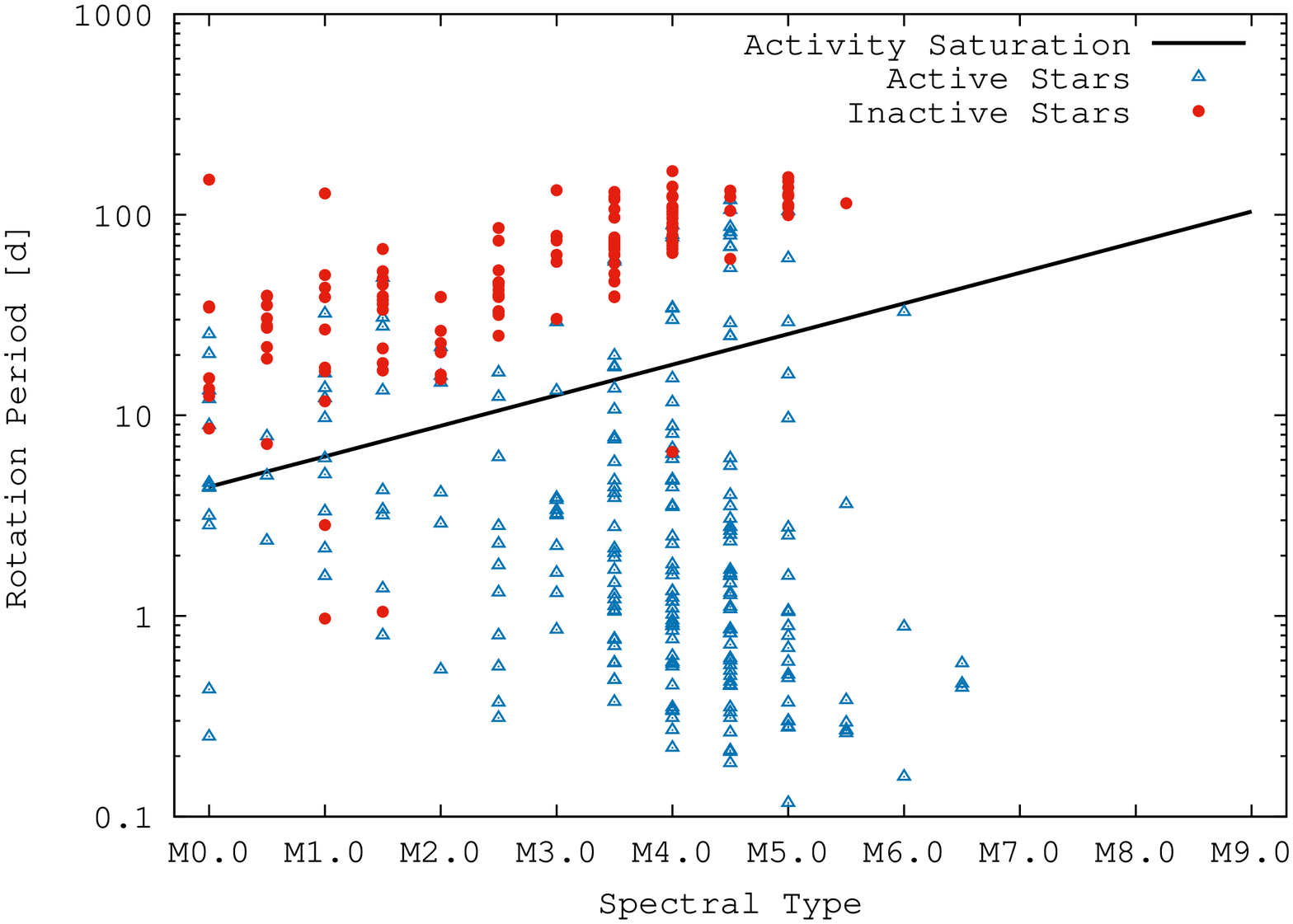}} &    
  \imagetop{\includegraphics[angle=270,width=0.48\textwidth]{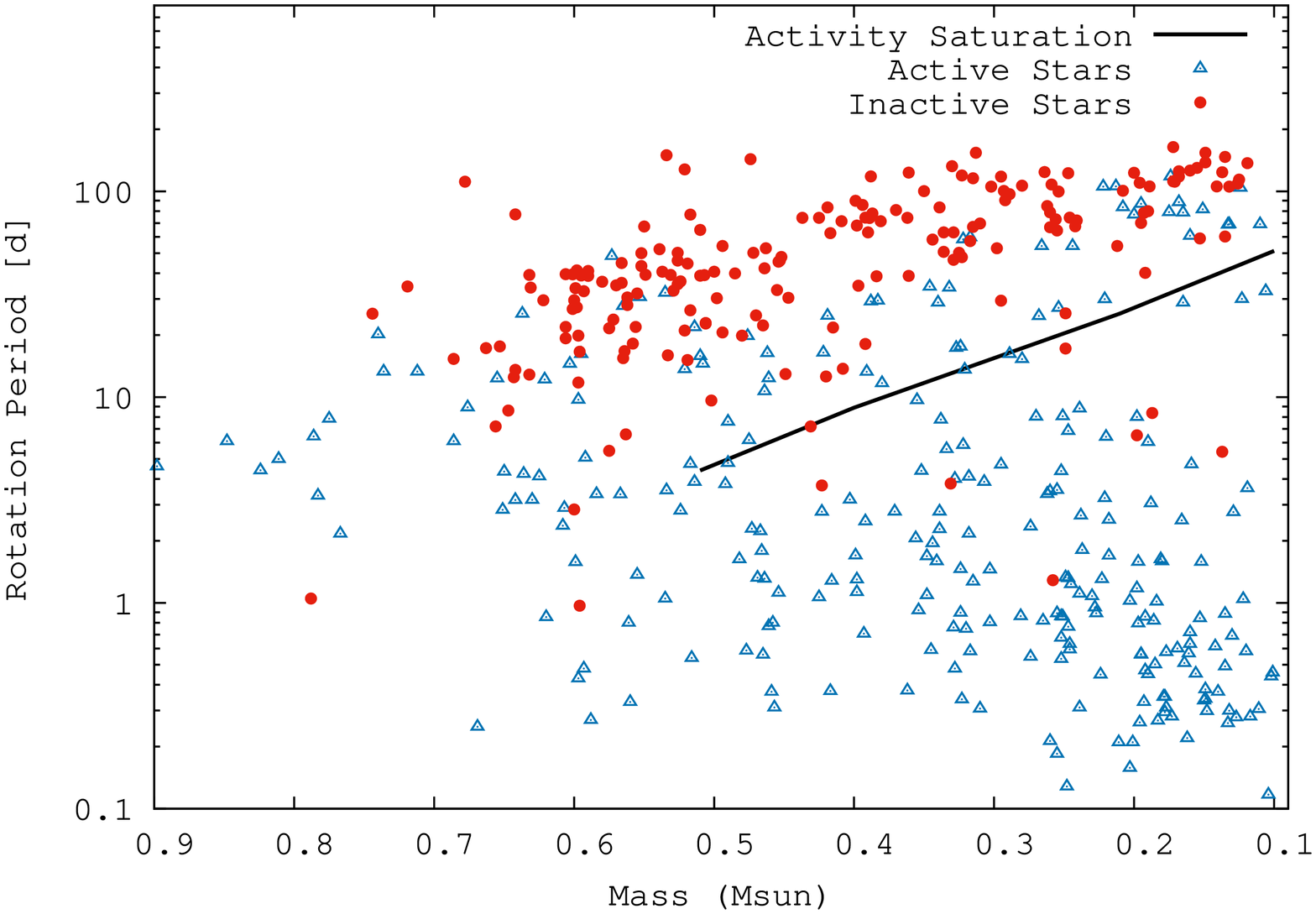}} \\   
 \end{tabular} 
 \caption{Rotation period as a function of spectral type (left-panel) and mass (right panel). The solid black lines indicate the rotation periods at which X-ray saturation sets in (from equation 10 in Reiners et al. (2014)) and are consistent with the results shown in Fig. 8.}
\protect\label{f-RTN_Psat_SPT_MASS}
\end{center}
\end{figure*}

For stars that rotate faster than a certain rotation period, H$\alpha$ is always saturated \citep{Delfosse1998, Mohanty2003, Reiners2012, Douglas2014}.  The rotation period at which saturation occurs, $P_{sat}$, was investigated by \cite{Reiners2014}, where it can be computed as a function of the stellar bolometric luminosity ($L_{\rm bol}$) \citep[Equation 10 in][]{Reiners2014}.  The rotation periods in CARMENCITA as a function of spectral type are shown in Fig.~\ref{f-RTN_Psat_SPT_MASS}, where the left panel shows the correlation as a function of spectral type and the right panel shows the dependence on stellar mass (as determined using the method of Reiners et al. 2017, submitted).  In both Figures, the model rotation period at which saturation occurs is indicated by a solid black line.  The vast majority of the  H$\alpha$ inactive stars have rotation periods longer than the saturation period and the discrepant points that are below the black line are discussed later. There are also a small number of  H$\alpha$ active stars with rotation periods greater than the saturation period. 

\subsubsection{X-ray saturation}

We computed the saturation of stellar activity by investigating the correlation of $L_{\rm x}$/$L_{\rm bol}$ as a function of the generalised Rossby scaling from \cite{Reiners2014}, i.e. $P_{\rm rot}^{-2} \times R_{\star}^{-4}$, where P$_{\rm rot}$ is the rotation period and R$_{\star}$ is the stellar radius.  The results are shown in Fig.~\ref{f-Sat_Halpha_Lx} in the upper left panel. In the saturated regime the $L_{\rm x}$/$L_{\rm bol}$ values remain constant followed by a sharp power-law decrease in $L_{\rm x}$/$L_{\rm bol}$ values in the unsaturated regime.  The stars that comprise the  saturated regime are classified as `X-ray active' stars.  In the unsaturated regime,  the stars are typically  H$\alpha$ inactive, although there are several  H$\alpha$ active stars.  There are indications of a dependence on spectral type.  For spectral types M4.5\,V and earlier, the stars occupy both the saturated and unsaturated regions of the plot.  However, for stars with later spectral types for example M5.0\,V and later, there are no X-ray detections for  H$\alpha$ inactive stars.  This could be a selection effect of the sample for which X-ray detections are available. 

\subsubsection{H$\alpha$ saturation}

The correlation of normalised H$\alpha$ luminosity with the generalised Rossby scaling from \cite{Reiners2014} is shown in the upper right panel of Fig. ~\ref{f-Sat_Halpha_Lx}.  The global behaviour of the correlation is similar to the $L_{\rm x}$/$L_{\rm bol}$ saturation plot and to \cite{Newton2017} (Fig. 11) in comprising a saturated and an unsaturated regime.   The main difference is that the transition from these two regimes is much more gradual compared to the sharp transition seen in the $L_{\rm x}$/$L_{\rm bol}$ saturation plot and in the results of \cite{Newton2017}.  In this transition region, the H$\alpha$ values slowly decrease, or fill in, before reaching the same steep decline for the unsaturated regime.  We conclude that the saturation of normalised H$\alpha$ luminosity behaves differently compared to the $L_{\rm x}$/$L_{\rm bol}$ saturation. In contrast to the results of \cite{Newton2017} we do not see a dependence on either spectral type or stellar mass. 

\subsubsection{Correlation of H$\alpha$ and $L_{\rm x}$/$L_{\rm bol}$}

The correlation of normalised H$\alpha$ luminosity and $L_{\rm x}$/$L_{\rm bol}$ is shown in Fig.~\ref{f-Sat_Halpha_Lx}.  There is a correlation between normalised H$\alpha$ luminosity and $L_{\rm x}$/$L_{\rm bol}$ with high values of $L_{\rm x}$/$L_{\rm bol}$ (i.e. $>10^{-4}$) spanning the full range of normalised H$\alpha$ luminosities from  H$\alpha$ inactive to -3.5.  For lower $L_{\rm x}$/$L_{\rm bol}$ values (e.g.$<10^{-5}$), the stars are generally H$\alpha$ inactive.

\subsubsection{Correlation of $v \sin{i}$ and $L_{\rm x}$/$L_{\rm bol}$}

\begin{figure*}
\def\imagetop#1{\vtop{\null\hbox{#1}}}
\begin{center}
\begin{tabular}[h]{c c}               
  \imagetop{\includegraphics[angle=270,width=0.48\textwidth]{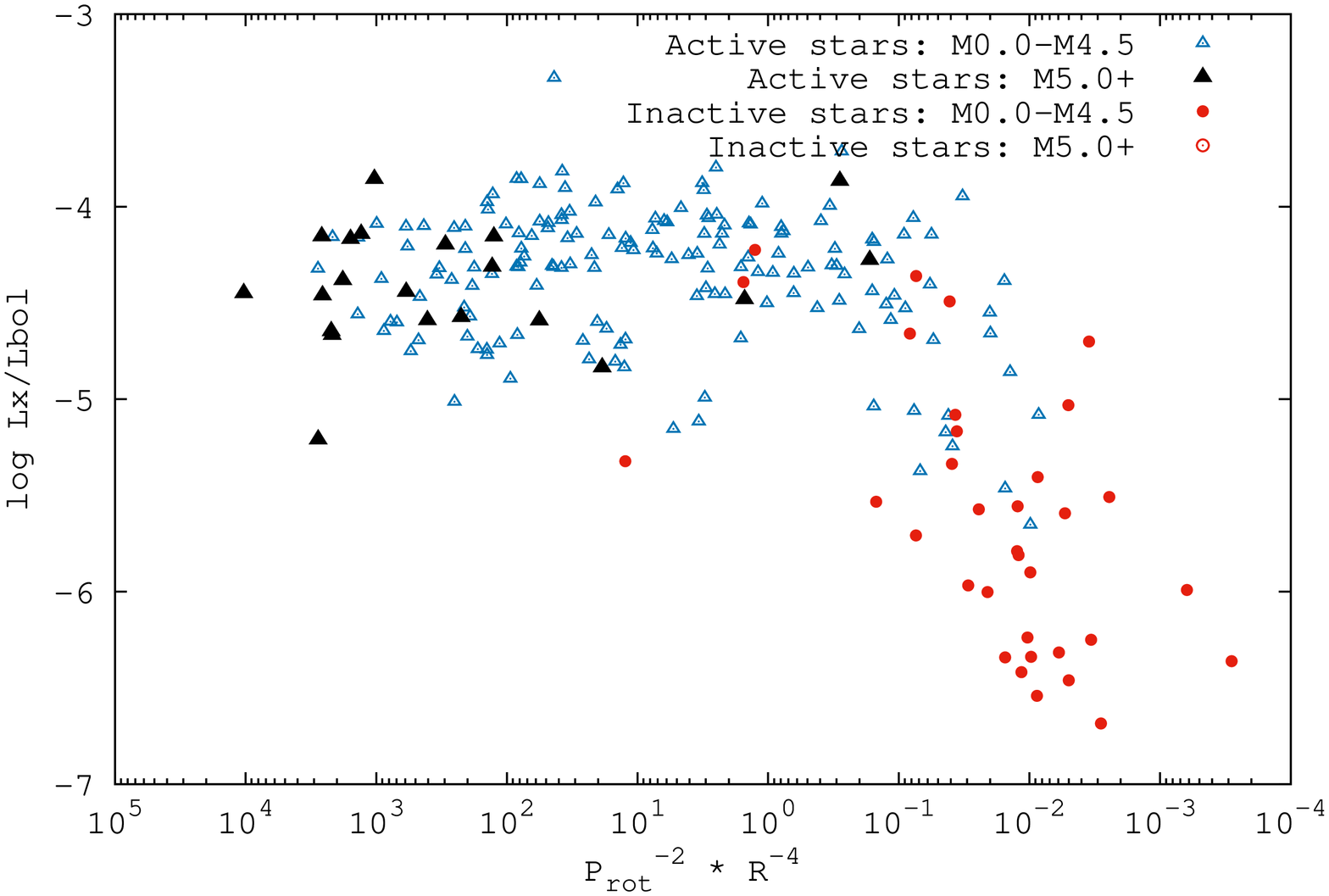}} &    
  \imagetop{\includegraphics[angle=270,width=0.48\textwidth]{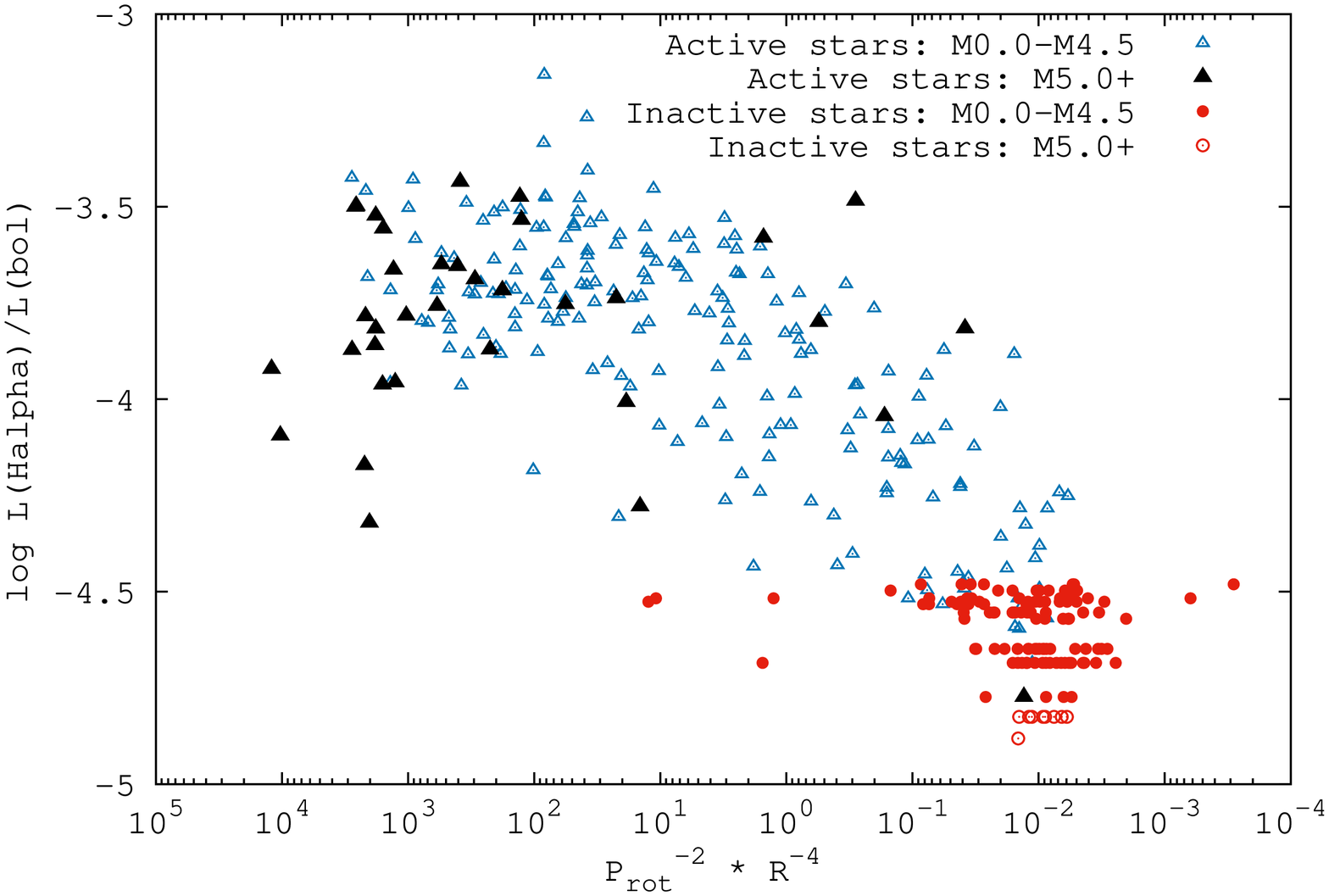}} \\
  \imagetop{\includegraphics[angle=270,width=0.48\textwidth]{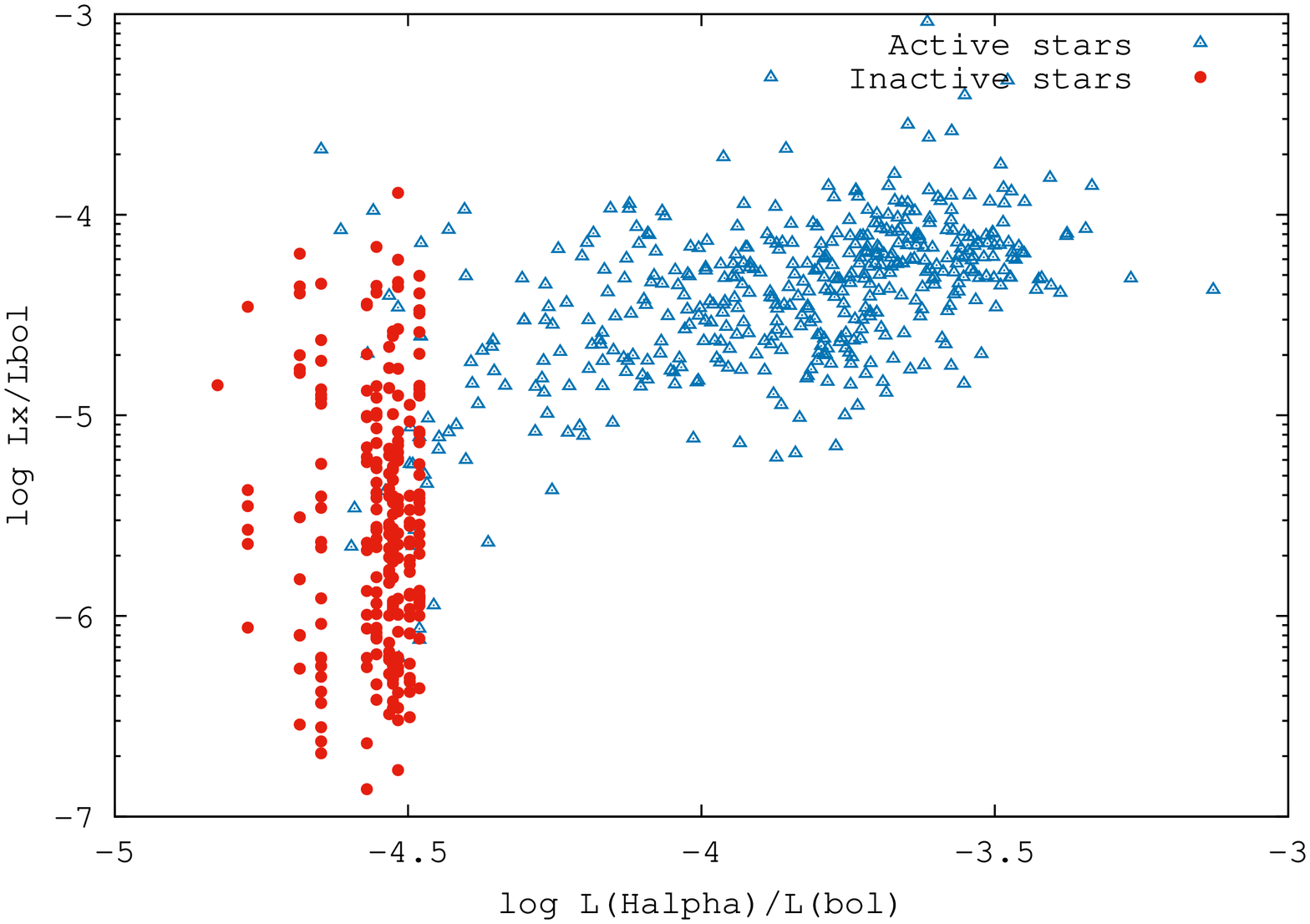}} &    
\imagetop{\includegraphics[angle=270,width=0.48\textwidth]{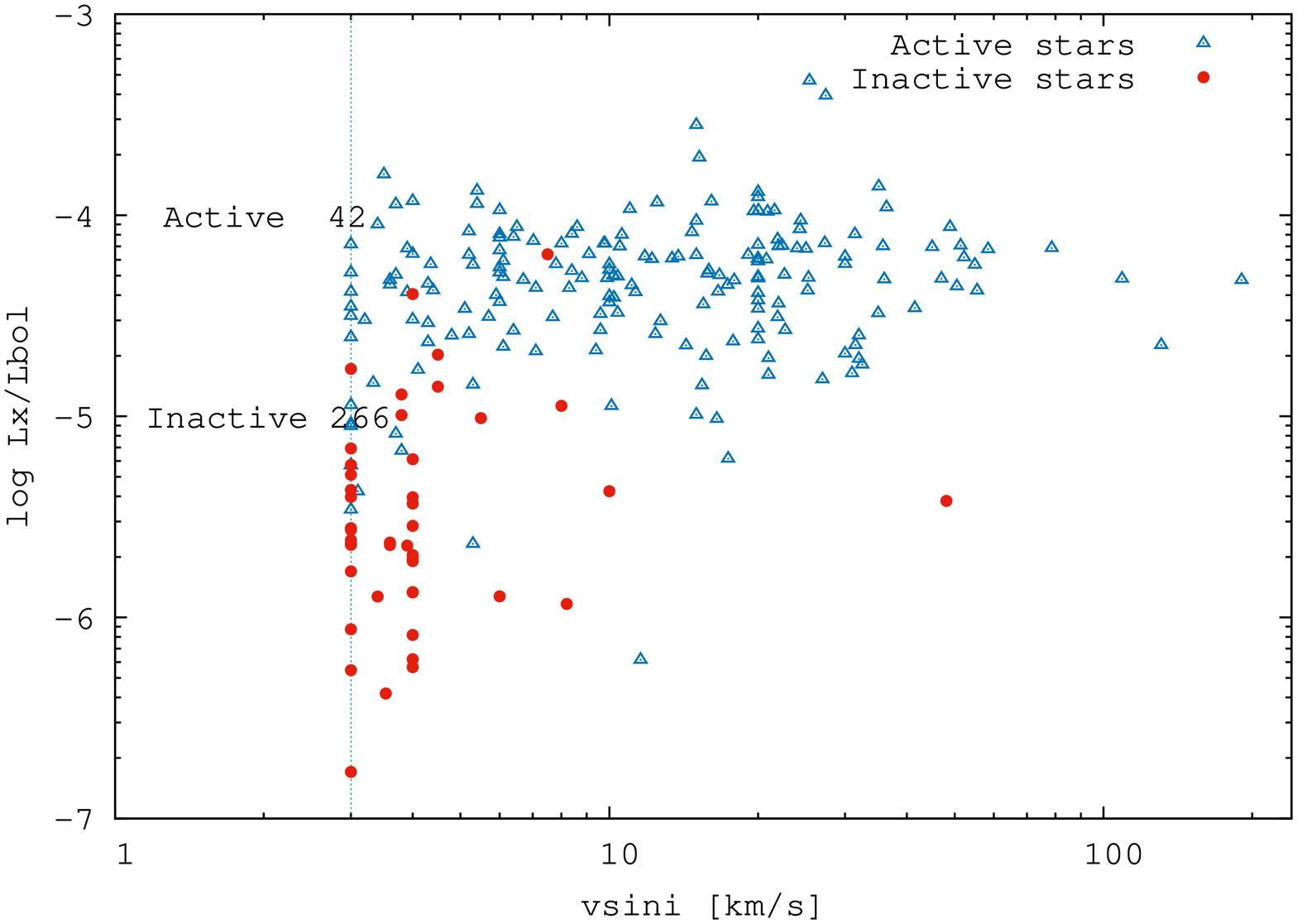}} \\
 \end{tabular} 
\caption{Activity-rotation relation using the generalised Rossby scaling from Reiners et al. (2014) for (upper left panel) $L_{\rm x}$/$L_{\rm bol}$ and (upper right panel) normalised H$\alpha$ luminosity.  The lower left plot shows the correlation of $L_{\rm x}$/$L_{\rm bol}$ and normalised H$\alpha$ luminosity and the lower right plots shows the correlation of $L_{\rm x}$/$L_{\rm bol}$ with $v \sin{i}$, where only detectable $v \sin{i}$ values are shown and the number of points not plotted is shown.  For  H$\alpha$ inactive stars, the value of $L_{{\rm H}\alpha}/L_{\rm bol}$ where the numbers are placed in the figures are at the $L_{{\rm H}\alpha}/L_{\rm bol}$ values that correspond to the detection limit of pEW(H$\alpha$)=-0.5 \AA.}
\protect\label{f-Sat_Halpha_Lx} 
\end{center}
\end{figure*}

The stellar $v \sin{i}$ also shows a correlation with $L_{\rm x}$/$L_{\rm bol}$ (Fig.~\ref{f-Sat_Halpha_Lx}).  In the saturated $L_{\rm x}$/$L_{\rm bol}$ regime, the  $L_{\rm x}$/$L_{\rm bol}$ values remain constant with increasing  $v \sin{i}$ values ranging from our detection limit of 3 km s$^{-1}$ to values $>$ 100 km s$^{-1}$.  Approximately 10\% of the stars in the unsaturated regime have  $v \sin{i}$ $>$ 3 km s$^{-1}$ and are shown in Fig.~\ref{f-Sat_Halpha_Lx}.  Both saturated and unsaturated M stars have detectable $v \sin{i}$ values with values $>$ 5 km s$^{-1}$ always occurring in the saturated regime, whereas values $<$ 5 km s$^{-1}$ can occur in both the saturated and unsaturated regime. 

\subsection{Small populations of stars}

While the correlation of rotation and activity is valid for a population of stars, it may not be valid for all stars individually.  The measurement of $v \sin{i}$ is a projected measurement of rotational velocity of the star and, consequently, has a dependence on the inclination angle of the star.  Additionally, it is important to understand the correlation of the  H$\alpha$ line with $v \sin{i}$. In this section, we investigate the following categories of stars the Carmencita sample:($i$) very  H$\alpha$ active M dwarfs, ($ii$)  H$\alpha$ active early-M dwarfs, ($iii$)  H$\alpha$ active late-M dwarfs, ($iv$) slowly rotating  H$\alpha$ active stars, and ($v$) rapidly rotating  H$\alpha$ inactive M dwarfs.  

\subsubsection{Very active M dwarfs} 

We investigated the 58 very  H$\alpha$ active stars that were identified in Fig.~\ref{f-pEW_SpT} (pEW(H$\alpha$) $<$ -8.0 \AA) in more detail.  The translation of this into normalised H$\alpha$ luminosity includes the constant $\chi$ factor, which depends on spectral type.  As indicated in Fig.~\ref{f-pEW_SpT} the same normalised H$\alpha$ luminosity (e.g. -4) can indicate either a moderately  H$\alpha$ active star for early spectral types ($<$M3.0) or a very  H$\alpha$ active star for late spectral types ($>$M6.0).  We also investigate whether these very  H$\alpha$ active stars are also very young stars.  The stellar spectral type, $v \sin{i}$, pEW(H$\alpha$), X-ray luminosity, and Galactic population membership are shown in Table~\ref{tab-veryactive}.  Of the 58 very active M dwarfs, 45 have X-ray detections and  40 have indications about their age from kinematics.  In particular, 24 belong to the (Galactic) young disc, 14 to the thin disc, and three to the thin-disc/thick-disc transition.  The remaining 18 stars do not have information about their location.  We conclude that most of these stars are very young, which explains their increased activity levels.

\begin{table*}
\centering
\caption{Very  H$\alpha$ active mid- to late-M dwarfs in Carmencita.}
\protect\label{tab-veryactive}
\begin{tabular}{ll c c c c l }
\hline
\hline
   \noalign{\smallskip}
Karmn           & Name                  & SpT   & $v \sin{i}$   & pEW(H$\alpha)$ & Lx/Lbol       & Population$^a$         \\
                &                       & M \,V & km s$^{-1}$ & \AA & \\
\hline
J00122+304       &  2M J00121341+3028443 &  4.5          &               &  -8.70         & 3.74 $\times$ 10$^{-5}$ &    \\
J00245+300       &  G 130-068            &  4.5          &  20.0                 &  -9.98         & 4.84 $\times$ 10$^{-5}$ &  D \\
J01033+623       &  V388 Cas             &  5.0          &               &  -10.10                & 6.98 $\times 10^{-5}$ &  YD  \\
J01114+154       &  LP 467-016 AB        &  5.0          &  17.1                 &  -8.27                 & 4.75 $\times 10^{-5}$ &  YD  \\
J01567+305       &  NLTT 6496            &  4.5          &               &  -16.00                & 4.81$\times 10^{-5}$ &       \\
J02142-039       &  LP 649-072           &  5.5          &  12.0                 &  -12.30                & ... &  D     \\
J02171+354       &  LP 245-010           &  5.0          &  28.2                 &  -10.00        & ... &  YD    \\
J03510+142       &  2M J03510078+1413398 &  4.5          &               &  -13.40                & 8.49$\times 10^{-5}$ &  D    \\
J03548+163       &  LP 413-108           &  4.0          &               &  -8.60                 & 1.15$\times 10^{-4}$ &       \\
J04059+712W      &  LP 031-302 BC                &  5.0          &               &  -8.90                 & 1.24$\times 10^{-4}$ &       \\
J04173+088       &  LTT 11392            &  4.5          &  190.28       &  -11.15        & 4.76$\times 10^{-5}$ &  YD   \\
J04206+272       &  XEST 16-045          &  4.5          &               &  -8.90                 & 2.02$\times 10^{-5}$ &       \\
J04234+809       &  1RXS J042323.2+805511        &  4.0          &               &  -8.20                 & 6.47$\times 10^{-5}$ &       \\
J04238+149       &  IN Tau               &  3.5          &               &  -9.10                 & 4.08$\times 10^{-5}$ &       \\
J04313+241       &  V927 Tau AB          &  4.5          &  13.4                 &  -9.40                 & ... &        \\
J04373+193       &  LP 416-1644          &  4.0          &               &  -8.30                 & 5.96$\times 10^{-5}$ &       \\
J04393+335       &  RX J0439.4+3332B     &  4.0          &               &  -11.20                & 1.39$\times 10^{-4}$ &  D    \\
J04472+206       &  RX J0447.2+2038      &  5.0          &               &  -14.40                & 4.83$\times 10^{-5}$ &  YD   \\
J05072+375       &  1RXS J050714.8+373103        &  5.0          &               &  -9.00                 & 5.15$\times 10^{-5}$ &       \\
J05084-210       &  2M J05082729-2101444         &  5.0          &               &  -24.90        & 4.22$\times 10^{-5}$ &  YD   \\
J05152+236       &  2M J05151753+2336260         &  5.0          &               &  -8.40                 & ... &        \\
J05187+464       &  2M J05184455+4629597         &  4.5          &               &  -10.26        & ... &  YD    \\
J05243-160       &  1RXS J052419.1-160117        &  4.5          &               &  -11.70        & 4.43$\times 10^{-5}$ &  YD   \\
J05394+406       &  LSR J0539+4038       &  8.0          &               &  -9.54         & ... &        \\
J06054+608       &  LP 086-173           &  4.5          &               &  -9.30                 & 8.15$\times 10^{-5}$ &  TD   \\
J06075+472       &  LSPM J0607+4712      &  4.5          &               &  -9.20                 & 1.16$\times 10^{-4}$ &  YD   \\
J06318+414       &  LP 205-044           &  5.0          &  39.6                 &  -9.26         & 6.78$\times 10^{-5}$ &  YD   \\
J07310+460       &  1RXS J073101.9+460030        &  4.0          &               &  -9.50                 & 1.52$\times 10^{-4}$ &  YD   \\
J07523+162       &  LP 423-031           &  6.0          &  9.0                  &  -25.40                & ... &  D     \\
J07591+173       &  2M J07590718+1719474         &  4.0          &               &  -9.20                 & 4.83$\times 10^{-5}$ &       \\
J08404+184       &  AZ Cnc               &  6.0          &               &  -22.68        & 1.35$\times 10^{-3}$ &  D    \\
J08536-034       &  LP 666-009           &  9.0          &  13.5                 &  -8.08         & 2.12$\times 10^{-5}$ &  YD   \\
J09003+218       &  LP 368-128           &  6.5          &  20.0                 &  -10.00                & 2.25$\times 10^{-5}$ &       \\
J09449-123       &  G 161-071            &  5.0          & 30.0 &  -17.10                & 8.03$\times 10^{-5}$ &  YD    \\
J09593+438W      &  G 116-072 A          &  3.5          &               &  -15.49        & ... &  D     \\
J10028+484       &  G 195-055            &  5.5          &  22.0                 &  -12.00                & 7.01$\times 10^{-5}$ &  YD   \\
J10564+070       &  CN Leo               &  6.0          &  2.82                 &  -9.06                 & 1.72$\times 10^{-5}$ &  D    \\
J11055+435       &  WX UMa               &  5.5          &  7.7                  &  -10.20                & 4.34$\times 10^{-5}$ &  TD-D \\
J11474+667       &  1RXS J114728.8+664405       &  5.0   &               &  -9.20                 & 1.43$\times 10^{-5}$ &       \\
J13143+133       &  NLTT 33370 AB               &  6.0   &  45.0                 &  -8.30                 & ... &  YD    \\
J13317+292       &  DG CVn AB            &  4.0          &  55.5                 &  -9.00                 & 4.22$\times 10^{-5}$ &  YD   \\
J14130-120       &  GQ Vir               &  4.5          &               &  -9.69                 & 9.14$\times 10^{-5}$ &  TD-D \\
J14142-153       &  LTT 5581             &  3.5          &               &  -12.13        & ... &  YD    \\
J14279-003S      &  GJ 1183A             &  4.5          &               &  -12.42        & 7.83$\times 10^{-5}$ &  D    \\
J14472+570       &  RX J1447.2+5701      &  4.0          &  4.30                 &  -8.17                 & 1.30$\times 10^{-4}$ &  D    \\
J15079+762       &  LSPM J1507+7613      &  4.5          &               &  -9.00                 & ... &        \\
J15166+391       &  LP 222-065           &  6.5          &               &  -10.71        & ... &  YD    \\
J15238+584       &  G 224-065            &  4.0          &               &  -8.20                 & 6.09$\times 10^{-5}$ &  D    \\
J15349-143       &  2MUCD 11346          &  7.0          &               &  -11.80                & ... &  D     \\
J16403+676       &  LP 069-457           &  5.5          &               &  -8.24         & 7.20$\times 10^{-5}$ &  YD   \\
J17198+265       &  V639 Her             &  4.5          &  6.79                 &  -9.07                 & 4.92$\times 10^{-5}$ &  YD   \\
J17338+169       &  1RXS J173353.5+165515        &  5.5          &  106.32       &  -12.48        & 3.45$\times 10^{-5}$ &  YD   \\
J19169+051S      &  V1298 Aql (vB 10)    &  8.0          &  6.5                  &  -9.50                 & 5.05$\times 10^{-6}$ &  D    \\
J19312+361       &  G 125-015            &  4.5          &               &  -8.50                 & 1.24$\times 10^{-4}$ &  YD   \\
J20093-012       &  2M J20091824-0113377         &  5.0          &               &  -8.60                 & 2.91$\times 10^{-5}$ &       \\
J21173+640       &  G 262-038            &  5.0          &               &  -8.77                 & 3.30$\times 10^{-5}$ &  D    \\
J21376+016       &  GSC 00543-00620      &  4.5          &  25.01                &  -10.31        & 6.95$\times 10^{-5}$ &  YD   \\
J23228+787       &  NLTT 56725           &  5.0          &               &  -11.20                & 4.86$\times 10^{-5}$ &       \\
\noalign{\smallskip}
\hline
\hline
\end{tabular}
\begin{list}{}{}
\item[$^{a}$] Galactic population according to \cite{Cortes2016}.  YD: young disc, D: thin disc, TD-D: thin disc-thick disc transition
\end{list}
\end{table*}

\begin{table*}
\centering
\caption{H$\alpha$ active early-M dwarfs in Carmencita with $v \sin{i}$ measurements.}
\protect\label{tab-feactive}
\begin{tabular}{ll c c c c l c}
\hline
\hline
   \noalign{\smallskip}
Karmn           & Name                  & SpT   & $v \sin{i}$   & pEW(H$\alpha)$         & Lx/Lbol & Stellar kinematic   & Ref$^{a}$.  \\
                        &                               & (M\,V)        & [km\,s$^{-1}$] & [\AA]          &       & group or age          &       \\
   \noalign{\smallskip}
\hline
   \noalign{\smallskip}
J00428+355  &  FF And  & 1.0 & 10.6 & -2.13 & 7.99$\times$10$^{-5}$ &  AB Doradus & a \\
J03233+116  &  G 005-032  & 2.5 & $<$3.0 & -2.36 & 2.48$\times$10$^{-5}$ & $\tau$ $>$ 1 Gyr & b \\
J03332+462  &  BD+45 784B  & 0.0 & 15.0 & -3.40 & 2.81$\times$10$^{-4}$ &  AB Doradus & c \\
J03574-011  &  BD-01 565B  & 2.5 & 10.0 & -2.20 & 5.35$\times$10$^{-5}$ & $\beta$ Pictoris  & d \\
J04376-024  &  StKM 1-497 Cab  & 1.0 & 15.2 & -1.79 & 1.94$\times$10$^{-4}$ & $\beta$ Pictoris  & d \\
J04595+017  &  V1005 Ori  & 0.0 & 10.4 & -1.26 & 4.96$\times$10$^{-5}$ & $\beta$ Pictoris  & d \\
J05068-215E  &  BD-21 1074 A  & 1.5 & 8.6 & -1.93 & 8.73$\times$10$^{-5}$ & $\beta$ Pictoris  & e \\
J05337+019  &  V371 Ori  & 2.5 & 9.8 & -5.29 & 7.22$\times$10$^{-5}$ & $\beta$ Pictoris  & d \\
J06171+751  &  TYC 4525-194-1  & 2.0 & 22.4 & -3.82 & 7.05$\times$10$^{-5}$ & $\beta$ Pictoris  & d \\
J07295+359  &  1RXS J072931.4+355607  & 1.5 & 15.0 & -1.92 & 6.34$\times$10$^{-5}$ & $\beta$ Pictoris  & f \\
J07346+318  &  YY Gem CD  & 1.0 & 36.4 & -2.20 & 1.10$\times$10$^{-4}$ &  Castor & g \\
J09177+462  &  RX J0917.7+4612  & 2.5 & 35.1 & -3.73 & 1.39$\times$10$^{-4}$ &   Castor & h \\
J10043+503  &  G 196-003A  & 2.5 & 20.0 & -3.31 & 1.23$\times$10$^{-4}$ &  $\tau$ = 10--150 Myr & f \\
J10143+210  &  DK Leo  & 0.5 & 7.7 & -1.12 & 3.11$\times$10$^{-5}$ & Pleidaes Supercluster & i \\
J11159+553  &  StKM 1-932  & 0.5 & 30.0 & -3.40 & 5.72$\times$10$^{-5}$ &    &  \\
J11201-104  &  LP 733-099  & 2.0 & 3.6 & -3.30 & 4.52$\times$10$^{-5}$ &    &  \\
J12576+352E  &  BF CVn  & 1.5 & 8.4 & -1.36 & 8.08$\times$10$^{-5}$ & $\tau$ = 10--150 Myr & j \\
J13007+123  &  DT Vir AB  & 2.0 & 9.8 & -1.09 & 7.24$\times$10$^{-5}$ & $\tau$ = 10--150 Myr & j \\
J13518+127  &  RX J1351.8+1247  & $<$2.0 & 20.0 & -1.18 & 4.10$\times$10$^{-5}$ &    &  \\
J14200+390  &  StKM 1- 1145  & 2.5 & 109.0 & -5.64 & 4.83$\times$10$^{-5}$ &    &  \\
J14544+161  &  CE Boo A  & 1.0 & 2.5 & -1.40 & 2.03$\times$10$^{-5}$ &  Ursa Major & j \\
J15218+209  &  OT Ser  & 1.5 & 4.3 & -3.01 & 4.57$\times$10$^{-5}$ &  $\tau$ = 10--150 Myr & k \\
J15273+415  &  TYC 3055-1525-1  & 1.5 & 7.0 & -5.30 & ...   &  &  \\
J15480+043  &  RX J1548.0+0421  & 2.5 & 20.0 & -4.70 & 6.07$\times$10$^{-5}$ &    &  \\
J15597+440  &  RX J1559.7+4403  & 2.0 & 54.9 & -1.93 & 5.67$\times$10$^{-5}$ & $\tau$ = 10--150 Myr & j \\
J16170+552  &  CR Dra AB  & 1.0 & 17.4 & -1.93 & 4.50$\times$10$^{-5}$ &    &  \\
J18174+483  &  TYC 3529-1437-1  & 2.0 & $<$2.0 & -1.51 & 1.67$\times$10$^{-5}$ &    &  \\
J20435+240  &  Wolf 1360  & 2.5 & 24.3 & -3.91 & 8.53$\times$10$^{-5}$ & Field & l \\
J21000+400  &  V1396 Cyg AB  & 1.5 & 4.3 & -2.50 & ...   &  Ursa Major & j \\
J21100-193  &  BPS CS 22898-0065  & 2.0 & 9.4 & -5.40 & ...   & $\beta$ Pictoris  & m \\
J21185+302  &  GSC 02703-00706  & 1.5 & 10.0 & -1.84 & 3.68$\times$10$^{-5}$ & $\beta$ Pictoris & n \\
J22387-206S  &  FK Aqr  & 1.5 & 7.0 & -2.20 & 7.47$\times$10$^{-5}$ &  Ursa Major & j \\
J23060+639  &  MCC 858  & 0.0 & 8.0 & -2.16 & 7.24$\times$10$^{-5}$ &  AB Doradus & o \\
J23083-154  &  HK Aqr  & 0.0 & 78.6 & -1.80 & 6.87$\times$10$^{-5}$ &  Castor & h \\
J23327-167  &  HD 221503 BC  & 2.0 & 7.1 & -2.40 & 4.35$\times$10$^{-5}$ &  $\tau$ = 10--150 Myr & j \\
  \noalign{\smallskip}
\hline
\hline
\end{tabular}
\begin{list}{}{}
\item[$^{a}$] References for stellar kinematic groups -- a: \cite{Schlieder2010}; b: \cite{Li2000}; c: \cite{Zuckerman2011}; d: \cite{Malo2013};  e: \cite{Riedel2014}; f: \cite{Schlieder2012b}; g: \cite{Torres2002}, h: \cite{Caballero2010};  i: \cite{Klutsch2014}; j: \cite{Cortes2016}; k: \cite{Shkolnik2009} ; l: \cite{Wright2011}; m: \cite{Malo2014a}; n: \cite{Schlieder2012}; o: \cite{Lopez2006}.  
\end{list}
\end{table*}

\subsubsection{H$\alpha$ active early-M dwarfs} 

\begin{table*}
\centering
\caption{Rapidly rotating  H$\alpha$ inactive stars (using $v \sin{i}$).}
\protect\label{tab-rrINactive}
\begin{centering}
\begin{tabular}{l l c c c c c l }
\hline
\hline
   \noalign{\smallskip}
Karmn.           &  Name                 &  SpT          & pEW(H$\alpha$)&  $v \sin{i}$  &  Inst.   & L$_{\rm x}$/L$_{\rm bol}$ &  Multiplicity \\
                 &                       & (M\,V)        & [\AA]         & [km\,s$^{-1}$]&                &      \\
\noalign{\smallskip}
\hline
\noalign{\smallskip}
J00279+223 & LP 349-025 AB      & 8.0   & 0.00  & 48.10 &  & 3.79$\times$10$^{-6}$& Resolved Physical \\
J02465+164 & LP 411-006         & 6.0   & 0.00  & 6.20  &       & ...   & Single \\
J03018-165N & LP 771-095        & 3.0   & 0.37  & 5.50  &       & 9.79$\times$10$^{-6}$& Wide + Resolved Physical \\
J04252+080N & HG 7-207          & 4.0   & -0.10 & 7.50  & CAFE  & 6.39$\times$10$^{-5}$& Wide \\
J04480+170 & LP 416-043         & 0.5   & -0.01 & 8.00  &  & 1.13$\times$10$^{-5}$ & SB \\
J07232+460 & G 107-065          & 0.5   & 0.25  & 6.00  &       & 1.27$\times$10$^{-6}$ & Single \\
J09231+223 & BD+22 2086B        & 0.0   & 0.88  & 7.00  & CAFE & ... & Wide \\
J12199+364 & G 123-036          & 1.0   & -0.48 & 20.33 &  & ... & Single \\
J12274+374 & G 148-061          & 1.5   & -0.47 & 23.90 &  & ... & Single \\
J13378+481 & BD+48 2138         & 0.0   & -0.33 & 8.20  &  & 1.17$\times$10$^{-6}$ & Wide + Resolved Physical \\
J14155+046 & GJ 1182            & 5.0   & 0.10  & 6.80  &  & ... & SB2 \\
J15011+071 & Ross 1030a         & 3.5   & 0.12  & 6.30  & FEROS &  & Single \\
J18395+301 & LP 335-013         & 0.0   & -0.44 & 23.45 &  & ... & Single \\
J18409+315 & BD+31 3330B        & 1.0   & -0.36 & 7.55  &  & ... & Wide \\
J20574+223 & Wolf 3273          & 2.0   & -0.25 & 10.42 &  & ... & Single \\
J21313-097 & BB Cap AB          & 4.5   & -0.18 & 10.00 &  & 4.23$\times$10$^{-6}$ & Resolved Physical (SB?) \\
J23577+233 & GJ 1292            & 3.5   & 0.12  & 5.20  & FEROS & ... & Resolved Visual \\
\hline
\hline
\end{tabular}
\end{centering}
\end{table*}
 
As previously discussed, there is a population of stars in the Carmencita sample with early spectral types ($<$M3\,V) that also have H$\alpha$ in emission.  The most active of these stars (a total of 35 stars) with $v \sin{i}$ measurements are tabulated in Table~\ref{tab-feactive} and can be identified in the upper left area of Fig.~\ref{f-vsini_SpT}.  Of this population, 32 stars are considered to have reliable $v \sin{i}$ detections (i.e. $v \sin{i} >$ 3\,km\,s$^{-1}$).  The occurrence of $v \sin{i}$ measurements that are below our detection limit, combined with H$\alpha$ in emission, suggests that the rotation axes of these stars have a small inclination.  As suggested by \cite{Reiners2012}, inclination angles below $\sim$60\,deg could be sufficient to push the $v \sin{i}$ value below the detection limit.   

The vast majority of the stars classified as ``Active early-M dwarfs" are young stars that are members of young moving groups with reliable ages: $\beta$ Pictoris, 20\,Myr ~\citep{Bell2015};  Carina, 40\,Myr ~\citep{Bell2015}; AB Doradus, 150\,Myr \citep{Bell2015}; Castor, 300\,Myr ~\citep{Mamajek2003}; andUrsa Major, 300\,Myr ~\citep{Soderblom1993}.  The young ages of these stars can explain their high rotation rates.  Since rotation is a key process in the generation of magnetic activity, high rotation rates are also expected to result in increased activity levels.  The remaining two stars (G~005--032 and \object{Wolf~1360}) are likely to be field stars.  

\subsubsection{H$\alpha$ active late-M dwarfs} 

The high levels of activity of later M dwarfs (i.e. cooler than the boundary to full convection) occur because these stars have longer activity lifetimes \citep{Hawley1996,Gizis2002,Silvestri2005,West2008}.  The increased fraction of late-M dwarfs that are  H$\alpha$ active can be explained by longer activity lifetimes of stars later than spectral type of approximately M3.0 \citep{West2008, ReinersM2012}. The reason for longer activity lifetimes in mid- and late-M stars, in contrast to earlier stars, is likely the non-existence of a radiative core, which leads to a dramatic decrease in radius where the percentage of active stars increases; stellar angular momentum loss shows a strong dependence on radius, which can explain the change in activity lifetime \citep[see][]{ReinersM2012}.  We note that sometimes the change in activity lifetime is attributed to a switch between dynamo modes at the boundary from partial to full convection, suggesting that (turbulent) dynamos in fully convective stars exhibit a weaker dependence on rotation.  While this scenario cannot be fully excluded by the observations, we emphasise that in mean-field theory both the $\Omega$ and the $\alpha$ effect depend on rotation \citep[with similar dependence on rotation, see][]{Durney1978, Noyes1984}, and that a sudden increase in activity lifetimes is therefore not a necessary consequence of a switch between dynamo modes at the threshold to full convection.

\subsubsection{Slowly rotating  H$\alpha$ active stars}

From the stars that are classified as  H$\alpha$ active in the Carmencita catalogue (with pEW(H$\alpha) < -0.5\,\AA$), approximately 25\% show low $v \sin{i}$ values with $\le$ 3 km s$^{-1}$ (see Fig~\ref{f-Ha_vsini}).  The measurement of rotational velocity has a dependence on $\sin{i}$.  
As discussed by \cite{Reiners2012}, in a sample of stars with randomly orientated rotation axes, approximately 19\,\% are observed at inclination angles $<$ 36\,deg with $\sin{i}$ $<$ 0.59.  This is slightly below the value of 25\,\% that we find in our results, but is consistent when Poisson measurement errors ($\pm$ 6\,\%) are also included.  However, when we examine the smaller subsample of stars with photometric periods, which do not have a dependence on stellar inclination, there are also several  H$\alpha$ active stars with long rotational periods at all M-dwarf spectral types (see Fig.~\ref{f-RTN_Halpha}). The spread in rotation and activity for stars with similar spectral types shows that the rotational evolution of these stars is complicated and does not follow a single path.  The rotational evolution of M dwarfs is discussed in more detail by ~\cite{Irwin2011}.

\subsubsection{Rapidly rotating  H$\alpha$ inactive stars}
\label{sec:Ha_act_vsini}

Another interesting population of stars in our analysis are the rapidly rotating stars, which are also H$\alpha$ inactive. A total of 17 stars (2\% of stars with $v \sin{i}$ measurements) in the Carmencita catalogue have rotation detections $v \sin{i}$ $>$ 5\,km\,s$^{-1}$ and are also  H$\alpha$ inactive.  These stars were previously shown in Fig.~\ref{f-Ha_vsini} and are listed in Table~\ref{tab-rrINactive}.   This subgroup of rapidly rotating  H$\alpha$ inactive stars has a very high multiplicity fraction and nearly 60\% of stars are in multiple systems. A total of seven stars have X-ray detections that occur in the unsaturated regime, however all of these stars are known binary systems with the exception of \object{G 107-065}. Since both X-ray and H$\alpha$ values were not secured simultaneously, the variable nature of the magnetic activity of the star could explain why there are X-ray detections for some stars that are H$\alpha$ inactive.

The sample of rapidly rotating inactive stars in general have earlier spectral types ($\leq$ M3.5\,V), in contrast to the results of \cite{West2009}, who reported the existence of three  H$\alpha$ inactive rapid rotators with much later spectral types (M6--7\,V).  Additionally, \cite{Jenkins2009} also detected rotation for several  H$\alpha$ inactive stars with spectral types between M3.0\,V and M4.5\,V.  However, \cite{Reiners2012} questioned the reliability of the $v \sin{i}$ values of these stars, as they were very close to, or lower than, the minimum $v \sin{i}$ value that can be reliably detected given the spectral resolution of the instrument that secured the observations (as previously discussed in Section 4.3.4). \cite{Reiners2012} further concluded that all stars with reliable detections of rotation ($v \sin{i}$ $>$ 3\,km\,s$^{-1}$) with spectral types M0--4.5\,V show chromospheric emission.  

As it is necessary to investigate the measurement of $v \sin{i}$ in more detail and to check that H$\alpha$ is actually in emission, we focus on discussing the four stars in our observing sample, namely HG 7-207, BD+22 2086B, Ross 1030a, and GJ 1292.  The $v \sin{i}$ values are well above our minimum detection limit, with a medium S/N, and are considered to be reliable measurements. The metallicity of the rapidly rotating  H$\alpha$ inactive stars in our sample is normal (Passegger et al. 2017, submitted).  The CCF  of the four stars is as expected for a single star, and we cannot exclude that this results from a chance alignment of a binary system.  There are two wide binary systems: \object{BD+22 2086B,} which is a companion orbiting at 8.2\,arcsec of the K star \object{BD+22 2086A}, and \object{HG 7-207,} which forms a wide ($\rho$ = 73.4\,arcsec) pair with the early M-dwarf, SB star HG 7-206. However, there is currently no information on close binarity for either of these systems.  The star \object{GJ~1292} is a visual binary.  The remaining star, \object{Ross~1030}, is observed as an unresolved or single star on high-resolution images ~\citep{Cortes2017}.

The results obtained in this analysis for the stars shown in Table~\ref{tab-rrINactive} do not confirm the conclusion of \cite{Reiners2012}, where all stars with spectral types M0--4.5\,V and with a reliable detection of rotation should show chromospheric emission.  However, as several known binary systems in our new observations show a CCF profile that is indistinguishable from that of a single star, we also cannot exclude the conclusion of \cite{Reiners2012}.  We conclude that the spectral resolution and S/N of our observations, and the observations by ~\cite{West2009} and ~\cite{Jenkins2009} are insufficient to definitely confirm or refute the existence of rapidly rotating  H$\alpha$ inactive stars.   A definitive conclusion can only be made with additional observations of all stars at different rotational phases to exclude binarity and flaring with sufficient spectral resolution and S/N.  

Additionally, there are also three stars as shown in Fig~\ref{f-RTN_Halpha} that are  H$\alpha$ inactive but with fast rotation periods $<$7 days.   For these stars there are no $v \sin{i}$ values available.  To understand if they are a notable class of objects or discrepant points further high cadence measurements over their suspected photometric periods and $v \sin{i}$ values are required.

\begin{table*}
\centering
\caption{Rapidly rotating  H$\alpha$ inactive stars from photometric periods}
\protect\label{tab-rrINactive_photom}
\begin{centering} 
\begin{tabular}{l l c c c c c l }
\hline
   \noalign{\smallskip}
\hline
   \noalign{\smallskip}
Karmn.           &  Name        &  SpT         & $P_{\rm rot}$           & pEW(H$\alpha$)&  $v \sin{i}$  & $L_{\rm x}$/$L_{\rm bol}$ &  Multiplicity \\
                 &              & (M\,V)       & [d]             & [\AA]         & [km\,s$^{-1}$]&                &      \\
\noalign{\smallskip}
\hline
\noalign{\smallskip}
J01531-210       &  BD-21 332 AB         &  1.0          &  2.84$^a$             &  -0.380                &        ...       & 5.96 $\times$10$^{-5}$ &  SB2 \\              
J04413+327       &  NLTT 13733           &  4.0          &  6.57$^b$             &  -0.4          &       ...        & 4.05$\times$10$^{-5}$ &  Resolved Physical \\
J08178+311       &  G 090-052            &  1.0          &  0.97$^b$             &  0.27          &       ...        & ... &  Single        \\              
   \noalign{\smallskip}
\hline
\hline
\end{tabular}
\begin{list}{}{}
\item[] References for rotation periods -- a: ~\cite{Kiraga2012}; b:~\cite{Hartman2011}.  
\end{list}
\end{centering}
\end{table*}


\section{Summary}
\label{sec:magactivityM}

The CARMENES exoplanet survey is monitoring approximately 300 M dwarfs with the aim of detecting small rocky planets orbiting in the HZ of their host stars.  The targets have been carefully selected from an original list of $\sim$2200 stars contained in the Carmencita catalogue of M dwarfs, which comprises normalised H$\alpha$ emission, stellar rotational velocities (including new observations of 480 stars using high-resolution spectroscopy), R$'_{\rm HK}$ measurements, stellar rotation periods, and X-ray luminosities.  In this paper we investigated correlations between these parameters. The vast number of stars are assembled only using limitations on stellar declination ($>$ -23$^\circ$).  In this paper we compiled the largest catalogue of the magnetic activity of M dwarfs to date.  The new results from our analysis can be summarised as follows:

\begin{enumerate}
\item We do not find conclusive evidence that confirms the existence of single  H$\alpha$ inactive stars with detectable rotational velocities. \\

\item The detection of H$\alpha$ in emission closely tracks the saturation rotational velocity observed in X-rays. Our results, together with the results of \cite{West2015}, show that the saturation period grows from about 10d in early-M to {\bf{40 days at mid-M, and to}} 90d in late-M type stars. Stars rotating faster than this show H$\alpha$ in emission, while slower rotators do not show H$\alpha$ in emission.\\

\item The most  H$\alpha$ active and rapidly rotating stars are kinematically young.\\

\item For rotation periods greater than five days, there is a decrease in the strength of normalised H$\alpha$ emission.  There is a gradual transition from  H$\alpha$ active to  H$\alpha$ inactive stars. \\

\item We confirm that normalised X-ray activity is observed at the saturation level for rapid rotators and falls off steeply at the saturation rotation. We show that the transition from saturated normalised H$\alpha$ luminosity to non-saturated emission is a much more gradual decline.  \\

\item Stars with $v \sin{i}$ values $>$ 5 km s$^{-1}$ are in the saturated regime of $L_{\rm x}$/$L_{\rm bol}$, whereas stars with $v \sin{i}$ values less than this can be both in the saturated and unsaturated regime.  \\

\end{enumerate}

These points are valid for the vast majority of the stars in the Carmencita sample.  However, it should also be noted that there exists a small population of stars that comprise $<$2\% of the total number of stars and do not follow these trends.  This is because they are H$\alpha$ inactive with detectable $v \sin{i}$ or short rotation periods.  We have discussed these discrepant stars in the text and can only definitely confirm or refute their status by securing additional high-resolution and high S/N observations over several epochs to exclude binarity and flaring and with high-cadence photometric observations to measure their photometric periods.

\subsection{Subsamples}

For the stellar parameters where there are a high number of values in the Carmencita catalogue, i.e. for pEW(H$\alpha$) and stellar rotational velocities, we also investigated these parameters for (1) the Carmencita catalogue (2) for the CARMENES GTO sample (approximately 300 stars), (3) for the Volume 14 sample (86\% complete for M0 to M5 and comprising approximately 440 stars), and (4) the Volume 7 sample (100\% complete for M0 to M5 and comprising approximately 60 stars).   While the total number of stars in each sample varies significantly, the proportional distribution of spectral types is similar; the majority of stars in each sample have a spectral type of M~4.0\,V to M~4.5\,V.  The detailed analysis of the CARMENES GTO sample in relation to the other samples is important for understanding any biases in the selection effects and stellar activity of the sample that can improve our detection capabilities while also understanding the environments of detected planets.
 
\subsubsection{Normalised H$\alpha$ luminosity}

The dependence of normalised H$\alpha$ luminosity with spectral type varies significantly between the four samples for earlier spectral types (M$<$3.0\,V), while for later spectral types it follows the same trends.  For the early spectral types, the Carmencita sample includes a population of H$\alpha$ active stars that exhibit a large range in activity levels (upper left panel of Fig.~\ref{f-pEW_SpT}) that are not present in the other three samples.  In the Carmencita sample there is also a higher fraction of H$\alpha$ active stars ($\sim$25\%) compared to the other three samples ($\sim$5\%) at these early spectral types (upper left panel of Fig.~\ref{f-frac_active}). When the ages of  stars are investigated we find that the vast majority of the stars are young star candidates (as shown in Table~\ref{tab-feactive}).

At later spectral types, all four samples show the same distributions in H$\alpha$ activity and the fractions of active stars are comparable for each of these samples (Fig.~\ref{f-frac_active}).  In particular, there is an additional small population that numbers a few percent of the total population of very active stars that are present in the Carmencita (4.5\%), CARMENES GTO (4.0\%), and Volume 14 (3.0\%) samples.  They are not present in the Volume 7 sample most likely owing to the small number of stars in this sample.  These very active stars are also shown to be young stars.  

\subsubsection{Stellar rotational velocity}

The correlation of $v \sin{i}$ also shows a similar behaviour as the normalised H$\alpha$ luminosity with many rapidly rotating early-M stars in the Carmencita sample that are not in the other three samples and can be attributed to the young age of these stars.  For later spectral types, all four samples show a wide range of $v \sin{i}$ values, which can range from just over the detection threshold of 3 km s$^{-1}$ to many 10's of  km s$^{-1}$.   

The correlation of normalised H$\alpha$ luminosity with $v \sin{i}$ summarises the above correlations, where similar trends are shown for all four samples.  Stars with later spectral types are shown as H$\alpha$ active stars below H$\alpha$ luminosities of -4.5.  This is due to the physics of the stars, which is accounted for in the spectral type dependence of the $\chi$ value used to convert pEW(H$\alpha$) into normalised H$\alpha$ luminosity (Section 4.2.1).  The presence of a few fast rotating inactive stars in the Carmencita sample is discussed in Section 6.3.4, where we conclude that additional observations are required to definitely confirm or refute their existence.

\subsection{Rotational velocity versus period}

We show a correlation of measured $v \sin{i}$ with the rotation period of the star. These results are particularly important as they comprise results using stellar rotational velocities and rotational periods.  It is important to consider both of these parameters as each has its own intrinsic biases.  For example, the rotational velocities can only be reliably detected for a value of $>$3 km s$^{-1}$, which is equivalent to a rotational period of approximately 9-10 days (excluding the effects of inclination).  This is shown in Fig. ~\ref{f-vsini_Period}.  The advantage is also that the rotational velocity of the star can be obtained directly from one spectrum with sufficient S/N.  On the other hand, the rotational periods can only be measured for stars where there is sufficient stellar activity to result in the presence of starspots that cause the rotational modulation and require many more epochs of observations.  

\subsection{New binary systems}

The radial velocity of the new observations of 480 M dwarfs was measured and from multiple RV measurements of individual stars we identified 44 binaries comprising 11 SB1 type and 32 SB2 type binaries and 1 SB3 system.  The total binarity fraction of the new observations is approximately 9\%.   A total of 28 (7 SB1s, 21 SB2s) of these binary systems are new discoveries.



\begin{acknowledgements}

CARMENES is an instrument for the Centro Astron\'omico Hispano-Alem\'an de Calar Alto (CAHA, Almer\'{\i}a, Spain). CARMENES is funded by the German Max-Planck-Gesellschaft (MPG), the Spanish Consejo Superior de Investigaciones Cient\'{\i}ficas (CSIC), the European Union through FEDER/ERF FICTS-2011-02 funds, and the members of the CARMENES Consortium (Max-Planck-Institut f\"ur Astronomie, Instituto de Astrof\'{\i}sica de Andaluc\'{\i}a, Landessternwarte K\"onigstuhl, Institut de Ci\`encies de l'Espai, Insitut f\"ur Astrophysik G\"ottingen, Universidad Complutense de Madrid, Th\"uringer Landessternwarte Tautenburg, Instituto de Astrof\'{\i}sica de Canarias, Hamburger Sternwarte, Centro de Astrobiolog\'{\i}a and Centro Astron\'omico Hispano-Alem\'an),  with additional contributions by the Spanish Ministry of Economy, the German Science Foundation through the Major Research Instrumentation Programme and DFG Research Unit FOR2544 ``Blue Planets around Red Stars'' (in particular project JE 701/3-1), the Klaus Tschira Stiftung, the states of Baden-W\"urttemberg and Niedersachsen, and by the Junta de Andaluc\'{\i}a.  I.~R., E.~H and J.~C.~M. acknowledge support from the Spanish Ministry of Economy and Competitiveness (MINECO) and the Fondo Europeo de Desarrollo Regional (FEDER) through grant ESP2016-80435-C2-1-R, as well as the support of the Generalitat de Catalunya/CERCA programme. We thank Martin K\"urster, Rafael Luque, Jorge Sanz-Forcada, and Esther Gonz\'alez-\'Alvarez who made contributions to this work. 

\end{acknowledgements}


\bibliographystyle{aa}
\bibliography{iau_journals,vsini_carmenes}


\appendix

\section{Long tables}

\include{vitastroph}

\end{document}

%% file: vitastroph.tex

\begin{longtable}{l cc l cc c c c}
\label{tab-journal_obs}\\
\caption[]{Journal of new high-resolution spectroscopic observations$^{a}$ of Carmencita stars (1 page only -- full table is available online).}\\
   \hline
   \hline
   \noalign{\smallskip}
Karmn	        		& $\alpha$	& $\delta$	& Instr.		& Obs.	& Obs.	& $ t_{\rm exp}$ 	& RV   			& S/N	\\
				& (J2000)		& (J2000)	& 			& date	& UT		& [s]				& [km\,s$^{-1}$] 	&		\\
\noalign{\smallskip}
    \hline
    \noalign{\smallskip}		
 \endfirsthead
\caption[]{Journal of new high-resolution spectroscopic observations$^{a}$ (cont.).}\\ 
  \hline
  \hline
  \noalign{\smallskip}		
Karmn	        		& $\alpha$	& $\delta$	& Instr.		& Obs.	& Obs.	& $ t_{\rm exp}$ 	& RV   			& S/N	\\
				& (J2000)		& (J2000)	& 			& date	& UT		& [s]				& [km\,s$^{-1}$] 	&		\\
  \noalign{\smallskip}
  \hline
  \noalign{\smallskip}
  \endhead
  \noalign{\smallskip}
  \hline
  \endfoot
%
%
J00051+457 & 00:05:10.78 & +45:47:11.6 & CAFE & 2013-06-02 & 03:01:43 & 1200 & --0.58 $\pm$ 0.31 & 95\\
 &  &  & CAFE & 2013-10-27 & 23:52:15 & 1300 & --0.11 $\pm$ 0.40 & 75\\
 &  &  & CAFE & 2013-12-30 & 19:40:00 & 1500 & --0.37 $\pm$ 0.64 & 55\\
J00056+458 & 00:05:40.90 & +45:48:37.5 & CAFE & 2013-10-31 & 00:48:25 & 2700 & --1.30 $\pm$ 0.31 & 108\\
J00077+603 & 00:07:42.64 & +60:22:54.3 & CAFE & 2013-02-03 & 19:10:58 & 1500 & +1.87 $\pm$ 0.16 & 11\\
 &  &  & CAFE & 2013-02-03 & 19:40:49 & 1800 & +1.92 $\pm$ 0.19 & 16\\
J00088+208 & 00:08:53.92 & +20:50:25.2 & CAFE & 2013-10-28 & 00:28:53 & 2200 & +10.41 $\pm$ 0.53 & 30\\
 &  &  & CAFE & 2013-10-28 & 01:07:11 & 2200 & +10.44 $\pm$ 0.10 & 29\\
J00162+198E & 00:16:16.08 & +19:51:51.5 & FEROS & 2013-07-24 & 09:43:11 & 1800 & --1.18 $\pm$ 0.19 & 52\\
 &  &  & FEROS & 2013-07-24 & 10:14:02 & 1800 & --1.33 $\pm$ 0.30 & 38\\
 &  &  & FEROS & 2013-11-20 & 02:35:16 & 1800 & --1.52 $\pm$ 0.07 & 52\\
 &  &  & FEROS & 2013-11-20 & 03:06:07 & 1800 & --1.40 $\pm$ 0.24 & 50\\
J00162+198W & 00:16:14.63 & +19:51:37.6 & FEROS & 2013-07-24 & 08:32:46 & 1600 & +22.87 $\pm$ 0.12 & 75\\
 &  &  & FEROS & 2013-07-24 & 09:00:17 & 1600 & +23.56 $\pm$ 0.07 & 70\\
 &  &  & FEROS & 2013-11-20 & 03:38:21 & 1500 & +24.27 $\pm$ 0.07 & 68\\
 &  &  & FEROS & 2013-11-20 & 04:04:14 & 1500 & +24.58 $\pm$ 0.07 & 72\\
J00286--066 & 00:28:39.48 & --06:39:48.1 & FEROS & 2013-09-16 & 04:47:17 & 1400 & --12.26 $\pm$ 0.25 & 52\\
 &  &  & FEROS & 2013-09-16 & 05:11:27 & 1400 & --12.09 $\pm$ 0.14 & 51\\
J00315--058 & 00:31:35.39 & --05:52:11.6 & FEROS & 2013-01-01 & 01:37:36 & 900 & --9.22 $\pm$ 0.34 & 63\\
 &  &  & FEROS & 2013-01-06 & 01:31:37 & 1800 & --9.36 $\pm$ 0.14 & 90\\
 &  &  & FEROS & 2013-09-16 & 08:36:16 & 1700 & --9.06 $\pm$ 0.12 & 69\\
 &  &  & FEROS & 2013-09-16 & 09:05:27 & 1700 & --9.61 $\pm$ 0.34 & 71\\
J00395+149S & 00:39:33.49 & +14:54:18.9 & HRS & 2012-12-02 & 04:59:32 & 600 & +4.15 $\pm$ 0.08 & 72\\
 &  &  & HRS & 2012-12-02 & 05:10:48 & 600 & +4.07 $\pm$ 0.06 & 71\\
 &  &  & HRS & 2012-12-02 & 05:22:04 & 600 & +4.20 $\pm$ 0.13 & 62\\
J00428+355 & 00:42:48.21 & +35:32:55.4 & CAFE & 2014-08-09 & 04:27:42 & 1800 & +33.11 $\pm$ 0.30 & 58\\
 &  &  & CAFE & 2014-09-14 & 01:48:51 & 1800 & --11.80 $\pm$ 0.56 & 55\\
J00443+126 & 00:44:19.34 & +12:37:02.7 & FEROS & 2013-07-23 & 10:02:51 & 1700 & +33.00 $\pm$ 0.36 & 60\\
 &  &  & FEROS & 2013-07-23 & 10:32:04 & 1700 & +32.61 $\pm$ 0.26 & 69\\
J00502+086 & 00:50:17.53 & +08:37:34.1 & HRS & 2012-09-19 & 09:46:00 & 600 & +19.01 $\pm$ 0.78 & 110\\
 &  &  & HRS & 2012-09-19 & 09:56:30 & 600 & +19.53 $\pm$ 0.87 & 102\\
 &  &  & HRS & 2012-09-19 & 10:07:01 & 600 & +20.12 $\pm$ 1.51 & 86\\
J00566+174 & 00:56:38.42 & +17:27:34.7 & FEROS & 2013-11-22 & 03:42:16 & 1800 & --25.94 $\pm$ 0.21 & 48\\
 &  &  & FEROS & 2013-11-22 & 04:13:08 & 1800 & --26.36 $\pm$ 0.13 & 55\\
 &  &  & FEROS & 2013-11-22 & 04:44:00 & 1800 & --26.63 $\pm$ 0.05 & 45\\
J00570+450 & 00:57:02.61 & +45:05:09.0 & CAFE & 2013-10-31 & 01:40:19 & 2700 & +6.82 $\pm$ 0.23 & 46\\
J00580+393 & 00:58:01.16 & +39:19:11.2 & HRS & 2011-10-11 & 09:00:16 & 600 & --1.40 $\pm$ 0.44 & 117\\
 &  &  & HRS & 2011-10-11 & 09:10:46 & 600 & --1.28 $\pm$ 0.50 & 123\\
 &  &  & HRS & 2011-10-11 & 09:21:15 & 600 & --1.34 $\pm$ 0.54 & 117\\
J01009--044 & 01:00:56.44 & --04:26:56.1 & FEROS & 2013-11-14 & 05:17:59 & 1800 & +28.75 $\pm$ 0.12 & 74\\
 &  &  & FEROS & 2013-11-14 & 05:48:51 & 1800 & +28.86 $\pm$ 0.26 & 68\\
 &  &  & FEROS & 2013-11-14 & 06:19:42 & 1800 & +28.63 $\pm$ 0.39 & 74\\
J01026+623 & 01:02:38.96 & +62:20:42.2 & CAFE & 2013-01-21 & 18:43:07 & 900 & --5.31 $\pm$ 0.22 & 55\\
J01256+097 & 01:25:36.66 & +09:45:24.4 & FEROS & 2013-01-01 & 01:59:49 & 1200 & --14.11 $\pm$ 0.28 & 55\\
 &  &  & FEROS & 2013-09-16 & 06:12:33 & 1800 & --14.76 $\pm$ 0.59 & 34\\
J01384+006 & 01:38:29.98 & +00:39:05.9 & FEROS & 2013-11-15 & 05:30:20 & 1800 & +0.28 $\pm$ 0.02 & 118\\
J01390--179 & 01:39:01.20 & --17:57:02.7 & FEROS & 2013-01-01 & 01:21:03 & 600 & +28.62 $\pm$ 4.14 & 138\\
J01449+163 & 01:44:58.52 & +16:20:39.7 & CAFE & 2014-02-13 & 22:44:55 & 900 & +5.04 $\pm$ 0.72 & 54\\

\end{longtable}
\begin{list}{}{}
\item[$^{a}$] Carmencita identifier, 2MASS equatorial coordinates, used instrument, observing night (from noon to noon), UT of mid-point of observation, exposure time, radial velocity, and signal-to-noise ratio of the corresponding spectrum.
\end{list}

%

\begin{longtable}{ll c cl cl l}
\label{tab-rotACT}\\
\caption[]{Name, spectral type, H$\alpha$ pseudo-equivalent width, and rotational velocity of stars observed with CAFE and FEROS (1 page only -- full table is available online).}\\
   \hline
   \hline
   \noalign{\smallskip}
Karmn	        	& Name			& SpT		& pEW(H$\alpha$)	& Ref. 	& $v \sin{i}$		& Ref. 	& Instrument\\
			& 				& (M\,V)		& [\AA]			&		& [km\,s$^{-1}$]	& 	& \\
\noalign{\smallskip}
    \hline
    \noalign{\smallskip}		
 \endfirsthead
\caption[]{Name, spectral type, H$\alpha$ pseudo-equivalent width, and rotational velocity of stars observed with CAFE and FEROS (cont.).}\\ 
  \hline
  \hline
  \noalign{\smallskip}		
Karmn	        	& Name			& SpT		& pEW(H$\alpha$)	& Ref. 	& $v \sin{i}$		& Ref. 	& Instrument\\
			& 				& (M\,V)		& [\AA]			&		& [km\,s$^{-1}$]	& 	& \\
  \noalign{\smallskip}
  \hline
  \noalign{\smallskip}
  \endhead
  \noalign{\smallskip}
  \hline
  \endfoot
%
%
J00051+457 	&  GJ 2 		&  1.0 		&   +0.3 $\pm$ 0.1 	&   	&   $<$ 3.0 &  	& CAFE\\
J00056+458 	&  HD 38B 		&  0.0 		&   +0.4 $\pm$  0.1 	&   	&   $<$ 3.0 &  	& CAFE\\
J00077+603 	&  G 217-032 		&  4.0 		&  --4.6 $\pm$ 0.7 	&   	&   7.8 &  	& CAFE\\
J00088+208 	&  LP 404-033 		&  4.5 		&  --5.1 $\pm$ 0.4 	&   	&   9.6 &  	& CAFE\\
J00162+198E 	&  LP 404-062 		&  4.0 		&   +0.1 $\pm$ 0.1 	&   	&   $<$ 3.0 &  	& FEROS\\
J00162+198W 	&  EZ Psc 		&  4.0 		&  --4.3 $\pm$ 0.2 	&   	&   3.6 &  	& FEROS\\
J00286-066 	&  GJ 1012 		&  4.0 		&   +0.1 $\pm$ 0.1 	&   	&   $<$ 3.0 &  	& FEROS\\
J00315-058 	&  GJ 1013 		&  3.5 		&   +0.1 $\pm$ 0.1 	&   	&   $<$ 3.0 &  	& FEROS\\
J00443+126 	&  G 032-044 		&  3.5 		&   +0.1 $\pm$ 0.1 	&   	&   $<$ 3.0 &  	& FEROS\\
J00566+174 	&  GJ 1024 		&  4.0 		&   +0.0 $\pm$ 0.1 	&   	&   $<$ 3.0 &  	& FEROS\\
J00570+450 	&  G 172-030 		&  3.0 		&   +0.1 $\pm$ 0.1 	&   	&   $<$ 3.0 &  	& CAFE\\
J01009-044 	&  GJ 1025 		&  4.0 		&   +0.1 $\pm$ 0.1 	&   	&   $<$ 3.0 &  	& FEROS\\
J01026+623 	&  BD+61 195 		&  1.5 		&   +0.2 $\pm$ 0.1 	&   	&   $<$ 3.0 &  	& CAFE\\
J01256+097 	&  Wolf 66 		&  4.0 		&   +0.1 $\pm$ 0.1 	&   	&   $<$ 3.0 &  	& FEROS\\
J01384+006 	&  G 071-024 		&  2.0 		&   +0.4 $\pm$ 0.1 	&   	&   $<$ 3.0 &  	& FEROS\\
J01390-179 	&  BL Cet + UV Cet 	&  5.0 		&  --4.4 $\pm$ 0.1 	&   	&  32.0 &  	& FEROS\\
J01449+163 	&  Wolf 1530 		&  4.0 		&   +0.1 $\pm$ 0.1 	&   	&   $<$ 3.0 &  	& CAFE\\
J01466-086 	&  LP 708-416 		&  4.0 		&   +0.2 $\pm$ 0.1 	&   	&   $<$ 3.0 &  	& FEROS\\
J02002+130 	&  TZ Ari 		&  3.5 		&  --2.2 $\pm$ 0.4 	&   	&   $<$ 3.0 &  	& CAFE\\
J02026+105 	&  RX J0202.4+1034 	&  4.5 		&  --4.1 $\pm$ 0.1 	&   	&   6.0 &  	& FEROS\\
J02050-176	&  BD-18 359 AB 	&  2.5 		&   +0.3 $\pm$ 0.1 	&   	&   $<$ 3.0 &  	& FEROS\\
J02070+496 	&  G 173-037 		&  3.5 		&  --0.3 $\pm$ 0.2 	&   	&   $<$ 3.0 &  	& CAFE\\
J02096-143 	&  LP 709-040 		&  2.5 		&   +0.3 $\pm$ 0.1 	&   	&   $<$ 3.0 &  	& FEROS\\
J02116+185 	&  G 035-032 		&  3.0 		&   +0.2 $\pm$ 0.1 	&   	&   $<$ 3.0 &  	& FEROS\\
J02123+035 	&  BD+02 348 		&  1.5 		&   +0.3 $\pm$0.1 	&   	&   $<$ 3.0 &  	& FEROS\\
J02129+000 	&  G 159-046 		&  4.0 		&  --2.3 $\pm$ 0.1 	&   	&   $<$ 3.0 &  	& FEROS\\
J02190+353 	&  Ross 19 		&  3.5 		&   +0.2 $\pm$ 0.4 	&   	&   $<$ 3.0 &  	& CAFE\\
J02222+478 	&  BD+47 612 		&  0.5 		&   +0.5 $\pm$  0.1 	&   	&   4.0 &  	& CAFE\\
J02336+249 	&  GJ 102 		&  4.0 		&  --2.7 $\pm$ 0.1 	&   	&   3.1 &  	& FEROS\\
J02358+202 	&  BD+19 381 		&  2.0 		&   +0.4 $\pm$ 0.1 	&   	&   1.6 &  	& FEROS\\
J02362+068 	&  BX Cet 		&  4.0 		&   +0.0 $\pm$ 0.1 	&   	&   $<$ 3.0 &  	& FEROS\\
J02534+174 	&  NLTT 9223 		&  3.5 		&   +0.3 $\pm$ 0.1 	&   	&   $<$ 3.0 &  	& FEROS\\
J02581-128 	&  LP 711-032 		&  2.5 		&   +0.2 $\pm$ 0.1 	&   	&   $<$ 3.0 &  	& FEROS\\
J03026-181 	&  GJ 121.1 		&  2.5 		&  --0.2 $\pm$ 0.1 	&   	&   $<$ 3.0 &  	& FEROS\\
J03040-203 	&  LP 771-077 		&  3.5 		&   +0.1 $\pm$ 0.1 	&   	&   $<$ 3.0 &  	& FEROS\\
J03102+059 	&  EK Cet 		&  2.0 		&   +0.3 $\pm$ 0.1 	&   	&   $<$ 3.0 &  	& FEROS\\
J03217-066 	&  G 077-046 		&  2.0 		&   +0.0 $\pm$ 0.1 	&   	&   $<$ 3.0 &  	& FEROS\\
J03233+116 	&  G 005-032 		&  2.5 		&  --2.4 $\pm$ 0.1 	&   	&   $<$ 3.0 &  	& FEROS\\
J03242+237 	&  GJ 140C 		&  2.0 		&   +0.2 $\pm$ 0.1 	&   	&   $<$ 3.0 &  	& CAFE\\
J03317+143 	&  GJ 143.3 		&  2.0 		&   +0.3 $\pm$ 0.1 	&   	&   $<$ 3.0 &  	& FEROS\\
J03366+034 	&  [R78b] 233 		&  4.5 		&  --8.1 $\pm$ 0.2 	&   	& 130.9 &  	& FEROS\\
J03438+166 	&  BD+16 502A 		&  0.0 		&   +0.5 $\pm$ 0.1 	&   	&   $<$ 3.0 &  	& FEROS\\
J03463+262 	&  HD 23453 		&  0.0 		&   +0.4 $\pm$ 0.1 	&   	&   $<$ 3.0 &  	& FEROS\\
J03473-019 	&  G 080-021 		&  3.0 		&  --3.7 $\pm$ 0.1 	&   	&   5.5 &  	& FEROS\\
J03507-060 	&  GJ 1065 		&  3.5 		&   +0.1 $\pm$ 0.1 	&   	&   $<$ 3.0 &  	& FEROS\\
J03526+170 	&  Wolf 227 		&  4.5 		&  --0.6 $\pm$ 0.2 	&   	&   $<$ 3.0 &  	& FEROS\\
J03531+625 	&  Ross 567 		&  3.0 		&   +0.2 $\pm$ 0.1 	&   	&   $<$ 3.0 &  	& CAFE\\
J03598+260 	&  Ross 873 		&  3.0 		&   +0.2 $\pm$ 0.2 	&   	&   $<$ 3.0 &  	& CAFE\\
J04148+277 	&  HG 8-1 		&  3.5 		&  --1.6 $\pm$ 0.3 	&   	&   $<$ 3.0 &  	& CAFE\\
J04153-076 	&  $\o^{02}$ Eri C 	&  4.5 		&  --4.1 $\pm$ 0.2 	&   	&   $<$ 3.0 &  	& FEROS\\
\end{longtable}

\begin{longtable}{ll c cl cll}
\label{tab-rotCAR}\\
\caption[]{Name, spectral type, H$\alpha$ pseudo-equivalent width, rotational velocity and log R'HK of the full Carmencita catalogue (1 page only -- full table is available online).}\\
   \hline
   \hline
   \noalign{\smallskip}
Karmn	        	& Name			& SpT		& pEW(H$\alpha$)	& Ref. 	& $v \sin{i}$		& Ref. 	& log R'HK \\
			& 				& (M\,V)		& [\AA]			&		& [km\,s$^{-1}$] & \\
\noalign{\smallskip}
    \hline
    \noalign{\smallskip}		
 \endfirsthead
\caption[]{Name, spectral type, H$\alpha$ pseudo-equivalent width, rotational velocity and log R'HK of the full Carmencita catalogue (cont.).}\\ 
  \hline
  \hline
  \noalign{\smallskip}		
Karmn	        	& Name			& SpT		& pEW(H$\alpha$)	& Ref. 	& $v \sin{i}$		& Ref. 	& log R'HK  \\
			& 				& (M\,V)		& [\AA]			&		& [km\,s$^{-1}$]	&\\
  \noalign{\smallskip}
  \hline
  \noalign{\smallskip}
  \endhead
  \noalign{\smallskip}
  \hline
  \endfoot
J00012+139S 	 & 	  BD+13 5195B 	 & 	 0.0 	 & 	 -0.41   	 & 	 a 	 & 	  	  	  	 & 	   \\
J00033+046 	 & 	  StKM 1-2199 	 & 	 1.5 	 & 	 -0.39   	 & 	 a 	 & 	  	  	  	 & 	   \\
J00051+457 	 & 	  GJ 2 	         & 	 1.0 	 & 	 -0.128 	 & 	 b 	 & 	$<$ 2.0 		  	 & 	 1  \\
J00056+458 	 & 	  HD 38B 	 & 	 0.0 	 & 	 0.43 $\pm$ 0.07 	 & 	 b 	 & 	 = 3.3 $\pm$ 2.3 	 & 	 2  \\
J00067-075 	 & 	  GJ 1002 	 & 	 5.5 	 & 	 0.0 $\pm$ 0.0 	 & 	 b 	 & 	$<$ 2.0 	 & 	  	  	 1  \\
J00077+603 	 & 	  G 217-032 	 & 	 4.0 	 & 	 -4.5 $\pm$ 0.6 	 & 	 b 	 & 	 = 7.8 	 	  	 & 	 3  \\
J00078+676 	 & 	  2M J00075079+6736255 	 & 	 2.0 	 & 	 -3.11  	 & 	 a 	  	  	 & 	  	 & 	   \\
J00081+479 	 & 	  1RXS J000806.3+475659 	 & 	 4.0 	 & 	 -2.93  	 & 	 a 	 	  	 & 	  	 & 	   \\
J00084+174 	 & 	  MCC 351 	 & 	 0.0 	 & 	 0.4 $\pm$ 0.07 	 & 	 d 	 & 	  	  	  	 & 	   \\
J00088+208 	 & 	  LP 404-033 	 & 	 4.5 	 & 	 -5.2 $\pm$ 0.3 	 & 	 b 	 & 	 = 9.6 	  	  	 & 	 3  \\
J00110+052 	 & 	  G 031-029 	 & 	 1.0 	 & 	 -0.35  	 & 	 a 	 & 	  	  	  	 & 	   \\
J00115+591 	 & 	  LSR J0011+5908 	 & 	 5.5 	 & 	 -1.6 $\pm$ 0.4 	 & 	 e 	 & 	  	  	  	 & 	   \\
J00118+229 	 & 	  LP 348-040 	 & 	 3.5 	 & 	 -0.5 $\pm$ 0.2 	 & 	 e 	 & 	  	  	  	 & 	   \\
J00119+330 	 & 	  G 130-053 	 & 	 3.5 	 & 	 -0.3 $\pm$ 0.2 	 & 	 e 	 & 	  	  	  	 & 	   \\
J00122+304 	 & 	  2M J00121341+3028443 	 & 	 4.5 	 & 	 -8.7 $\pm$ 0.5 	 & 	 e 	 & 	  	  	  	 & 	   \\
J00131+703 	 & 	  TYC 4298-613-1 	 & 	 2.0 	 & 	 -0.37  	 & 	 a 	 & 	  	  	  	 & 	   \\
J00132+693 	 & 	  GJ 11 AB 	 & 	 3.0 	 & 	 0.01 $\pm$ 0.13 	 & 	 d 	 & 	  	  	  	 & 	   \\
J00133+275 	 & 	  2M J00131951+2733310 	 & 	 4.5 	 & 	 -4.0 $\pm$ 0.4 	 & 	 e 	 & 	  	  	  	 & 	   \\
J00136+806 	 & 	  G 242-048 	 & 	 1.5 	 & 	 0.00 $\pm$ 0.2 	 & 	 e 	 & 	 = 2.47 $\pm$ 0.3 	 & 	 4  \\
J00137+806 	 & 	  LP 012-304 	 & 	 5.0 	 & 	 -3.20 	 & 	 c 	 & 	  	 & 	  	  	   \\
J00154-161 	 & 	  GJ 1005 AB 	 & 	 4.0 	 & 	 0.224 	 & 	 c 	 & 	$<$ 3.0 	 & 	  	  	 5  \\
J00156+722 	 & 	  G 242-049 	 & 	 2.0 	 & 	 -0.22 	 & 	 a 	 & 	  	 & 	  	  	   \\
J00158+135 	 & 	  GJ 12 	 & 	 3.0 	 & 	 0.326 	 & 	 c 	 & 	  	 & 	  	  	   \\
J00159-166 	 & 	  BPS CS 31060-0015 	 & 	 4.0 	 & 	 -4.36 	 & 	 f 	 & 	  	  	  	 & 	   \\
J00162+198E 	 & 	  LP 404-062 	 & 	 4.0 	 & 	 -0.035 $\pm$ 0.005 	 & 	 b 	 & 	$<$ 2.0 	  	  	 & 	 1  \\
J00162+198W 	 & 	  EZ Psc 	 & 	 4.0 	 & 	 -4.155 $\pm$ 0.064 	 & 	 b 	 & 	 = 3.6 	  	  	 & 	 3  \\
J00169+051 	 & 	  GJ 1007 	 & 	 4.5 	 & 	0.00	 & 	 c 	 & 	  	 & 	  	  	   \\
J00169+200 	 & 	  G 131-047 	 & 	 3.5 	 & 	 -1.31 	 & 	 c 	 & 	 = 22.0 	 & 	  	  	 6  \\
J00173+291 	 & 	  Ross 680 	 & 	 2.0 	 & 	 0.35 $\pm$ 0.08 	 & 	 d 	 & 	  	  	  	 & 	   \\
J00176-086 	 & 	  BD-09 40 	 & 	 0.0 	 & 	 0.23 $\pm$ 0.06 	 & 	 d 	 & 	  	  	  	 & 	   \\
J00179+209 	 & 	  LP 404-081 	 & 	 1.0 	 & 	 -0.52	 & 	 a 	 & 	  	  	  	 & 	   \\
J00182+102 	 & 	  GJ 16 	 & 	 1.5 	 & 	  	 & 	 c 	 & 	 = 2.34 $\pm$ 0.3 	 & 	 4  \\
J00183+440 	 & 	  GX And 	 & 	 1.0 	 & 	 -0.131 $\pm$ 0.004 	 & 	 b 	 & 	$<$ 2.0 	  	  	 & 	 1  \\
J00184+440 	 & 	  GQ And 	 & 	 3.5 	 & 	 0.028 $\pm$ 0.009 	 & 	 b 	 & 	$<$ 2.0 	  	  	 & 	 1  \\
J00188+278 	 & 	  LP 292-066 	 & 	 4.0 	 & 	 -0.89 	 & 	 c 	 & 	  	 & 	  	  	   \\
J00201-170 	 & 	  LP 764-108 	 & 	 1.0 	 & 	 0.3 $\pm$ 0.08 	 & 	 d 	 & 	$<$ 3.0 	  	  	 & 	 5  \\
J00204+330 	 & 	  LP 292-067 	 & 	 5.5 	 & 	 0.00	 & 	 c 	 & 	$<$ 4.5 	  	  	 & 	 7  \\
J00207+596 	 & 	  [I81] M 134 	 & 	 2.5 	 & 	 -0.20	 & 	 a 	 & 	  	  	  	 & 	   \\
J00209+176 	 & 	  StKM 1-25 	 & 	 0.0 	 & 	 -0.35	 & 	 a 	 & 	  	  	  	 & 	   \\
\end{longtable}
